\documentclass[german,english,article,nojss]{jss}
\usepackage[T1]{fontenc}
\usepackage[latin9]{inputenc}
\usepackage{booktabs}
\usepackage{units}
\usepackage{mathtools}
\usepackage{url}
\usepackage{amsmath}
\usepackage{amssymb}
\usepackage{graphicx}
\usepackage[authoryear]{natbib}

\makeatletter

\title{A guide on spectral methods applied to discrete data\\
Part I: One-dimensional signals}
\author{Martin Seilmayer\\
HZDR\And Matthias Ratajczak\\
HZDR}
\Abstract{Spectral analysis in conjunction with discrete data in one and more
dimensions can become a challenging task, because the methods are
sometimes difficult to understand. This paper intends to provide an
overview about the usage of the Fourier transform, its related methods
and focuses on the subtleties to which the users must pay attention.
Typical questions, which are often addressed to the data, will be
discussed. Such a problem can be the issue of frequency or band limitation
of the signal. Or the source of artifacts might be of interest, when
a Fourier transform is carried out. Another topic is the issue with
fragmented data. Here, the Lomb-Scargle method will be explained with
an illustrative example to deal with this special type of signal.
Furthermore, a challenge encountered very often is the time-dependent
spectral analysis, with which one can evaluate the point in time when
a certain frequency appears in the signal. The information to solve
such problems and to answer this questions is spread over many disciplines
ranging from mathematics, electrical engineering, economic science
to astrophysics. The goal of the first part of this paper is to collect
the important information about the common methods to give the reader
a guide on how to use these for application on one-dimensional data.
The second part of this paper will then address the two- and more-dimensional
data. The introduced methods are supported by the \pkg{spectral}
package, which has been published for the statistical environment
\proglang{R} prior this article.}
\Plainauthor{Martin Seilmayer, Matthias Ratajczak}
\Plainkeywords{spectral methods, Fourier transform, Hilbert transform, Lomb-Scargle,
R}
\Address{Martin Seilmayer\emph{ and} Matthias Ratajczak\\
Helmholtz-Zentrum Dresden - Rossendorf e.V.\\
Institut für Fluiddynamik\\
Bautzner Landstraße 400\\
01328 Dresden\\
E-mail: \email{m.seilmayer@hzdr.de}\\
E-mail: \email{m.ratajczak@hzdr.de}}
\providecommand{\tabularnewline}{\\}

\usepackage{tikz}
\renewcommand{\pgfimage}[2][]{\includegraphics[#1]{#2}}

\usepackage{letltxmacro}
\LetLtxMacro\latexincludegraphics\includegraphics
\renewcommand{\includegraphics}[2][]{\centering\latexincludegraphics[#1]{#2}}

\usepackage{placeins}
\AtBeginDocument{%
  \expandafter\renewcommand\expandafter\subsection\expandafter
    {\expandafter\@fb@secFB\subsection}%
  \newcommand\@fb@secFB{\FloatBarrier
    \gdef\@fb@afterHHook{\@fb@topbarrier \gdef\@fb@afterHHook{}}}%
  \g@addto@macro\@afterheading{\@fb@afterHHook}%
  \gdef\@fb@afterHHook{}%
}

\@ifundefined{showcaptionsetup}{}{%
 \PassOptionsToPackage{caption=false}{subfig}}
\usepackage{subfig}
\makeatother

\usepackage{babel}
\begin{document}
\global\long\def\diff{\mathop{}\!\mathrm{d}}

\global\long\def\e{\mathrm{e}}

\global\long\def\im{\mathrm{i}}

\global\long\def\pvec#1{\vec{#1}\mkern2mu\vphantom{#1}}

\binoppenalty=10000
\relpenalty=10000
\maketitle

\section[Introduction]{Introduction}

Because the field of interest is as wide as the amount of information
about spectral methods, it becomes necessary to compose this article
out of the most important methods. It is split into three main parts,
which point at different levels of abstraction and complexity. The
second chapter will give a brief and short overview over the mathematical
principles and ideas behind the common methods of spectral analysis.
For further reading the appendix contains a selection of derivations.
Remember, the intention of this script is not the completeness of
mathematical proofs. Its focus lies on the application of methods
on discrete data and the mistakes to be avoided. The third chapter
will focus on simple one-dimensional methods, their properties, advantages
and drawbacks. 

The individual sections are supported by examples which are programmed
in the statistical language \proglang{R} \citep{R} with the help
of the \pkg{spectral} \citep{spectral} package which has been published
prior this article.

\section[Mathematical concepts]{Mathematical concepts}

This chapter introduces the basic ideas and principles which lead
to the methods of spectral analysis for discrete data. Thereby, the
focus lays on measurements and time series of physical processes.
For now, utilizing the time as variable makes it easier to understand
the explained methods. Of course, the time variable $t$ might be
replaced by the space variable $x$, if a distribution of spatial
events is of interest. So in this sense space and time are meant to
be the location where the signal is defined.

The following definitions and explanations are short and sweet. For
further reading it is suggested to refer to ``The Scientist and Engineer's
Guide to Digital Signal Processing'' by \citet{smith1997thescientist}
or to ``Time-Frequency Analysis'' by \citet{cohen1995timefrequency}.

\subsection[Definition of a signal]{\label{subsec:Definition-of-a-Signal}Definition of a signal}

In the context of this article a signal $s(t)$ corresponds to a physical
measure $p(t)$ and is therefore real-valued and causal. This means,
that with the measurement of the process $p(t)$ the signal $s(t)$
starts to exist at a certain point in time and ends later when the
measurement is finished. 
\begin{figure}
\centering\input{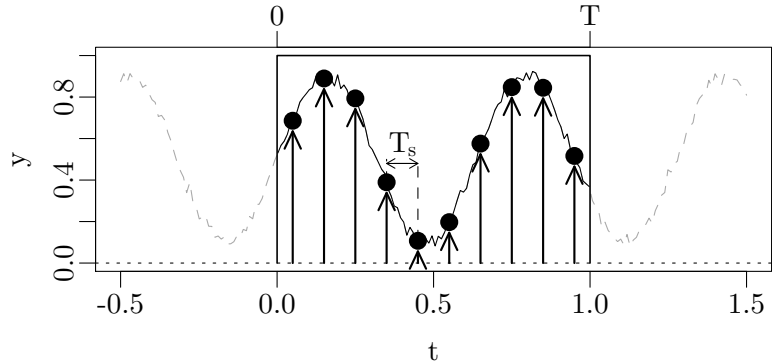}

\caption{\label{fig:Simple_signal}The arbitrary signal $p(t)=0.4\cos(2\pi\cdot1.5t-\pi/2)+\mathcal{N}(0,0.02)+0.5$
with overlapped normal distributed noise is shown here. Only within
the rectangular window the measurement takes place and the signal
$s(t)$ can exist. Outside this window the physical process might
go on, but is not recognized anymore. The points and arrows mark the
sampling of that signal.}
\end{figure}
With that in mind, the signal function $s(t)$ represents a slice
of the length $T$ for times $t>0$. This can be properly defined
as follows:
\begin{equation}
s(t)=\begin{cases}
p(t) & 0<t\leq T,\\
0 & \text{elsewhere.}
\end{cases}\label{eq:1_signalDef}
\end{equation}
The simple noisy signal in Figure~\ref{fig:Simple_signal} illustrates
that. For this example, the underlying physical process $p(t)$ could
be the temperature, which is measured in the time range between $0<t\leq1$.
Evidently the temperature of an object exists before and after the
measurement takes place, so the content of $s(t)$ only maps to a
time interval of this process.

With respect to a digital representation of information, the measurement
of $p(t)$ takes place by acquiring data in the form of sampling the
continuous signal at $N$ different instances in time. The resulting
sampled signal series 
\begin{eqnarray}
\hat{s}{}_{n} & = & \sum_{n=0}^{N-1}s(n\cdot T_{s})\cdot\vec{e}_{n}\label{eq:1_DataVector}
\end{eqnarray}
is the data vector $\hat{s}{}_{n}$. Here, the index $n$ addresses
the $n$\textsuperscript{th} element in this $N$-dimensional data
vector $\hat{s}$. In equation \eqref{eq:1_DataVector} the basis
vector $\vec{e}_{n}$ ensures that $\hat{s}$ becomes a vector in
a mathematical sense, which simply represents an ordered list of elements.
Remember, the absolute value $\left|\vec{e}_{n}\right|=1$ is always
equal to one, so the sum only helps to iterate along the continuous
signal $s(t)$. 

\subsection[The fundamental idea of frequency analysis]{\label{subsec:Frequency-analysis}The fundamental idea of frequency
analysis}

To explain the idea of frequency analysis right from the beginning,
it is necessary to redefine the signal function $s(t)$. From now
on, the signal represents a sum of sinusoidal functions in the form:
\begin{align}
s(t) & =\sum_{i}A_{i}\cdot\cos(\underset{\omega_{i}}{\underbrace{2\pi f_{i}}}t+\varphi_{i}),\label{eq:1_sinus_s}
\end{align}
which, in the final consequence, leads to the expansion into orthogonal
functions. 

The aim of frequency analysis is to calculate the amplitude and phase
spectrum of a signal. The result of this procedure is a mapping of
amplitudes to their corresponding frequencies. If a measured time
series of a physical process is the sum of many individual sinusoidal
functions, it could be of interest to estimate each of the amplitudes
$A_{i}$ and phases $\varphi_{i}$ for a given frequency $f_{i}$. 

The first approach to do that would be to estimate each $A$ and $\varphi$
by a least square fit against a sinusoidal model for each frequency
of interest. This is a legitimate attempt with certain drawbacks and
some outstanding advantages, which are discussed below in section
\ref{subsec:Gapped-and-irregularly-sampled-data}. 

The root of a more general method to determine the spectral components
of a signal are the two common addition theorems \citep{bronvstejn2001taschenbuch}
\begin{eqnarray}
\cos\alpha\cdot\cos\beta & = & \frac{1}{2}\left(\cos(\alpha-\beta)+\cos(\alpha+\beta)\right),\label{eq:trig_ident_1}\\
\cos\alpha\cdot\sin\beta & = & \frac{1}{2}\left(\sin(\beta-\alpha)+\sin(\alpha+\beta)\right),\label{eq:trig_ident_2}
\end{eqnarray}
which mark the starting point of the following explanations.

Given a signal in the continuous domain
\begin{equation}
s(t)=A\cdot\cos(\omega t+\varphi),
\end{equation}
then multiplying this with $\cos(\omega_{0}t)$ as well as $\sin(\omega_{0}t)$
and finally integrating over all times yields for the $\cos$-term:
\begin{eqnarray}
\int_{-\infty}^{\infty}s(t)\cdot\cos(\omega_{0}t)\diff t & = & \frac{A}{2}\left(\int_{-\infty}^{\infty}\cos\left((\omega-\omega_{0})t+\varphi\right)\diff t+\underset{=0}{\underbrace{\int_{-\infty}^{\infty}\cos\left((\omega+\omega_{0})t+\varphi\right)\diff t}}\right),\label{eq:1_QDT_cos_pre}\\
 & = & \begin{cases}
\frac{A}{2}\cos(\varphi) & ,\omega=\omega_{0}\\
\rightarrow0 & ,\omega\neq\omega_{0}
\end{cases},\label{eq:1_QDT_cos}\\
 & = & R\qquad\text{(real part \emph{or} inphase component)}.\nonumber 
\end{eqnarray}
which is going to be named $R$. Herein, the variable $\omega_{0}$
denotes the frequency of interest. The $\sin$-term is very similar
\begin{eqnarray}
\int_{-\infty}^{\infty}s(t)\cdot\sin(\omega_{0}t)\diff t & = & \frac{A}{2}\left(\int_{-\infty}^{\infty}\sin\left((\omega_{0}-\omega)t+\varphi\right)\diff t+\underset{=0}{\underbrace{\int_{-\infty}^{\infty}\sin\left((\omega+\omega_{0})t+\varphi\right)\diff t}}\right),\label{eq:1_QDT_sin_pre}\\
 & = & \begin{cases}
\frac{A}{2}\sin(\varphi) & ,\omega=\omega_{0}\\
\rightarrow0 & ,\omega\neq\omega_{0}
\end{cases},\label{eq:1_QDT_sin}\\
 & = & I\qquad\text{(imaginary part \emph{or} quadrature component)}\nonumber 
\end{eqnarray}
and is called $I$. In both equations on the right hand side the right
integral vanishes to zero, because each positive area element of a
sin-function can be mapped to a negative area element. Only in the
case of $\omega_{0}=\omega$ a phase-dependent part is left,otherwise
the limit of the integral converges to zero. 

With the identity $\sin^{2}\alpha+\cos^{2}\alpha=1$ the amplitude
can be calculated as
\begin{equation}
A=2\sqrt{R^{2}+I^{2}}.\label{eq:1_QDT_A}
\end{equation}
The phase information is then gained by 
\begin{equation}
\varphi=\arctan\left(\frac{I}{R}\right)\label{eq:1_QDT_phi}
\end{equation}
The introduced procedure is called quadrature demodulation technique
(QDT). Finally, by evaluating the amplitude $A$ at many different
frequencies $\omega_{0}$, a periodogram emerges. Note, the $R$ and
$I$ part can also be interpreted as the real and imaginary part of
a signal. The complex exponential formulation then corresponds to
\begin{equation}
s(t)=\mathfrak{Re}\left(A\cdot\e^{\im\omega t+\varphi_{i}}\right)
\end{equation}
with the identity $\e^{\im\phi}=\cos\phi+\im\sin\phi$.

One of the main disadvantages arise for finite time ranges $0<t\leq T$.
In that case, this method produces a truncation error $\varepsilon$,
which depends on the remainder after the modulus division 
\begin{equation}
n=(2\pi f\cdot T)\,\text{mod}\,(2\pi)\label{eq:1_Tmod2pi}
\end{equation}
with respect to the total time interval $T$. In other words, if the
period $2\pi/\omega$ of the signal does not fit completely in the
integration range, the integrals containing $\omega+\omega_{0}$ will
not vanish completely. However, remember the right term of equation
\eqref{eq:1_QDT_cos_pre}. Here the integral can be split into an
$n\cdot2\pi$ periodic part and a residue. The periodic integral of
$\cos\left((\omega+\omega_{0})t+\varphi\right)$ vanishes, 
\begin{align}
R & =\frac{A}{2}\left.\left(\int_{0}^{T}\cos\left((\omega-\omega_{0})t+\varphi\right)\diff t+\int_{0}^{T}\cos\left((\omega+\omega_{0})t+\varphi\right)\diff t\right)\right|_{\omega=\omega_{0}},\nonumber \\
 & =\frac{A}{2}\left(\cos(\varphi)+\underset{\rightarrow0}{\underbrace{\int_{0}^{\frac{n\cdot2\pi-\varphi}{\omega+\omega_{0}}}\cos\left((\omega+\omega_{0})t+\varphi\right)\diff t}}+\underset{=\varepsilon}{\underbrace{\int_{\frac{n\cdot2\pi-\varphi}{\omega+\omega_{0}}}^{T}\cos\left((\omega+\omega_{0})t+\varphi\right)\diff t}}\right),\nonumber \\
 & =\frac{A}{2}\left(\cos(\varphi)+\varepsilon\right),\label{eq:1_QDT_error2}
\end{align}
whereas a residue $\varepsilon$ remains in the second integral from
the last $n\cdot2\pi$ period up to the full length $T$. Obviously,
a minimum of $\varepsilon$ is reached if $T=n\cdot2\pi$. All this
holds if $\omega=\omega_{0}$. If this is not the case, also the integral
$\int\cos\left((\omega-\omega_{0})t+\varphi\right)$ splits into two
parts from $0$ to $n\cdot2\pi$ and a residue, which leads to a second
error $\varepsilon_{2}$. Finally, with $\omega\approx\omega_{0}$
this leakage effect will produce artificial amplitudes in the nearest
neighborhood of $\omega_{0}$. Any way, note that from inserting the
result \eqref{eq:1_QDT_error2} in equation \eqref{eq:1_QDT_A} or
\eqref{eq:1_QDT_phi}, mixed products arise, which makes it difficult
or even impossible to reject the error $\varepsilon$. All this is
far away from being mathematically complete, but it points in the
direction towards an error discussion, which is presented in more
detail by \citet[Chap. 6]{jerri1977theshannon}. 

At this point, the statistical approach ``fitting a sinusoidal model
to the data set of interest'' becomes attractive. A brief discussion
of this concept is given in ``Studies in astronomical time series
analysis. II'' by \citet{scargle1982studies}. Chapter \ref{subsec:Gapped-and-irregularly-sampled-data}
provides an introduction on the Lomb-Scargle method, which is equal
to the already mentioned least square fit.\clearpage{}

\subsection[Integral transforms]{\label{subsec:Integral-Transforms}Integral transforms}

In contrast to the previously described QDT, which is more or less
a statistical approach, the integral transforms are the mathematical
backbone of spectral analysis. An integral transform represents the
bijective mapping $s(t)\leftrightarrow S(\omega)$ between the time
and frequency domains, which states that all information present in
the first will be transformed to the second domain. In the following,
capital letters denote always the frequency domain, whereas lowercase
letters represent the time domain of the corresponding variable.
\begin{table}
\centering{}%
\begin{tabular}{ccc}
\toprule 
 & continuous & discrete\tabularnewline
\midrule
transform & ${\displaystyle S(\omega)=\int_{-\infty}^{\infty}s(t)\cdot\e^{-\im\omega t}\diff t}$ & ${\displaystyle \hat{S}(m)=\frac{1}{N}\sum_{n=0}^{N-1}\hat{s}\left(n\right)\e^{-\im m\frac{2\pi}{N}n}}$\tabularnewline
back transform & ${\displaystyle s(t)=\frac{1}{2\pi}\int_{-\infty}^{\infty}S(\omega)\cdot\e^{\im\omega t}\diff\omega}$ & ${\displaystyle \hat{s}(n)=\sum_{m=0}^{N-1}\hat{S}\left(m\right)\e^{\im m\frac{2\pi}{N}n}}$\tabularnewline
\bottomrule
\end{tabular}\caption{\label{tab:Continuous-and-discrete-Fourier}Continuous and discrete
Fourier transform.}
\end{table}

\paragraph{Fourier transform}

First of all the Fourier transform should be introduced. Thereby,
the definitions and some of the properties in the continuous and discrete
time domain are different. A comparison is given in table~\ref{tab:Continuous-and-discrete-Fourier}.
Nevertheless, the signal $s(t)$ forms a Fourier pair with $S(\omega)$.
The following notation describes a transform of the signal $s(t)$
into the frequency domain,
\begin{equation}
\mathcal{F}\left(s(t)\right)=S(\omega).\label{eq:1_FT_notation}
\end{equation}
This uses the Fourier operator $\mathcal{F}$, which expresses the
calculus according to table~\ref{tab:Continuous-and-discrete-Fourier}.
The inverse back transform is denoted by $\mathcal{F}^{-1}$. The
Fourier transform gains different features like the frequency shifting
\begin{equation}
\mathcal{F}\left(\e^{\im bt}f(t)\right)=F\left(\omega-b\right)\qquad b\in\mathbb{R},\label{eq:1_FT_shifting}
\end{equation}
which becomes important when amplitude modulation in chapter \ref{subsec:Calculating-the-envelope}
is discussed. All necessary properties are listed in Appendix~\ref{sec:Properties-of_FourierTransform}.

In addition to the characteristics of the continuous Fourier transform,
the discrete Fourier transform (DFT) has a $2\pi$-periodic spectrum.
As shown in Appendix \ref{sec:Derivation_Fourier}, the result of
the DFT is a complex valued data vector, in which the first element
$\hat{S}_{1}$ \emph{always} contains the mean value of the data series.
Subsequent bins hold the amplitude values up to $\hat{S}_{N/2}$.
From this position on the spectrum is mirrored and repeats up to the
sampling frequency. In chapter \ref{sec:Spectral-methods-in-one-dimension}
the symmetry of the DFT's and its consequences are discussed in more
detail. It will be shown how to take advantage of this to recover
the ``negative'' frequencies. In case of more dimensional data,
it turns out that the result of the DFT contains additional information,
where also the ``negative'' frequency components play a role. Further
insights on this will be given in part II of this paper.

\paragraph[Hilbert transform]{\label{par:Hilbert-transform}Hilbert transform}

The Hilbert transform (HT) is defined by the real-valued folding operation
on the real-valued signal $s(t)$
\begin{equation}
\mathcal{H}\left(s(t)\right)=\frac{1}{\pi}\int_{-\infty}^{\infty}\frac{s(t')}{t-t'}\diff t'.\label{eq:1_Hilbertransform}
\end{equation}
Additionally, the linear operator $\mathcal{H}$ is introduced to
execute the Hilbert transform on a function or data set. From the
representation in the Fourier space the constant phase shifting feature
becomes clear, 
\begin{equation}
\mathcal{F}\left(\mathcal{H}\left(s(t)\right)\right)=-\im\cdot\mathrm{sign}(\omega)\cdot\mathcal{F}\left(s(t)\right).\label{eq:1_F(H)}
\end{equation}
Hence, equation \eqref{eq:1_F(H)} can be used to easily calculate
$\mathcal{H}\left(s(t)\right)$ in terms of a Fourier transform. Remember,
here the $\mathrm{sign}(\omega)$ function is defined on $-\infty<\omega<\infty$.
This must be taken into account, when performing $\mathcal{H}(\hat{s}(n))$
on discrete data. In this case, the discrete Hilbert transform is
calculated by 
\begin{equation}
\mathcal{F}\left(\mathcal{H}(\hat{s}_{n})\right)=\begin{cases}
\im\cdot\mathcal{F}\left(\hat{s}_{n}\right) & 0\leq n<N/2\\
-\im\cdot\mathcal{F}\left(\hat{s}_{n}\right) & N/2\leq n<N.
\end{cases}\label{eq:1_Discete_Hilbert}
\end{equation}
This formalism considers the fact that ``negative'' frequencies
are mirrored into the upper parts of the DFT data vector. However,
the result of $\mathcal{H}\left(s(t)\right)$ is real-valued, with
each frequency component phase-shifted by $\pm\pi$. In conjunction
with a real-valued and causal signal $s(t)$, the HT helps to represent
the real physical character of measured data, which should only contain
positive frequencies, due to real worlds processes.From this, the
analytical signal 
\begin{equation}
a(t)=s(t)+\im\mathcal{H}\left(s(t)\right),\label{eq:1_analytical_Signal}
\end{equation}
follows, which is a complex representation of $s(t)$ containing a
one-sided Fourier spectrum. 

One derivation of the Hilbert transform in conjunction with the analytical
signal is given in appendix \ref{sec:Derivation_Hilbert} and its
application in terms of data filtering is briefly discussed in section
\ref{subsec:The-analytic-signal} and the following. 

In addition to that, the concept of instantaneous frequency can be
explained with the help of the HT. This topic it discussed in detail
in ``The Empirical Mode Decomposition and the Hilbert Spectrum for
Nonlinear and Non-Stationary Time Series Analysis'' by \citet{huang1998theempirical}.
For further reading on that ``Amplitude, phase, frequency\textemdash fundamental
concepts of oscillation theory'' by \citet{vakman1977amplitude}
should be mentioned.

\section[Spectral methods in one dimension]{\label{sec:Spectral-methods-in-one-dimension}Spectral methods in
one dimension}

The application of the introduced Fourier and Hilbert transform will
be discussed in detail in this chapter. Each of the following sections
includes a theoretical part, which is finally supported by examples.
The focus lies on the application of the corresponding methods. 

First of all, let the signal function
\begin{equation}
s(t)=\sum_{i=0}^{m}A_{i}\cdot\cos(\omega_{i}t+\varphi_{i})+\mathcal{N}(0,\sigma)\label{eq:2_signal_def}
\end{equation}
be a set of sinusoidal functions plus an optional noise term. This
signal is sampled with an equidistant spacing $T_{\mathrm{s}}$, so
the discrete data vector 
\[
\hat{s}{}_{n}=\sum_{n=0}^{N-1}s(n\cdot T_{\mathrm{s}})\cdot\vec{e}_{n}
\]
of length $N$ represents the sampling series. Using the discrete
Fourier transform, see table~\ref{tab:Continuous-and-discrete-Fourier},
the spectrum of the data vector $\hat{s}$ 
\[
\hat{S}{}_{m}=\mathcal{F}(\hat{s}_{n})
\]
can be calculated.

For further reading a brief introduction can be found in ``The Shannon
sampling theorem\textemdash Its various extensions and applications:
A tutorial review'' by \citet{jerri1977theshannon}. A detailed comprehensive
mathematical work is given in the book ``Fourier Analysis'' by \citet{duoandikoetxeazuazo2001fourier}.

\subsection[The essence of band limitation and the Nyquist condition]{\label{subsec:Bandlimitation}The essence of band limitation and
the Nyquist condition}

The question of band limitation is a question of the unique representation
of a function $s(t)$ in the frequency space $S(\omega)$. In the
simplest case, a band limited signal has a spectrum
\begin{equation}
S(\omega)=\begin{cases}
\neq0 & \left|\omega\right|\leq\omega_{\mathrm{max}}\\
0 & \text{elsewhere}
\end{cases}\label{eq:2_Nyquist_cond}
\end{equation}
which is only non-zero within the range $\left|\omega\right|\leq\omega_{\mathrm{max}}$.
Outside this interval the spectrum is equal to zero. The sampling
theorem by \citet{shannon1949communication} and the proof of \citet{whittaker1935interpolatory}
states, that in such a situation even a time-limited continuous function
is completely defined by a discrete and finite set of samples, if
all the samples outside the time range are exactly zero (see the exemplary
function in Figure~\ref{fig:Simple_signal}). Closely related to
that is the minimal required sampling frequency (twice the ``Nyquist
frequency'') $1/T_{s}=f_{s}\geq2f_{\mathrm{max}}$, which is necessary
to obtain all information about the signal. 

The consequence is that there should not be any frequency present
in the data set higher than $f_{s}/2$ to achieve a unique sampling
in application. Theoretically, this means a function $s(t)$ is completely
determined by its discrete spectrum $S(n\cdot2\pi f_{s})$ if the
signal is bandlimited to $\left|f\right|<f_{s}/2$. Therefore the
discrete spectrum of the discrete sampled signal $\hat{s}{}_{n}$
provides a unique spectral representation of its continuous counterpart
$s(t)$. 

\paragraph{Application}

\begin{figure}
\subfloat[]{\centering\input{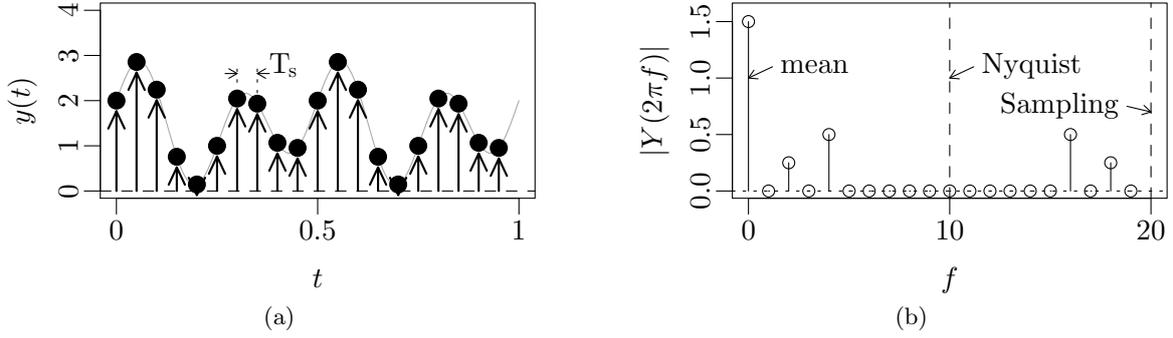}

}\hfill{}\subfloat[]{\centering
\begin{tikzpicture}[x=1pt,y=1pt]
\definecolor{fillColor}{RGB}{255,255,255}
\path[use as bounding box,fill=fillColor,fill opacity=0.00] (0,0) rectangle (198.74,115.63);
\begin{scope}
\path[clip] ( 36.00, 36.00) rectangle (198.74,109.63);
\definecolor{drawColor}{RGB}{0,0,0}

\path[draw=drawColor,line width= 0.4pt,line join=round,line cap=round] ( 42.03,102.64) circle (  2.25);

\path[draw=drawColor,line width= 0.4pt,line join=round,line cap=round] ( 49.56, 38.73) circle (  2.25);

\path[draw=drawColor,line width= 0.4pt,line join=round,line cap=round] ( 57.10, 49.38) circle (  2.25);

\path[draw=drawColor,line width= 0.4pt,line join=round,line cap=round] ( 64.63, 38.73) circle (  2.25);

\path[draw=drawColor,line width= 0.4pt,line join=round,line cap=round] ( 72.17, 60.03) circle (  2.25);

\path[draw=drawColor,line width= 0.4pt,line join=round,line cap=round] ( 79.70, 38.73) circle (  2.25);

\path[draw=drawColor,line width= 0.4pt,line join=round,line cap=round] ( 87.23, 38.73) circle (  2.25);

\path[draw=drawColor,line width= 0.4pt,line join=round,line cap=round] ( 94.77, 38.73) circle (  2.25);

\path[draw=drawColor,line width= 0.4pt,line join=round,line cap=round] (102.30, 38.73) circle (  2.25);

\path[draw=drawColor,line width= 0.4pt,line join=round,line cap=round] (109.84, 38.73) circle (  2.25);

\path[draw=drawColor,line width= 0.4pt,line join=round,line cap=round] (117.37, 38.73) circle (  2.25);

\path[draw=drawColor,line width= 0.4pt,line join=round,line cap=round] (124.91, 38.73) circle (  2.25);

\path[draw=drawColor,line width= 0.4pt,line join=round,line cap=round] (132.44, 38.73) circle (  2.25);

\path[draw=drawColor,line width= 0.4pt,line join=round,line cap=round] (139.97, 38.73) circle (  2.25);

\path[draw=drawColor,line width= 0.4pt,line join=round,line cap=round] (147.51, 38.73) circle (  2.25);

\path[draw=drawColor,line width= 0.4pt,line join=round,line cap=round] (155.04, 38.73) circle (  2.25);

\path[draw=drawColor,line width= 0.4pt,line join=round,line cap=round] (162.58, 60.03) circle (  2.25);

\path[draw=drawColor,line width= 0.4pt,line join=round,line cap=round] (170.11, 38.73) circle (  2.25);

\path[draw=drawColor,line width= 0.4pt,line join=round,line cap=round] (177.65, 49.38) circle (  2.25);

\path[draw=drawColor,line width= 0.4pt,line join=round,line cap=round] (185.18, 38.73) circle (  2.25);
\end{scope}
\begin{scope}
\path[clip] (  0.00,  0.00) rectangle (198.74,115.63);
\definecolor{drawColor}{RGB}{0,0,0}

\path[draw=drawColor,line width= 0.4pt,line join=round,line cap=round] ( 36.00, 38.73) -- ( 36.00,102.64);

\path[draw=drawColor,line width= 0.4pt,line join=round,line cap=round] ( 36.00, 38.73) -- ( 30.00, 38.73);

\path[draw=drawColor,line width= 0.4pt,line join=round,line cap=round] ( 36.00, 60.03) -- ( 30.00, 60.03);

\path[draw=drawColor,line width= 0.4pt,line join=round,line cap=round] ( 36.00, 81.34) -- ( 30.00, 81.34);

\path[draw=drawColor,line width= 0.4pt,line join=round,line cap=round] ( 36.00,102.64) -- ( 30.00,102.64);

\node[text=drawColor,rotate= 90.00,anchor=base,inner sep=0pt, outer sep=0pt, scale=  1.00] at ( 27.60, 38.73) {0.0};

\node[text=drawColor,rotate= 90.00,anchor=base,inner sep=0pt, outer sep=0pt, scale=  1.00] at ( 27.60, 60.03) {0.5};

\node[text=drawColor,rotate= 90.00,anchor=base,inner sep=0pt, outer sep=0pt, scale=  1.00] at ( 27.60, 81.34) {1.0};

\node[text=drawColor,rotate= 90.00,anchor=base,inner sep=0pt, outer sep=0pt, scale=  1.00] at ( 27.60,102.64) {1.5};

\path[draw=drawColor,line width= 0.4pt,line join=round,line cap=round] ( 36.00, 36.00) --
	(198.74, 36.00) --
	(198.74,109.63) --
	( 36.00,109.63) --
	( 36.00, 36.00);
\end{scope}
\begin{scope}
\path[clip] (  0.00,  0.00) rectangle (198.74,115.63);
\definecolor{drawColor}{RGB}{0,0,0}

\node[text=drawColor,anchor=base,inner sep=0pt, outer sep=0pt, scale=  1.00] at (117.37,  2.40) {$f$};

\node[text=drawColor,rotate= 90.00,anchor=base,inner sep=0pt, outer sep=0pt, scale=  1.00] at (  9.60, 72.82) {$\left|Y(2\pi f)\right|$};
\end{scope}
\begin{scope}
\path[clip] (  0.00,  0.00) rectangle (198.74,115.63);
\definecolor{drawColor}{RGB}{0,0,0}

\path[draw=drawColor,line width= 0.4pt,line join=round,line cap=round] ( 42.03, 36.00) -- (192.71, 36.00);

\path[draw=drawColor,line width= 0.4pt,line join=round,line cap=round] ( 42.03, 36.00) -- ( 42.03, 30.00);

\path[draw=drawColor,line width= 0.4pt,line join=round,line cap=round] (117.37, 36.00) -- (117.37, 30.00);

\path[draw=drawColor,line width= 0.4pt,line join=round,line cap=round] (192.71, 36.00) -- (192.71, 30.00);

\node[text=drawColor,anchor=base,inner sep=0pt, outer sep=0pt, scale=  1.00] at ( 42.03, 20.40) {0};

\node[text=drawColor,anchor=base,inner sep=0pt, outer sep=0pt, scale=  1.00] at (117.37, 20.40) {10};

\node[text=drawColor,anchor=base,inner sep=0pt, outer sep=0pt, scale=  1.00] at (192.71, 20.40) {20};
\end{scope}
\begin{scope}
\path[clip] ( 36.00, 36.00) rectangle (198.74,109.63);
\definecolor{drawColor}{RGB}{0,0,0}

\path[draw=drawColor,line width= 0.4pt,line join=round,line cap=round] ( 42.03, 38.73) -- ( 42.03,102.64);

\path[draw=drawColor,line width= 0.4pt,line join=round,line cap=round] ( 49.56, 38.73) -- ( 49.56, 38.73);

\path[draw=drawColor,line width= 0.4pt,line join=round,line cap=round] ( 57.10, 38.73) -- ( 57.10, 49.38);

\path[draw=drawColor,line width= 0.4pt,line join=round,line cap=round] ( 64.63, 38.73) -- ( 64.63, 38.73);

\path[draw=drawColor,line width= 0.4pt,line join=round,line cap=round] ( 72.17, 38.73) -- ( 72.17, 60.03);

\path[draw=drawColor,line width= 0.4pt,line join=round,line cap=round] ( 79.70, 38.73) -- ( 79.70, 38.73);

\path[draw=drawColor,line width= 0.4pt,line join=round,line cap=round] ( 87.23, 38.73) -- ( 87.23, 38.73);

\path[draw=drawColor,line width= 0.4pt,line join=round,line cap=round] ( 94.77, 38.73) -- ( 94.77, 38.73);

\path[draw=drawColor,line width= 0.4pt,line join=round,line cap=round] (102.30, 38.73) -- (102.30, 38.73);

\path[draw=drawColor,line width= 0.4pt,line join=round,line cap=round] (109.84, 38.73) -- (109.84, 38.73);

\path[draw=drawColor,line width= 0.4pt,line join=round,line cap=round] (117.37, 38.73) -- (117.37, 38.73);

\path[draw=drawColor,line width= 0.4pt,line join=round,line cap=round] (124.91, 38.73) -- (124.91, 38.73);

\path[draw=drawColor,line width= 0.4pt,line join=round,line cap=round] (132.44, 38.73) -- (132.44, 38.73);

\path[draw=drawColor,line width= 0.4pt,line join=round,line cap=round] (139.97, 38.73) -- (139.97, 38.73);

\path[draw=drawColor,line width= 0.4pt,line join=round,line cap=round] (147.51, 38.73) -- (147.51, 38.73);

\path[draw=drawColor,line width= 0.4pt,line join=round,line cap=round] (155.04, 38.73) -- (155.04, 38.73);

\path[draw=drawColor,line width= 0.4pt,line join=round,line cap=round] (162.58, 38.73) -- (162.58, 60.03);

\path[draw=drawColor,line width= 0.4pt,line join=round,line cap=round] (170.11, 38.73) -- (170.11, 38.73);

\path[draw=drawColor,line width= 0.4pt,line join=round,line cap=round] (177.65, 38.73) -- (177.65, 49.38);

\path[draw=drawColor,line width= 0.4pt,line join=round,line cap=round] (185.18, 38.73) -- (185.18, 38.73);

\path[draw=drawColor,line width= 0.4pt,dash pattern=on 1pt off 3pt ,line join=round,line cap=round] ( 36.00, 38.73) -- (198.74, 38.73);

\path[draw=drawColor,line width= 0.4pt,dash pattern=on 4pt off 4pt ,line join=round,line cap=round] (117.37, 36.00) -- (117.37,109.63);

\path[draw=drawColor,line width= 0.4pt,dash pattern=on 4pt off 4pt ,line join=round,line cap=round] (192.71, 36.00) -- (192.71,109.63);

\path[draw=drawColor,line width= 0.4pt,line join=round,line cap=round] ( 51.03, 85.60) -- ( 42.03, 81.34);

\path[draw=drawColor,line width= 0.4pt,line join=round,line cap=round] ( 44.08, 84.31) --
	( 42.03, 81.34) --
	( 45.63, 81.04);

\path[draw=drawColor,line width= 0.4pt,line join=round,line cap=round] (126.37, 85.60) -- (117.37, 81.34);

\path[draw=drawColor,line width= 0.4pt,line join=round,line cap=round] (119.43, 84.31) --
	(117.37, 81.34) --
	(120.97, 81.04);

\path[draw=drawColor,line width= 0.4pt,line join=round,line cap=round] (183.71, 70.69) -- (192.71, 68.55);

\path[draw=drawColor,line width= 0.4pt,line join=round,line cap=round] (189.25, 67.52) --
	(192.71, 68.55) --
	(190.09, 71.03);

\node[text=drawColor,anchor=base west,inner sep=0pt, outer sep=0pt, scale=  1.00] at ( 54.03, 83.30) {mean};

\node[text=drawColor,anchor=base west,inner sep=0pt, outer sep=0pt, scale=  1.00] at (129.37, 83.30) {Nyquist};

\node[text=drawColor,anchor=base east,inner sep=0pt, outer sep=0pt, scale=  1.00] at (180.71, 68.39) {Sampling};
\end{scope}
\end{tikzpicture}

}

\caption{\label{fig:Over-sampled-simple-signal}Over sampled signal $y(t)$.
Band limitation is achieved by definition of the signal. Mention,
that the first sample and the $i_{\mathrm{th}}+1$ sample at $t=1$
would be the same, because of the periodicity of the signal.}
\end{figure}

The artificial function 
\begin{equation}
y(t)=\sin(2\pi\cdot4t)+0.5\cdot\cos(2\pi\cdot2t)+1.5\label{eq:2-1_Artificial_Function}
\end{equation}
is given in Figure~\ref{fig:Over-sampled-simple-signal}a. 
It consists of two functions and an offset value. The sampling took
place at a rate of $T_{s}=0.05$, so that there are theoretically
$N=21$ possible samples. Note, $y(t)$ is perfectly periodic, so
the first sample would equal the last sample. Because the latter marks
the first point of the next sampling period, it must be removed from
the data set, which now has $N^{*}=20$ elements. Ignoring this fact
would lead to an error (leakage effect) with the result that the elements
to the left and to the right of the frequency peaks are not zero anymore.
The root of this behavior is discussed in section~\ref{subsec:Frequency-analysis}
and becomes visible in the subsequent figures in this section. The
resulting frequency resolution in this example is 
\begin{eqnarray}
\mathit{\delta f} & = & \frac{1}{T_{s}N^{*}},\label{eq:frequency-resolution}\\
 & = & 1\nonumber 
\end{eqnarray}
so the signal's frequencies fit perfectly into this grid. The absolute
amplitude spectrum $\left|\hat{Y}_{n}\right|$ is given in Figure~\ref{fig:Over-sampled-simple-signal}b.
Here, one can see that the mean, saved in the first bin, has its original
value of $A_{0}=1.5$, whereas the magnitude of the amplitudes at
$f_{n=2}=2$ and $f_{n=4}=4$ are split into the upper and lower frequency
domain. To retrieve the correct signal amplitudes the individual values
in both half-planes must be added. Remember, the DFT is mirror-symmetric
to the Nyquist frequency $f_{\mathrm{s}}/2=10$ so that the first
amplitude $A_{1}$ becomes 
\begin{equation}
A_{1}=\left|\hat{Y}_{n=4}\right|+\left|\hat{Y}_{n=16}\right|.
\end{equation}

\begin{figure}
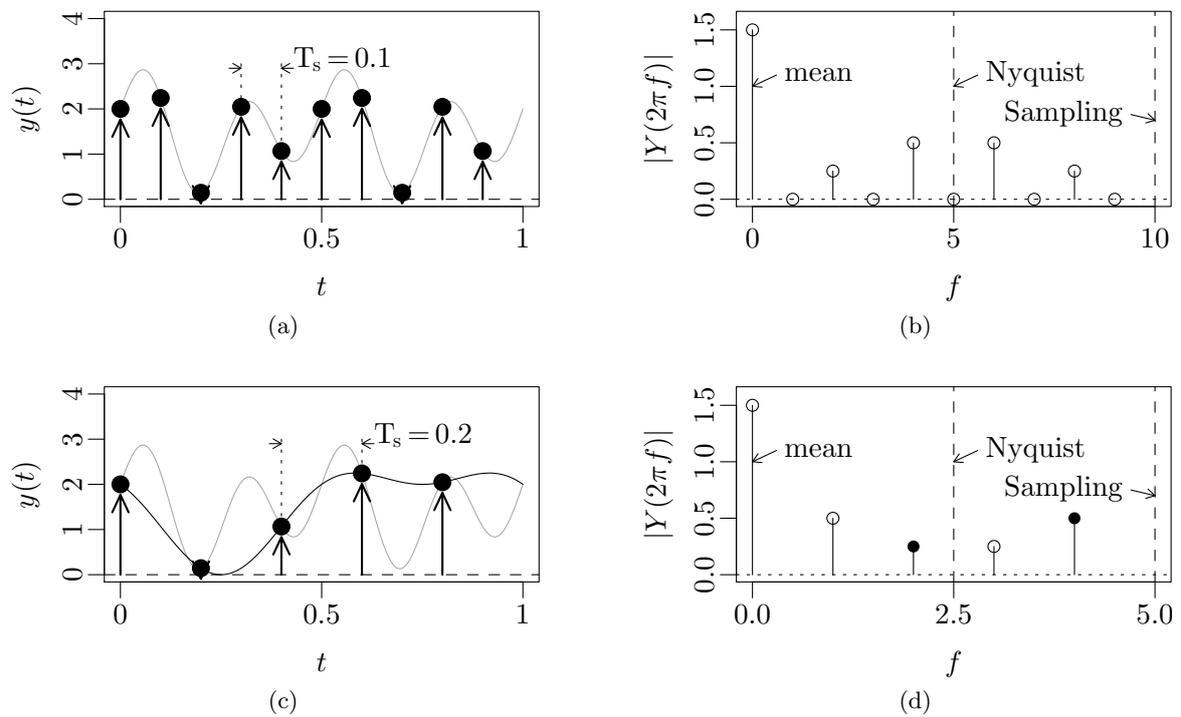

\subfloat[]{\centering\input{Pictures/Example02_marginal_Sampling.tex}

}\hfill{}\subfloat[]{\centering
\begin{tikzpicture}[x=1pt,y=1pt]
\definecolor{fillColor}{RGB}{255,255,255}
\path[use as bounding box,fill=fillColor,fill opacity=0.00] (0,0) rectangle (198.74,115.63);
\begin{scope}
\path[clip] ( 36.00, 36.00) rectangle (198.74,109.63);
\definecolor{drawColor}{RGB}{0,0,0}

\path[draw=drawColor,line width= 0.4pt,line join=round,line cap=round] ( 42.03,102.64) circle (  2.25);

\path[draw=drawColor,line width= 0.4pt,line join=round,line cap=round] ( 57.10, 38.73) circle (  2.25);

\path[draw=drawColor,line width= 0.4pt,line join=round,line cap=round] ( 72.16, 49.38) circle (  2.25);

\path[draw=drawColor,line width= 0.4pt,line join=round,line cap=round] ( 87.23, 38.73) circle (  2.25);

\path[draw=drawColor,line width= 0.4pt,line join=round,line cap=round] (102.30, 60.03) circle (  2.25);

\path[draw=drawColor,line width= 0.4pt,line join=round,line cap=round] (117.37, 38.73) circle (  2.25);

\path[draw=drawColor,line width= 0.4pt,line join=round,line cap=round] (132.44, 60.03) circle (  2.25);

\path[draw=drawColor,line width= 0.4pt,line join=round,line cap=round] (147.51, 38.73) circle (  2.25);

\path[draw=drawColor,line width= 0.4pt,line join=round,line cap=round] (162.58, 49.38) circle (  2.25);

\path[draw=drawColor,line width= 0.4pt,line join=round,line cap=round] (177.65, 38.73) circle (  2.25);
\end{scope}
\begin{scope}
\path[clip] (  0.00,  0.00) rectangle (198.74,115.63);
\definecolor{drawColor}{RGB}{0,0,0}

\path[draw=drawColor,line width= 0.4pt,line join=round,line cap=round] ( 36.00, 38.73) -- ( 36.00,102.64);

\path[draw=drawColor,line width= 0.4pt,line join=round,line cap=round] ( 36.00, 38.73) -- ( 30.00, 38.73);

\path[draw=drawColor,line width= 0.4pt,line join=round,line cap=round] ( 36.00, 60.03) -- ( 30.00, 60.03);

\path[draw=drawColor,line width= 0.4pt,line join=round,line cap=round] ( 36.00, 81.34) -- ( 30.00, 81.34);

\path[draw=drawColor,line width= 0.4pt,line join=round,line cap=round] ( 36.00,102.64) -- ( 30.00,102.64);

\node[text=drawColor,rotate= 90.00,anchor=base,inner sep=0pt, outer sep=0pt, scale=  1.00] at ( 27.60, 38.73) {0.0};

\node[text=drawColor,rotate= 90.00,anchor=base,inner sep=0pt, outer sep=0pt, scale=  1.00] at ( 27.60, 60.03) {0.5};

\node[text=drawColor,rotate= 90.00,anchor=base,inner sep=0pt, outer sep=0pt, scale=  1.00] at ( 27.60, 81.34) {1.0};

\node[text=drawColor,rotate= 90.00,anchor=base,inner sep=0pt, outer sep=0pt, scale=  1.00] at ( 27.60,102.64) {1.5};

\path[draw=drawColor,line width= 0.4pt,line join=round,line cap=round] ( 36.00, 36.00) --
	(198.74, 36.00) --
	(198.74,109.63) --
	( 36.00,109.63) --
	( 36.00, 36.00);
\end{scope}
\begin{scope}
\path[clip] (  0.00,  0.00) rectangle (198.74,115.63);
\definecolor{drawColor}{RGB}{0,0,0}

\node[text=drawColor,anchor=base,inner sep=0pt, outer sep=0pt, scale=  1.00] at (117.37,  2.40) {$f$};

\node[text=drawColor,rotate= 90.00,anchor=base,inner sep=0pt, outer sep=0pt, scale=  1.00] at (  9.60, 72.82) {$\left|Y(2\pi f)\right|$};
\end{scope}
\begin{scope}
\path[clip] (  0.00,  0.00) rectangle (198.74,115.63);
\definecolor{drawColor}{RGB}{0,0,0}

\path[draw=drawColor,line width= 0.4pt,line join=round,line cap=round] ( 42.03, 36.00) -- (192.71, 36.00);

\path[draw=drawColor,line width= 0.4pt,line join=round,line cap=round] ( 42.03, 36.00) -- ( 42.03, 30.00);

\path[draw=drawColor,line width= 0.4pt,line join=round,line cap=round] (117.37, 36.00) -- (117.37, 30.00);

\path[draw=drawColor,line width= 0.4pt,line join=round,line cap=round] (192.71, 36.00) -- (192.71, 30.00);

\node[text=drawColor,anchor=base,inner sep=0pt, outer sep=0pt, scale=  1.00] at ( 42.03, 20.40) {0};

\node[text=drawColor,anchor=base,inner sep=0pt, outer sep=0pt, scale=  1.00] at (117.37, 20.40) {5};

\node[text=drawColor,anchor=base,inner sep=0pt, outer sep=0pt, scale=  1.00] at (192.71, 20.40) {10};
\end{scope}
\begin{scope}
\path[clip] ( 36.00, 36.00) rectangle (198.74,109.63);
\definecolor{drawColor}{RGB}{0,0,0}

\path[draw=drawColor,line width= 0.4pt,line join=round,line cap=round] ( 42.03, 38.73) -- ( 42.03,102.64);

\path[draw=drawColor,line width= 0.4pt,line join=round,line cap=round] ( 57.10, 38.73) -- ( 57.10, 38.73);

\path[draw=drawColor,line width= 0.4pt,line join=round,line cap=round] ( 72.16, 38.73) -- ( 72.16, 49.38);

\path[draw=drawColor,line width= 0.4pt,line join=round,line cap=round] ( 87.23, 38.73) -- ( 87.23, 38.73);

\path[draw=drawColor,line width= 0.4pt,line join=round,line cap=round] (102.30, 38.73) -- (102.30, 60.03);

\path[draw=drawColor,line width= 0.4pt,line join=round,line cap=round] (117.37, 38.73) -- (117.37, 38.73);

\path[draw=drawColor,line width= 0.4pt,line join=round,line cap=round] (132.44, 38.73) -- (132.44, 60.03);

\path[draw=drawColor,line width= 0.4pt,line join=round,line cap=round] (147.51, 38.73) -- (147.51, 38.73);

\path[draw=drawColor,line width= 0.4pt,line join=round,line cap=round] (162.58, 38.73) -- (162.58, 49.38);

\path[draw=drawColor,line width= 0.4pt,line join=round,line cap=round] (177.65, 38.73) -- (177.65, 38.73);

\path[draw=drawColor,line width= 0.4pt,dash pattern=on 1pt off 3pt ,line join=round,line cap=round] ( 36.00, 38.73) -- (198.74, 38.73);

\path[draw=drawColor,line width= 0.4pt,dash pattern=on 4pt off 4pt ,line join=round,line cap=round] (117.37, 36.00) -- (117.37,109.63);

\path[draw=drawColor,line width= 0.4pt,dash pattern=on 4pt off 4pt ,line join=round,line cap=round] (192.71, 36.00) -- (192.71,109.63);

\path[draw=drawColor,line width= 0.4pt,line join=round,line cap=round] ( 51.03, 85.60) -- ( 42.03, 81.34);

\path[draw=drawColor,line width= 0.4pt,line join=round,line cap=round] ( 44.08, 84.31) --
	( 42.03, 81.34) --
	( 45.63, 81.04);

\path[draw=drawColor,line width= 0.4pt,line join=round,line cap=round] (126.37, 85.60) -- (117.37, 81.34);

\path[draw=drawColor,line width= 0.4pt,line join=round,line cap=round] (119.43, 84.31) --
	(117.37, 81.34) --
	(120.97, 81.04);

\path[draw=drawColor,line width= 0.4pt,line join=round,line cap=round] (183.71, 70.69) -- (192.71, 68.55);

\path[draw=drawColor,line width= 0.4pt,line join=round,line cap=round] (189.25, 67.52) --
	(192.71, 68.55) --
	(190.09, 71.03);

\node[text=drawColor,anchor=base west,inner sep=0pt, outer sep=0pt, scale=  1.00] at ( 54.03, 83.30) {mean};

\node[text=drawColor,anchor=base west,inner sep=0pt, outer sep=0pt, scale=  1.00] at (129.37, 83.30) {Nyquist};

\node[text=drawColor,anchor=base east,inner sep=0pt, outer sep=0pt, scale=  1.00] at (180.71, 68.39) {Sampling};
\end{scope}
\end{tikzpicture}

}

\subfloat[]{\centering\input{Pictures/Example02_under_Sampling.tex}

}\hfill{}\subfloat[]{\centering
\begin{tikzpicture}[x=1pt,y=1pt]
\definecolor{fillColor}{RGB}{255,255,255}
\path[use as bounding box,fill=fillColor,fill opacity=0.00] (0,0) rectangle (198.74,115.63);
\begin{scope}
\path[clip] ( 36.00, 36.00) rectangle (198.74,109.63);
\definecolor{drawColor}{RGB}{0,0,0}

\path[draw=drawColor,line width= 0.4pt,line join=round,line cap=round] ( 42.03,102.64) circle (  2.25);

\path[draw=drawColor,line width= 0.4pt,line join=round,line cap=round] ( 72.16, 60.03) circle (  2.25);
\definecolor{fillColor}{RGB}{0,0,0}

\path[fill=fillColor] (102.30, 49.38) circle (  2.25);

\path[draw=drawColor,line width= 0.4pt,line join=round,line cap=round] (132.44, 49.38) circle (  2.25);

\path[fill=fillColor] (162.58, 60.03) circle (  2.25);
\end{scope}
\begin{scope}
\path[clip] (  0.00,  0.00) rectangle (198.74,115.63);
\definecolor{drawColor}{RGB}{0,0,0}

\path[draw=drawColor,line width= 0.4pt,line join=round,line cap=round] ( 36.00, 38.73) -- ( 36.00,102.64);

\path[draw=drawColor,line width= 0.4pt,line join=round,line cap=round] ( 36.00, 38.73) -- ( 30.00, 38.73);

\path[draw=drawColor,line width= 0.4pt,line join=round,line cap=round] ( 36.00, 60.03) -- ( 30.00, 60.03);

\path[draw=drawColor,line width= 0.4pt,line join=round,line cap=round] ( 36.00, 81.34) -- ( 30.00, 81.34);

\path[draw=drawColor,line width= 0.4pt,line join=round,line cap=round] ( 36.00,102.64) -- ( 30.00,102.64);

\node[text=drawColor,rotate= 90.00,anchor=base,inner sep=0pt, outer sep=0pt, scale=  1.00] at ( 27.60, 38.73) {0.0};

\node[text=drawColor,rotate= 90.00,anchor=base,inner sep=0pt, outer sep=0pt, scale=  1.00] at ( 27.60, 60.03) {0.5};

\node[text=drawColor,rotate= 90.00,anchor=base,inner sep=0pt, outer sep=0pt, scale=  1.00] at ( 27.60, 81.34) {1.0};

\node[text=drawColor,rotate= 90.00,anchor=base,inner sep=0pt, outer sep=0pt, scale=  1.00] at ( 27.60,102.64) {1.5};

\path[draw=drawColor,line width= 0.4pt,line join=round,line cap=round] ( 36.00, 36.00) --
	(198.74, 36.00) --
	(198.74,109.63) --
	( 36.00,109.63) --
	( 36.00, 36.00);
\end{scope}
\begin{scope}
\path[clip] (  0.00,  0.00) rectangle (198.74,115.63);
\definecolor{drawColor}{RGB}{0,0,0}

\node[text=drawColor,anchor=base,inner sep=0pt, outer sep=0pt, scale=  1.00] at (117.37,  2.40) {$f$};

\node[text=drawColor,rotate= 90.00,anchor=base,inner sep=0pt, outer sep=0pt, scale=  1.00] at (  9.60, 72.82) {$\left|Y(2\pi f)\right|$};
\end{scope}
\begin{scope}
\path[clip] (  0.00,  0.00) rectangle (198.74,115.63);
\definecolor{drawColor}{RGB}{0,0,0}

\path[draw=drawColor,line width= 0.4pt,line join=round,line cap=round] ( 42.03, 36.00) -- (192.71, 36.00);

\path[draw=drawColor,line width= 0.4pt,line join=round,line cap=round] ( 42.03, 36.00) -- ( 42.03, 30.00);

\path[draw=drawColor,line width= 0.4pt,line join=round,line cap=round] (117.37, 36.00) -- (117.37, 30.00);

\path[draw=drawColor,line width= 0.4pt,line join=round,line cap=round] (192.71, 36.00) -- (192.71, 30.00);

\node[text=drawColor,anchor=base,inner sep=0pt, outer sep=0pt, scale=  1.00] at ( 42.03, 20.40) {0.0};

\node[text=drawColor,anchor=base,inner sep=0pt, outer sep=0pt, scale=  1.00] at (117.37, 20.40) {2.5};

\node[text=drawColor,anchor=base,inner sep=0pt, outer sep=0pt, scale=  1.00] at (192.71, 20.40) {5.0};
\end{scope}
\begin{scope}
\path[clip] ( 36.00, 36.00) rectangle (198.74,109.63);
\definecolor{drawColor}{RGB}{0,0,0}

\path[draw=drawColor,line width= 0.4pt,line join=round,line cap=round] ( 42.03, 38.73) -- ( 42.03,102.64);

\path[draw=drawColor,line width= 0.4pt,line join=round,line cap=round] ( 72.16, 38.73) -- ( 72.16, 60.03);

\path[draw=drawColor,line width= 0.4pt,line join=round,line cap=round] (102.30, 38.73) -- (102.30, 49.38);

\path[draw=drawColor,line width= 0.4pt,line join=round,line cap=round] (132.44, 38.73) -- (132.44, 49.38);

\path[draw=drawColor,line width= 0.4pt,line join=round,line cap=round] (162.58, 38.73) -- (162.58, 60.03);

\path[draw=drawColor,line width= 0.4pt,dash pattern=on 1pt off 3pt ,line join=round,line cap=round] ( 36.00, 38.73) -- (198.74, 38.73);

\path[draw=drawColor,line width= 0.4pt,dash pattern=on 4pt off 4pt ,line join=round,line cap=round] (117.37, 36.00) -- (117.37,109.63);

\path[draw=drawColor,line width= 0.4pt,dash pattern=on 4pt off 4pt ,line join=round,line cap=round] (192.71, 36.00) -- (192.71,109.63);

\path[draw=drawColor,line width= 0.4pt,line join=round,line cap=round] ( 51.03, 85.60) -- ( 42.03, 81.34);

\path[draw=drawColor,line width= 0.4pt,line join=round,line cap=round] ( 44.08, 84.31) --
	( 42.03, 81.34) --
	( 45.63, 81.04);

\path[draw=drawColor,line width= 0.4pt,line join=round,line cap=round] (126.37, 85.60) -- (117.37, 81.34);

\path[draw=drawColor,line width= 0.4pt,line join=round,line cap=round] (119.43, 84.31) --
	(117.37, 81.34) --
	(120.97, 81.04);

\path[draw=drawColor,line width= 0.4pt,line join=round,line cap=round] (183.71, 70.69) -- (192.71, 68.55);

\path[draw=drawColor,line width= 0.4pt,line join=round,line cap=round] (189.25, 67.52) --
	(192.71, 68.55) --
	(190.09, 71.03);

\node[text=drawColor,anchor=base west,inner sep=0pt, outer sep=0pt, scale=  1.00] at ( 54.03, 83.30) {mean};

\node[text=drawColor,anchor=base west,inner sep=0pt, outer sep=0pt, scale=  1.00] at (129.37, 83.30) {Nyquist};

\node[text=drawColor,anchor=base east,inner sep=0pt, outer sep=0pt, scale=  1.00] at (180.71, 68.39) {Sampling};
\end{scope}
\end{tikzpicture}

}

\caption{\label{fig:Nyquist-condition-signal}(a, b) The signal $y(t)$ is
sampled marginally. In (c, d) it is under-sampled so the upper and
lower parts of the spectrum infiltrate each other. Compare the filled
symbols, which correspond to the points in the lower half plane of
(b). The black line in c) indicates the alternative reconstruction.}
\end{figure}
Figure~\ref{fig:Over-sampled-simple-signal} and Figure~\ref{fig:Nyquist-condition-signal}a
display an over-sampled signal, where the required Nyquist condition
is fulfilled. The result is a unique mapping of the time domain into
frequency domain. In other words, the corresponding frequency vector
$\hat{f}{}_{n}$ is only valid for $0\leq n<N/2$, because it repeats
in the inverse order for $N/2\leq n<N$.

However, as Figure~\ref{fig:Nyquist-condition-signal}c indicates,
if the requirement of band limitation ($f_{s}\geq2f_{\mathrm{max}})$
is violated, the mapping is not unique anymore. Remember, the frequency
resolution in all given examples is $\delta f=1$. In comparison to
Figure~\ref{fig:Nyquist-condition-signal}b, where the condition
of band limitation is complied with, Figure~\ref{fig:Nyquist-condition-signal}d
illustrates that the upper and lower parts of the frequency bands
now infiltrate each other. This makes it even more complicated to
distinguish between certain frequencies and their physical correspondence.
In such a scenario it becomes completely unclear which of the two
functions belong to the sampling points.

\paragraph{Example \textendash{} Under sampling}

The conditions described above can be extended to the requirement
\[
f_{\mathrm{max}}-f_{\mathrm{min}}=\Delta f\leq f_{\mathrm{s}},
\]
which means that a unique mapping is still possible, yet only if the
bandwidth $\Delta f$ is not larger than the sampling frequency. In
addition to this the frequency range must be known. This is automatically
fulfilled in the most common case, whenever the frequency ranges from
$0\leq f\leq f_{\mathrm{s}}$.
\begin{figure}
\subfloat[]{\centering\input{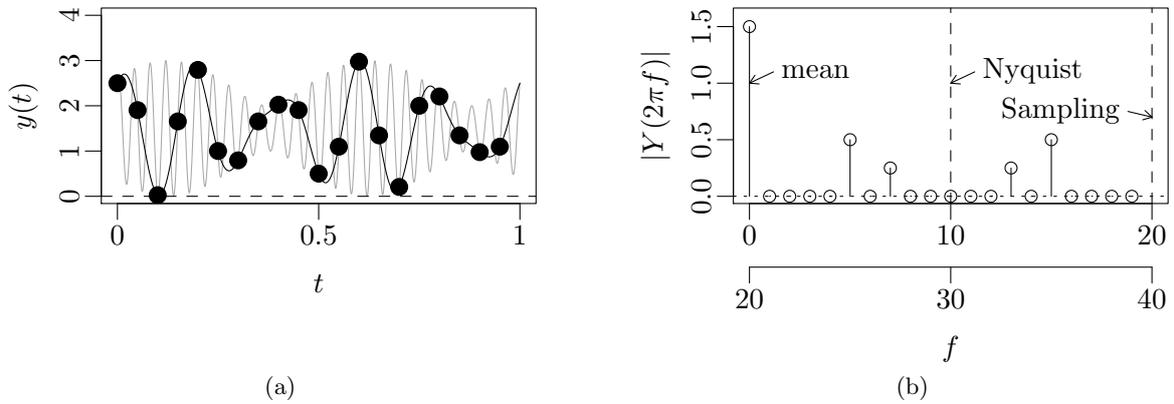}

}\hfill{}\subfloat[]{\centering
\begin{tikzpicture}[x=1pt,y=1pt]
\definecolor{fillColor}{RGB}{255,255,255}
\path[use as bounding box,fill=fillColor,fill opacity=0.00] (0,0) rectangle (198.74,139.74);
\begin{scope}
\path[clip] ( 36.00, 60.00) rectangle (198.74,133.74);
\definecolor{drawColor}{RGB}{0,0,0}

\path[draw=drawColor,line width= 0.4pt,line join=round,line cap=round] ( 42.03,126.74) circle (  2.25);

\path[draw=drawColor,line width= 0.4pt,line join=round,line cap=round] ( 49.56, 62.73) circle (  2.25);

\path[draw=drawColor,line width= 0.4pt,line join=round,line cap=round] ( 57.10, 62.73) circle (  2.25);

\path[draw=drawColor,line width= 0.4pt,line join=round,line cap=round] ( 64.63, 62.73) circle (  2.25);

\path[draw=drawColor,line width= 0.4pt,line join=round,line cap=round] ( 72.17, 62.73) circle (  2.25);

\path[draw=drawColor,line width= 0.4pt,line join=round,line cap=round] ( 79.70, 84.07) circle (  2.25);

\path[draw=drawColor,line width= 0.4pt,line join=round,line cap=round] ( 87.23, 62.73) circle (  2.25);

\path[draw=drawColor,line width= 0.4pt,line join=round,line cap=round] ( 94.77, 73.40) circle (  2.25);

\path[draw=drawColor,line width= 0.4pt,line join=round,line cap=round] (102.30, 62.73) circle (  2.25);

\path[draw=drawColor,line width= 0.4pt,line join=round,line cap=round] (109.84, 62.73) circle (  2.25);

\path[draw=drawColor,line width= 0.4pt,line join=round,line cap=round] (117.37, 62.73) circle (  2.25);

\path[draw=drawColor,line width= 0.4pt,line join=round,line cap=round] (124.91, 62.73) circle (  2.25);

\path[draw=drawColor,line width= 0.4pt,line join=round,line cap=round] (132.44, 62.73) circle (  2.25);

\path[draw=drawColor,line width= 0.4pt,line join=round,line cap=round] (139.97, 73.40) circle (  2.25);

\path[draw=drawColor,line width= 0.4pt,line join=round,line cap=round] (147.51, 62.73) circle (  2.25);

\path[draw=drawColor,line width= 0.4pt,line join=round,line cap=round] (155.04, 84.07) circle (  2.25);

\path[draw=drawColor,line width= 0.4pt,line join=round,line cap=round] (162.58, 62.73) circle (  2.25);

\path[draw=drawColor,line width= 0.4pt,line join=round,line cap=round] (170.11, 62.73) circle (  2.25);

\path[draw=drawColor,line width= 0.4pt,line join=round,line cap=round] (177.65, 62.73) circle (  2.25);

\path[draw=drawColor,line width= 0.4pt,line join=round,line cap=round] (185.18, 62.73) circle (  2.25);
\end{scope}
\begin{scope}
\path[clip] (  0.00,  0.00) rectangle (198.74,139.74);
\definecolor{drawColor}{RGB}{0,0,0}

\path[draw=drawColor,line width= 0.4pt,line join=round,line cap=round] ( 36.00, 62.73) -- ( 36.00,126.74);

\path[draw=drawColor,line width= 0.4pt,line join=round,line cap=round] ( 36.00, 62.73) -- ( 30.00, 62.73);

\path[draw=drawColor,line width= 0.4pt,line join=round,line cap=round] ( 36.00, 84.07) -- ( 30.00, 84.07);

\path[draw=drawColor,line width= 0.4pt,line join=round,line cap=round] ( 36.00,105.40) -- ( 30.00,105.40);

\path[draw=drawColor,line width= 0.4pt,line join=round,line cap=round] ( 36.00,126.74) -- ( 30.00,126.74);

\node[text=drawColor,rotate= 90.00,anchor=base,inner sep=0pt, outer sep=0pt, scale=  1.00] at ( 27.60, 62.73) {0.0};

\node[text=drawColor,rotate= 90.00,anchor=base,inner sep=0pt, outer sep=0pt, scale=  1.00] at ( 27.60, 84.07) {0.5};

\node[text=drawColor,rotate= 90.00,anchor=base,inner sep=0pt, outer sep=0pt, scale=  1.00] at ( 27.60,105.40) {1.0};

\node[text=drawColor,rotate= 90.00,anchor=base,inner sep=0pt, outer sep=0pt, scale=  1.00] at ( 27.60,126.74) {1.5};

\path[draw=drawColor,line width= 0.4pt,line join=round,line cap=round] ( 36.00, 60.00) --
	(198.74, 60.00) --
	(198.74,133.74) --
	( 36.00,133.74) --
	( 36.00, 60.00);
\end{scope}
\begin{scope}
\path[clip] (  0.00,  0.00) rectangle (198.74,139.74);
\definecolor{drawColor}{RGB}{0,0,0}

\node[text=drawColor,rotate= 90.00,anchor=base,inner sep=0pt, outer sep=0pt, scale=  1.00] at (  9.60, 96.87) {$\left|Y(2\pi f)\right|$};
\end{scope}
\begin{scope}
\path[clip] (  0.00,  0.00) rectangle (198.74,139.74);
\definecolor{drawColor}{RGB}{0,0,0}

\path[draw=drawColor,line width= 0.4pt,line join=round,line cap=round] ( 42.03, 60.00) -- (192.71, 60.00);

\path[draw=drawColor,line width= 0.4pt,line join=round,line cap=round] ( 42.03, 60.00) -- ( 42.03, 54.00);

\path[draw=drawColor,line width= 0.4pt,line join=round,line cap=round] (117.37, 60.00) -- (117.37, 54.00);

\path[draw=drawColor,line width= 0.4pt,line join=round,line cap=round] (192.71, 60.00) -- (192.71, 54.00);

\node[text=drawColor,anchor=base,inner sep=0pt, outer sep=0pt, scale=  1.00] at ( 42.03, 44.40) {0};

\node[text=drawColor,anchor=base,inner sep=0pt, outer sep=0pt, scale=  1.00] at (117.37, 44.40) {10};

\node[text=drawColor,anchor=base,inner sep=0pt, outer sep=0pt, scale=  1.00] at (192.71, 44.40) {20};

\path[draw=drawColor,line width= 0.4pt,line join=round,line cap=round] ( 42.03, 36.00) -- (192.71, 36.00);

\path[draw=drawColor,line width= 0.4pt,line join=round,line cap=round] ( 42.03, 36.00) -- ( 42.03, 30.00);

\path[draw=drawColor,line width= 0.4pt,line join=round,line cap=round] (117.37, 36.00) -- (117.37, 30.00);

\path[draw=drawColor,line width= 0.4pt,line join=round,line cap=round] (192.71, 36.00) -- (192.71, 30.00);

\node[text=drawColor,anchor=base,inner sep=0pt, outer sep=0pt, scale=  1.00] at ( 42.03, 20.40) {20};

\node[text=drawColor,anchor=base,inner sep=0pt, outer sep=0pt, scale=  1.00] at (117.37, 20.40) {30};

\node[text=drawColor,anchor=base,inner sep=0pt, outer sep=0pt, scale=  1.00] at (192.71, 20.40) {40};

\node[text=drawColor,anchor=base,inner sep=0pt, outer sep=0pt, scale=  1.00] at (117.37,  2.40) {$f$};
\end{scope}
\begin{scope}
\path[clip] ( 36.00, 60.00) rectangle (198.74,133.74);
\definecolor{drawColor}{RGB}{0,0,0}

\path[draw=drawColor,line width= 0.4pt,line join=round,line cap=round] ( 42.03, 62.73) -- ( 42.03,126.74);

\path[draw=drawColor,line width= 0.4pt,line join=round,line cap=round] ( 49.56, 62.73) -- ( 49.56, 62.73);

\path[draw=drawColor,line width= 0.4pt,line join=round,line cap=round] ( 57.10, 62.73) -- ( 57.10, 62.73);

\path[draw=drawColor,line width= 0.4pt,line join=round,line cap=round] ( 64.63, 62.73) -- ( 64.63, 62.73);

\path[draw=drawColor,line width= 0.4pt,line join=round,line cap=round] ( 72.17, 62.73) -- ( 72.17, 62.73);

\path[draw=drawColor,line width= 0.4pt,line join=round,line cap=round] ( 79.70, 62.73) -- ( 79.70, 84.07);

\path[draw=drawColor,line width= 0.4pt,line join=round,line cap=round] ( 87.23, 62.73) -- ( 87.23, 62.73);

\path[draw=drawColor,line width= 0.4pt,line join=round,line cap=round] ( 94.77, 62.73) -- ( 94.77, 73.40);

\path[draw=drawColor,line width= 0.4pt,line join=round,line cap=round] (102.30, 62.73) -- (102.30, 62.73);

\path[draw=drawColor,line width= 0.4pt,line join=round,line cap=round] (109.84, 62.73) -- (109.84, 62.73);

\path[draw=drawColor,line width= 0.4pt,line join=round,line cap=round] (117.37, 62.73) -- (117.37, 62.73);

\path[draw=drawColor,line width= 0.4pt,line join=round,line cap=round] (124.91, 62.73) -- (124.91, 62.73);

\path[draw=drawColor,line width= 0.4pt,line join=round,line cap=round] (132.44, 62.73) -- (132.44, 62.73);

\path[draw=drawColor,line width= 0.4pt,line join=round,line cap=round] (139.97, 62.73) -- (139.97, 73.40);

\path[draw=drawColor,line width= 0.4pt,line join=round,line cap=round] (147.51, 62.73) -- (147.51, 62.73);

\path[draw=drawColor,line width= 0.4pt,line join=round,line cap=round] (155.04, 62.73) -- (155.04, 84.07);

\path[draw=drawColor,line width= 0.4pt,line join=round,line cap=round] (162.58, 62.73) -- (162.58, 62.73);

\path[draw=drawColor,line width= 0.4pt,line join=round,line cap=round] (170.11, 62.73) -- (170.11, 62.73);

\path[draw=drawColor,line width= 0.4pt,line join=round,line cap=round] (177.65, 62.73) -- (177.65, 62.73);

\path[draw=drawColor,line width= 0.4pt,line join=round,line cap=round] (185.18, 62.73) -- (185.18, 62.73);

\path[draw=drawColor,line width= 0.4pt,dash pattern=on 1pt off 3pt ,line join=round,line cap=round] ( 36.00, 62.73) -- (198.74, 62.73);

\path[draw=drawColor,line width= 0.4pt,dash pattern=on 4pt off 4pt ,line join=round,line cap=round] (117.37, 60.00) -- (117.37,133.74);

\path[draw=drawColor,line width= 0.4pt,dash pattern=on 4pt off 4pt ,line join=round,line cap=round] (192.71, 60.00) -- (192.71,133.74);

\path[draw=drawColor,line width= 0.4pt,line join=round,line cap=round] ( 51.03,109.67) -- ( 42.03,105.40);

\path[draw=drawColor,line width= 0.4pt,line join=round,line cap=round] ( 44.08,108.38) --
	( 42.03,105.40) --
	( 45.63,105.11);

\path[draw=drawColor,line width= 0.4pt,line join=round,line cap=round] (126.37,109.67) -- (117.37,105.40);

\path[draw=drawColor,line width= 0.4pt,line join=round,line cap=round] (119.42,108.38) --
	(117.37,105.40) --
	(120.97,105.11);

\path[draw=drawColor,line width= 0.4pt,line join=round,line cap=round] (183.71, 94.74) -- (192.71, 92.60);

\path[draw=drawColor,line width= 0.4pt,line join=round,line cap=round] (189.25, 91.57) --
	(192.71, 92.60) --
	(190.09, 95.08);

\node[text=drawColor,anchor=base west,inner sep=0pt, outer sep=0pt, scale=  1.00] at ( 54.03,107.37) {mean};

\node[text=drawColor,anchor=base west,inner sep=0pt, outer sep=0pt, scale=  1.00] at (129.37,107.37) {Nyquist};

\node[text=drawColor,anchor=base east,inner sep=0pt, outer sep=0pt, scale=  1.00] at (180.71, 92.44) {Sampling};
\end{scope}
\end{tikzpicture}

}

\caption{\label{fig:Under-sampling}A signal with a limited bandwidth is sampled
times slower than the signal frequency. The black line in (a) displays
again the second possible curve, which would map to the spectrum in
(b).}
\end{figure}

The example given in Figure~\ref{fig:Under-sampling} shows the artificial
function 
\begin{equation}
y(t)=\cos(2\pi\cdot25\cdot t)+0.5\cdot\sin(2\pi\cdot27\cdot t)+1.5,
\end{equation}
which is sampled with a sampling frequency of $f_{\mathrm{s}}=20$.
The amplitude spectrum is given in Figure~\ref{fig:Under-sampling}b.
Obviously $y(t)$ is under-sampled. Remember, the discrete Fourier
transform calculates a periodic spectrum, which projects the high-frequency
parts into the lower frequency range between 0 and $f_{\mathrm{s}}$.
Without any additional information a mapping from spectral domain
to time domain will fail. In the present case we assume that the signal
frequency is in the range $25\leq f\leq27$, so the whole spectrum
can be shifted by an offset of $f_{\mathrm{o}}=f_{\mathrm{s}}$. The
second $f$-axis indicates this. The upper and lower frequency bands
can be clearly distinguished, so the mapping is unique. In contrast
to that, the interprenetation of the upper and lower bands prevents
such a mapping in the example in Figure~\ref{fig:Nyquist-condition-signal}d.

\paragraph[Example \textendash{} The derivative of a function]{Example \textendash{} \label{par:derivative_of_a_function}The derivative
of a function}

One important property of the Fourier transform is the change of mathematical
operations in the spectral domain. According to Appendix~\ref{sec:Properties-of_FourierTransform},
taking the derivative of a function reduces to a multiplication with
$\im\omega$
\begin{equation}
\mathcal{F}\left(\frac{\diff}{\diff x}y(x)\right)=\im\omega\cdot\underset{_{Y(\omega)}}{\underbrace{\mathcal{F}\left(y(x)\right)}}
\end{equation}
in the spectral domain. In this manner, differential equations can
be reduced to simpler linear algebraic representations. 

In contrast to the continuous case, discrete sampled functions must
be periodic with a finite bandwidth to fulfill the requirements for
the DFT. In case the function is non-periodic within the sampled window,
the application of a proper window function is necessary to minimize
leakage. The following code explains how to estimate the derivative
of the non-band-limited function like
\begin{eqnarray}
y(x) & = & -x^{3}+3\cdot x,\\
\text{with\,}\frac{\diff}{\diff x}y(x) & = & -3\cdot x^{2}+3
\end{eqnarray}
which is going to be sampled with $N=40$ discrete points. 

\begin{CodeChunk}
\begin{Code}
require(spectral)

x <- seq(-2.5,2.5,length.out = 40)  # 40 sample positions
y <- win.tukey(x,0.2) * (-x^3+3*x)  # window the function
                                    # and generate the 
                                    # sample points
  Y <- spec.fft(y,x,center = T)     # doing the DFT
Y$A <- 1i * 2*pi * Y$fx * Y$A       # calc. deriv.
# the realpart of the back transform
# contains the signal of interesst
dy <- base::Re(spec.fft(Y,inverse = T)$y)
\end{Code}
\end{CodeChunk}

\begin{figure}
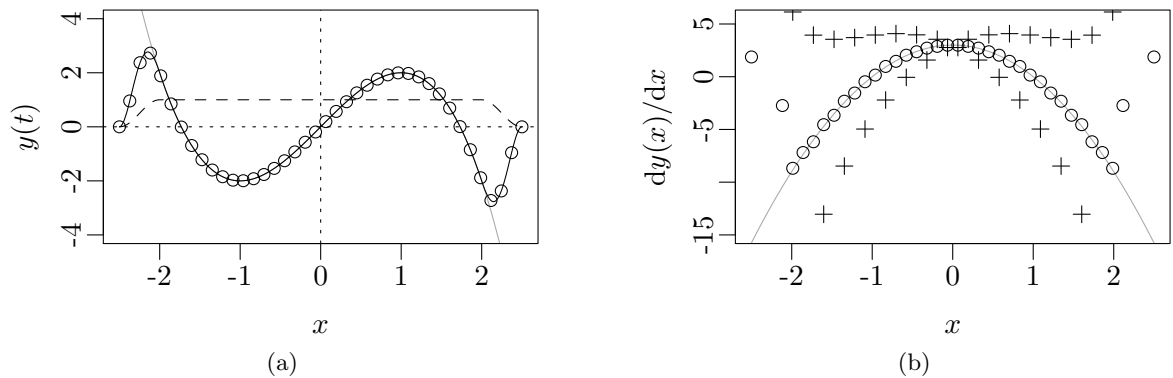

\subfloat[]{\input{Pictures/Example03_function.tex}

}\hfill{}\subfloat[]{\input{Pictures/Example03_derivative.tex}

}

\caption{\label{fig:The-derivative}Calculating the derivative of $y(x)$.
(a) displays the original function (gray), the windowed function (black)
and the sample points. The dashed line defines the window function.
In (b) the analytic (gray line) and estimated numeric (points) derivative
$\protect\diff y(t)/\protect\diff x$ is shown. The crosses indicate
the result of the calculation without utilizing the window function.
Note, that some crosses are vertically outside the window for $x<-2$
and $x>2$.}
\end{figure}
It must be mentioned that the function $y(x)$ must be windowed in
advance to achieve the periodic property. In this example this is
done via the ``Tukey''-window, which rises and falls with a quarter
cycle of a cosine function (compare dashed curve). Figure~\ref{fig:The-derivative}
shows the output of the listed code above. Pay special attention to
the windowed version of $y(x)$ (black curve) and its almost perfect
derivative. Here, the effect of the windowing process is to achieve
$Y(\left|\omega\right|>0.5\cdot\omega_{\mathrm{s}})=0$, which relaxes
the function at the right and left side. In consequence, the resulting
derivative equals the analytic solution in the mid range, but also
deviates at the boundaries. In comparison to that, the unweighted
function in (a) (gray line) produces an unusable faulty derivative
in (b) (crosses). The reason for that is the discussed missing band
limitation.

With a look into line 7 of the above code, notice that the additional
parameter \code{center = T} is provided to the \code{spec.fft()}
function. The effect is a zero-symmetric spectrum with the corresponding
frequencies ranging from $-f_{\mathrm{s}}/2\leq f<f_{\mathrm{s}}$,
which enables the direct computation of $\im\omega\cdot Y(\omega)$
utilizing the elements of the list variable \code{Y}. Remember, a
pure DFT would map to frequencies in the range of $0\leq f<f_{\mathrm{s}}$.
It follows that the negative frequencies $-f_{\mathrm{s}}/2\leq f<0$
were projected into the upper half of the frequency domain $f_{\mathrm{s}}/2\leq f<f_{\mathrm{s}}$.
In case of \code{center = F}, the latter frequency range would be
generated and used in further calculations, which must be considered
to prevent wrong results.

\subsection[Center a discrete Spectrum]{\label{subsec:Center-a-spectrum}Center a discrete spectrum}

As initially explained in section~\ref{subsec:Integral-Transforms}
the spectrum of a continuous function $s(t)$ is defined in the range
of $-\infty<\omega<\infty$. In contrast to that, its counterpart
the discrete Fourier transform produces a $2\pi$-periodic spectrum,
which is defined for a positive number of samples $0\leq n<N$ and
$0\leq m$ frequencies. The negative frequency components are projected
into the the range of $f_{\mathrm{s}}/2\leq f<f_{\mathrm{s}}$, so
the spectrum becomes mirror-symmetric with respect to the Nyquist
frequency $f_{\mathrm{s}}/2$.
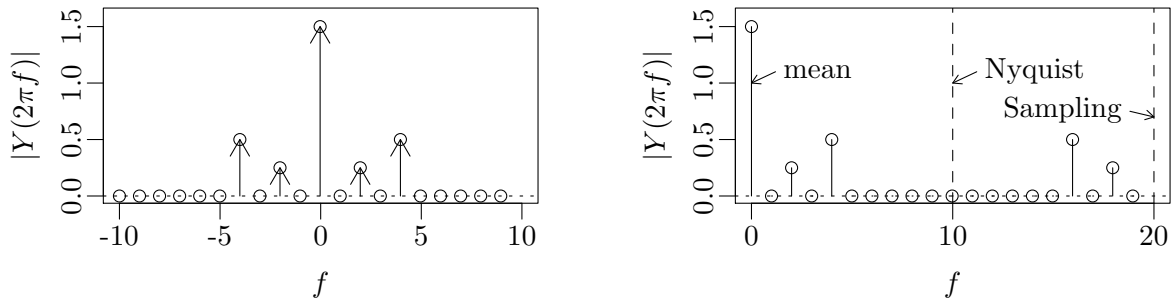
\begin{figure}
\subfloat[Natural spectrum (arrows), DFT spectrum (circles)]{
\begin{tikzpicture}[x=1pt,y=1pt]
\definecolor{fillColor}{RGB}{255,255,255}
\path[use as bounding box,fill=fillColor,fill opacity=0.00] (0,0) rectangle (198.74,115.63);
\begin{scope}
\path[clip] ( 36.00, 36.00) rectangle (198.74,109.63);
\definecolor{drawColor}{RGB}{0,0,0}

\path[draw=drawColor,line width= 0.4pt,line join=round,line cap=round] ( 42.03, 38.73) circle (  2.25);

\path[draw=drawColor,line width= 0.4pt,line join=round,line cap=round] ( 49.54, 38.73) circle (  2.25);

\path[draw=drawColor,line width= 0.4pt,line join=round,line cap=round] ( 57.05, 38.73) circle (  2.25);

\path[draw=drawColor,line width= 0.4pt,line join=round,line cap=round] ( 64.57, 38.73) circle (  2.25);

\path[draw=drawColor,line width= 0.4pt,line join=round,line cap=round] ( 72.08, 38.73) circle (  2.25);

\path[draw=drawColor,line width= 0.4pt,line join=round,line cap=round] ( 79.60, 38.73) circle (  2.25);

\path[draw=drawColor,line width= 0.4pt,line join=round,line cap=round] ( 87.11, 60.03) circle (  2.25);

\path[draw=drawColor,line width= 0.4pt,line join=round,line cap=round] ( 94.62, 38.73) circle (  2.25);

\path[draw=drawColor,line width= 0.4pt,line join=round,line cap=round] (102.14, 49.38) circle (  2.25);

\path[draw=drawColor,line width= 0.4pt,line join=round,line cap=round] (109.65, 38.73) circle (  2.25);

\path[draw=drawColor,line width= 0.4pt,line join=round,line cap=round] (117.16,102.64) circle (  2.25);

\path[draw=drawColor,line width= 0.4pt,line join=round,line cap=round] (124.68, 38.73) circle (  2.25);

\path[draw=drawColor,line width= 0.4pt,line join=round,line cap=round] (132.19, 49.38) circle (  2.25);

\path[draw=drawColor,line width= 0.4pt,line join=round,line cap=round] (139.70, 38.73) circle (  2.25);

\path[draw=drawColor,line width= 0.4pt,line join=round,line cap=round] (147.22, 60.03) circle (  2.25);

\path[draw=drawColor,line width= 0.4pt,line join=round,line cap=round] (154.73, 38.73) circle (  2.25);

\path[draw=drawColor,line width= 0.4pt,line join=round,line cap=round] (162.24, 38.73) circle (  2.25);

\path[draw=drawColor,line width= 0.4pt,line join=round,line cap=round] (169.76, 38.73) circle (  2.25);

\path[draw=drawColor,line width= 0.4pt,line join=round,line cap=round] (177.27, 38.73) circle (  2.25);

\path[draw=drawColor,line width= 0.4pt,line join=round,line cap=round] (184.78, 38.73) circle (  2.25);
\end{scope}
\begin{scope}
\path[clip] (  0.00,  0.00) rectangle (198.74,115.63);
\definecolor{drawColor}{RGB}{0,0,0}

\path[draw=drawColor,line width= 0.4pt,line join=round,line cap=round] ( 42.03, 36.00) -- (192.71, 36.00);

\path[draw=drawColor,line width= 0.4pt,line join=round,line cap=round] ( 42.03, 36.00) -- ( 42.03, 30.00);

\path[draw=drawColor,line width= 0.4pt,line join=round,line cap=round] ( 79.70, 36.00) -- ( 79.70, 30.00);

\path[draw=drawColor,line width= 0.4pt,line join=round,line cap=round] (117.37, 36.00) -- (117.37, 30.00);

\path[draw=drawColor,line width= 0.4pt,line join=round,line cap=round] (155.04, 36.00) -- (155.04, 30.00);

\path[draw=drawColor,line width= 0.4pt,line join=round,line cap=round] (192.71, 36.00) -- (192.71, 30.00);

\node[text=drawColor,anchor=base,inner sep=0pt, outer sep=0pt, scale=  1.00] at ( 42.03, 20.40) {-10};

\node[text=drawColor,anchor=base,inner sep=0pt, outer sep=0pt, scale=  1.00] at ( 79.70, 20.40) {-5};

\node[text=drawColor,anchor=base,inner sep=0pt, outer sep=0pt, scale=  1.00] at (117.37, 20.40) {0};

\node[text=drawColor,anchor=base,inner sep=0pt, outer sep=0pt, scale=  1.00] at (155.04, 20.40) {5};

\node[text=drawColor,anchor=base,inner sep=0pt, outer sep=0pt, scale=  1.00] at (192.71, 20.40) {10};

\path[draw=drawColor,line width= 0.4pt,line join=round,line cap=round] ( 36.00, 38.73) -- ( 36.00,102.64);

\path[draw=drawColor,line width= 0.4pt,line join=round,line cap=round] ( 36.00, 38.73) -- ( 30.00, 38.73);

\path[draw=drawColor,line width= 0.4pt,line join=round,line cap=round] ( 36.00, 60.03) -- ( 30.00, 60.03);

\path[draw=drawColor,line width= 0.4pt,line join=round,line cap=round] ( 36.00, 81.34) -- ( 30.00, 81.34);

\path[draw=drawColor,line width= 0.4pt,line join=round,line cap=round] ( 36.00,102.64) -- ( 30.00,102.64);

\node[text=drawColor,rotate= 90.00,anchor=base,inner sep=0pt, outer sep=0pt, scale=  1.00] at ( 27.60, 38.73) {0.0};

\node[text=drawColor,rotate= 90.00,anchor=base,inner sep=0pt, outer sep=0pt, scale=  1.00] at ( 27.60, 60.03) {0.5};

\node[text=drawColor,rotate= 90.00,anchor=base,inner sep=0pt, outer sep=0pt, scale=  1.00] at ( 27.60, 81.34) {1.0};

\node[text=drawColor,rotate= 90.00,anchor=base,inner sep=0pt, outer sep=0pt, scale=  1.00] at ( 27.60,102.64) {1.5};

\path[draw=drawColor,line width= 0.4pt,line join=round,line cap=round] ( 36.00, 36.00) --
	(198.74, 36.00) --
	(198.74,109.63) --
	( 36.00,109.63) --
	( 36.00, 36.00);
\end{scope}
\begin{scope}
\path[clip] (  0.00,  0.00) rectangle (198.74,115.63);
\definecolor{drawColor}{RGB}{0,0,0}

\node[text=drawColor,anchor=base,inner sep=0pt, outer sep=0pt, scale=  1.00] at (117.37,  2.40) {$f$};

\node[text=drawColor,rotate= 90.00,anchor=base,inner sep=0pt, outer sep=0pt, scale=  1.00] at (  9.60, 72.82) {$\left|Y(2\pi f)\right|$};
\end{scope}
\begin{scope}
\path[clip] ( 36.00, 36.00) rectangle (198.74,109.63);
\definecolor{drawColor}{RGB}{0,0,0}

\path[draw=drawColor,line width= 0.4pt,line join=round,line cap=round] ( 87.11, 38.73) -- ( 87.11, 60.03);

\path[draw=drawColor,line width= 0.4pt,line join=round,line cap=round] ( 90.72, 53.77) --
	( 87.11, 60.03) --
	( 83.50, 53.77);

\path[draw=drawColor,line width= 0.4pt,line join=round,line cap=round] (102.14, 38.73) -- (102.14, 49.38);

\path[draw=drawColor,line width= 0.4pt,line join=round,line cap=round] (105.75, 43.12) --
	(102.14, 49.38) --
	( 98.52, 43.12);

\path[draw=drawColor,line width= 0.4pt,line join=round,line cap=round] (117.16, 38.73) -- (117.16,102.64);

\path[draw=drawColor,line width= 0.4pt,line join=round,line cap=round] (120.78, 96.39) --
	(117.16,102.64) --
	(113.55, 96.39);

\path[draw=drawColor,line width= 0.4pt,line join=round,line cap=round] (132.19, 38.73) -- (132.19, 49.38);

\path[draw=drawColor,line width= 0.4pt,line join=round,line cap=round] (135.80, 43.12) --
	(132.19, 49.38) --
	(128.58, 43.12);

\path[draw=drawColor,line width= 0.4pt,line join=round,line cap=round] (147.22, 38.73) -- (147.22, 60.03);

\path[draw=drawColor,line width= 0.4pt,line join=round,line cap=round] (150.83, 53.77) --
	(147.22, 60.03) --
	(143.60, 53.77);

\path[draw=drawColor,line width= 0.4pt,dash pattern=on 1pt off 3pt ,line join=round,line cap=round] ( 36.00, 38.73) -- (198.74, 38.73);
\end{scope}
\end{tikzpicture}

}\hfill{}\subfloat[Discrete spectrum]{
\begin{tikzpicture}[x=1pt,y=1pt]
\definecolor{fillColor}{RGB}{255,255,255}
\path[use as bounding box,fill=fillColor,fill opacity=0.00] (0,0) rectangle (198.74,115.63);
\begin{scope}
\path[clip] ( 36.00, 36.00) rectangle (198.74,109.63);
\definecolor{drawColor}{RGB}{0,0,0}

\path[draw=drawColor,line width= 0.4pt,line join=round,line cap=round] ( 42.03,102.64) circle (  2.25);

\path[draw=drawColor,line width= 0.4pt,line join=round,line cap=round] ( 49.54, 38.73) circle (  2.25);

\path[draw=drawColor,line width= 0.4pt,line join=round,line cap=round] ( 57.05, 49.38) circle (  2.25);

\path[draw=drawColor,line width= 0.4pt,line join=round,line cap=round] ( 64.57, 38.73) circle (  2.25);

\path[draw=drawColor,line width= 0.4pt,line join=round,line cap=round] ( 72.08, 60.03) circle (  2.25);

\path[draw=drawColor,line width= 0.4pt,line join=round,line cap=round] ( 79.60, 38.73) circle (  2.25);

\path[draw=drawColor,line width= 0.4pt,line join=round,line cap=round] ( 87.11, 38.73) circle (  2.25);

\path[draw=drawColor,line width= 0.4pt,line join=round,line cap=round] ( 94.62, 38.73) circle (  2.25);

\path[draw=drawColor,line width= 0.4pt,line join=round,line cap=round] (102.14, 38.73) circle (  2.25);

\path[draw=drawColor,line width= 0.4pt,line join=round,line cap=round] (109.65, 38.73) circle (  2.25);

\path[draw=drawColor,line width= 0.4pt,line join=round,line cap=round] (117.16, 38.73) circle (  2.25);

\path[draw=drawColor,line width= 0.4pt,line join=round,line cap=round] (124.68, 38.73) circle (  2.25);

\path[draw=drawColor,line width= 0.4pt,line join=round,line cap=round] (132.19, 38.73) circle (  2.25);

\path[draw=drawColor,line width= 0.4pt,line join=round,line cap=round] (139.70, 38.73) circle (  2.25);

\path[draw=drawColor,line width= 0.4pt,line join=round,line cap=round] (147.22, 38.73) circle (  2.25);

\path[draw=drawColor,line width= 0.4pt,line join=round,line cap=round] (154.73, 38.73) circle (  2.25);

\path[draw=drawColor,line width= 0.4pt,line join=round,line cap=round] (162.24, 60.03) circle (  2.25);

\path[draw=drawColor,line width= 0.4pt,line join=round,line cap=round] (169.76, 38.73) circle (  2.25);

\path[draw=drawColor,line width= 0.4pt,line join=round,line cap=round] (177.27, 49.38) circle (  2.25);

\path[draw=drawColor,line width= 0.4pt,line join=round,line cap=round] (184.78, 38.73) circle (  2.25);
\end{scope}
\begin{scope}
\path[clip] (  0.00,  0.00) rectangle (198.74,115.63);
\definecolor{drawColor}{RGB}{0,0,0}

\path[draw=drawColor,line width= 0.4pt,line join=round,line cap=round] ( 36.00, 38.73) -- ( 36.00,102.64);

\path[draw=drawColor,line width= 0.4pt,line join=round,line cap=round] ( 36.00, 38.73) -- ( 30.00, 38.73);

\path[draw=drawColor,line width= 0.4pt,line join=round,line cap=round] ( 36.00, 60.03) -- ( 30.00, 60.03);

\path[draw=drawColor,line width= 0.4pt,line join=round,line cap=round] ( 36.00, 81.34) -- ( 30.00, 81.34);

\path[draw=drawColor,line width= 0.4pt,line join=round,line cap=round] ( 36.00,102.64) -- ( 30.00,102.64);

\node[text=drawColor,rotate= 90.00,anchor=base,inner sep=0pt, outer sep=0pt, scale=  1.00] at ( 27.60, 38.73) {0.0};

\node[text=drawColor,rotate= 90.00,anchor=base,inner sep=0pt, outer sep=0pt, scale=  1.00] at ( 27.60, 60.03) {0.5};

\node[text=drawColor,rotate= 90.00,anchor=base,inner sep=0pt, outer sep=0pt, scale=  1.00] at ( 27.60, 81.34) {1.0};

\node[text=drawColor,rotate= 90.00,anchor=base,inner sep=0pt, outer sep=0pt, scale=  1.00] at ( 27.60,102.64) {1.5};

\path[draw=drawColor,line width= 0.4pt,line join=round,line cap=round] ( 36.00, 36.00) --
	(198.74, 36.00) --
	(198.74,109.63) --
	( 36.00,109.63) --
	( 36.00, 36.00);
\end{scope}
\begin{scope}
\path[clip] (  0.00,  0.00) rectangle (198.74,115.63);
\definecolor{drawColor}{RGB}{0,0,0}

\node[text=drawColor,anchor=base,inner sep=0pt, outer sep=0pt, scale=  1.00] at (117.37,  2.40) {$f$};

\node[text=drawColor,rotate= 90.00,anchor=base,inner sep=0pt, outer sep=0pt, scale=  1.00] at (  9.60, 72.82) {$\left|Y(2\pi f)\right|$};
\end{scope}
\begin{scope}
\path[clip] (  0.00,  0.00) rectangle (198.74,115.63);
\definecolor{drawColor}{RGB}{0,0,0}

\path[draw=drawColor,line width= 0.4pt,line join=round,line cap=round] ( 42.03, 36.00) -- (192.71, 36.00);

\path[draw=drawColor,line width= 0.4pt,line join=round,line cap=round] ( 42.03, 36.00) -- ( 42.03, 30.00);

\path[draw=drawColor,line width= 0.4pt,line join=round,line cap=round] (117.37, 36.00) -- (117.37, 30.00);

\path[draw=drawColor,line width= 0.4pt,line join=round,line cap=round] (192.71, 36.00) -- (192.71, 30.00);

\node[text=drawColor,anchor=base,inner sep=0pt, outer sep=0pt, scale=  1.00] at ( 42.03, 20.40) {0};

\node[text=drawColor,anchor=base,inner sep=0pt, outer sep=0pt, scale=  1.00] at (117.37, 20.40) {10};

\node[text=drawColor,anchor=base,inner sep=0pt, outer sep=0pt, scale=  1.00] at (192.71, 20.40) {20};
\end{scope}
\begin{scope}
\path[clip] ( 36.00, 36.00) rectangle (198.74,109.63);
\definecolor{drawColor}{RGB}{0,0,0}

\path[draw=drawColor,line width= 0.4pt,line join=round,line cap=round] ( 42.03, 38.73) -- ( 42.03,102.64);

\path[draw=drawColor,line width= 0.4pt,line join=round,line cap=round] ( 49.54, 38.73) -- ( 49.54, 38.73);

\path[draw=drawColor,line width= 0.4pt,line join=round,line cap=round] ( 57.05, 38.73) -- ( 57.05, 49.38);

\path[draw=drawColor,line width= 0.4pt,line join=round,line cap=round] ( 64.57, 38.73) -- ( 64.57, 38.73);

\path[draw=drawColor,line width= 0.4pt,line join=round,line cap=round] ( 72.08, 38.73) -- ( 72.08, 60.03);

\path[draw=drawColor,line width= 0.4pt,line join=round,line cap=round] ( 79.60, 38.73) -- ( 79.60, 38.73);

\path[draw=drawColor,line width= 0.4pt,line join=round,line cap=round] ( 87.11, 38.73) -- ( 87.11, 38.73);

\path[draw=drawColor,line width= 0.4pt,line join=round,line cap=round] ( 94.62, 38.73) -- ( 94.62, 38.73);

\path[draw=drawColor,line width= 0.4pt,line join=round,line cap=round] (102.14, 38.73) -- (102.14, 38.73);

\path[draw=drawColor,line width= 0.4pt,line join=round,line cap=round] (109.65, 38.73) -- (109.65, 38.73);

\path[draw=drawColor,line width= 0.4pt,line join=round,line cap=round] (117.16, 38.73) -- (117.16, 38.73);

\path[draw=drawColor,line width= 0.4pt,line join=round,line cap=round] (124.68, 38.73) -- (124.68, 38.73);

\path[draw=drawColor,line width= 0.4pt,line join=round,line cap=round] (132.19, 38.73) -- (132.19, 38.73);

\path[draw=drawColor,line width= 0.4pt,line join=round,line cap=round] (139.70, 38.73) -- (139.70, 38.73);

\path[draw=drawColor,line width= 0.4pt,line join=round,line cap=round] (147.22, 38.73) -- (147.22, 38.73);

\path[draw=drawColor,line width= 0.4pt,line join=round,line cap=round] (154.73, 38.73) -- (154.73, 38.73);

\path[draw=drawColor,line width= 0.4pt,line join=round,line cap=round] (162.24, 38.73) -- (162.24, 60.03);

\path[draw=drawColor,line width= 0.4pt,line join=round,line cap=round] (169.76, 38.73) -- (169.76, 38.73);

\path[draw=drawColor,line width= 0.4pt,line join=round,line cap=round] (177.27, 38.73) -- (177.27, 49.38);

\path[draw=drawColor,line width= 0.4pt,line join=round,line cap=round] (184.78, 38.73) -- (184.78, 38.73);

\path[draw=drawColor,line width= 0.4pt,dash pattern=on 1pt off 3pt ,line join=round,line cap=round] ( 36.00, 38.73) -- (198.74, 38.73);

\path[draw=drawColor,line width= 0.4pt,dash pattern=on 4pt off 4pt ,line join=round,line cap=round] (117.37, 36.00) -- (117.37,109.63);

\path[draw=drawColor,line width= 0.4pt,dash pattern=on 4pt off 4pt ,line join=round,line cap=round] (192.71, 36.00) -- (192.71,109.63);

\path[draw=drawColor,line width= 0.4pt,line join=round,line cap=round] ( 51.03, 85.60) -- ( 42.03, 81.34);

\path[draw=drawColor,line width= 0.4pt,line join=round,line cap=round] ( 44.08, 84.31) --
	( 42.03, 81.34) --
	( 45.63, 81.04);

\path[draw=drawColor,line width= 0.4pt,line join=round,line cap=round] (126.37, 85.60) -- (117.37, 81.34);

\path[draw=drawColor,line width= 0.4pt,line join=round,line cap=round] (119.43, 84.31) --
	(117.37, 81.34) --
	(120.97, 81.04);

\path[draw=drawColor,line width= 0.4pt,line join=round,line cap=round] (183.71, 70.69) -- (192.71, 68.55);

\path[draw=drawColor,line width= 0.4pt,line join=round,line cap=round] (189.25, 67.52) --
	(192.71, 68.55) --
	(190.09, 71.03);

\node[text=drawColor,anchor=base west,inner sep=0pt, outer sep=0pt, scale=  1.00] at ( 54.03, 83.30) {mean};

\node[text=drawColor,anchor=base west,inner sep=0pt, outer sep=0pt, scale=  1.00] at (129.37, 83.30) {Nyquist};

\node[text=drawColor,anchor=base east,inner sep=0pt, outer sep=0pt, scale=  1.00] at (180.71, 68.39) {Sampling};
\end{scope}
\end{tikzpicture}

}

\caption{\label{fig:difference-natSpec_DFT}Two spectra of function \eqref{eq:2-1_Artificial_Function}.
The difference between the natural spectrum in (a) and the normal
discrete Fourier transform in (b) is the symmetry. In (a) there is
a symmetric spectrum around zero, whereas the in (b) the point of
symmetry is $f_{\mathrm{s}}/2$. The dots in panel (a) indicate the
result of the DFT of the modulated sampling series of \eqref{eq:2-1_Artificial_Function}.
Note, the exceptional first bin in (b), where the mean value is stored.}
\end{figure}

Figure~\ref{fig:difference-natSpec_DFT}a shows the result of the
continuous Fourier transform applied to the artificial example function
\eqref{eq:2-1_Artificial_Function}
\[
y(t)=\sin(2\pi\cdot4t)+0.5\cdot\cos(2\pi\cdot2t)+1.5.
\]
It is clear that the mean value $\overline{y(t)}$ is located at $f=0$
and the sinusoidal components are located symmetricly around zero.
In contrast to that, the discrete Fourier transform produces a data
vector $\hat{Y}{}_{m}$, to which only positive discrete frequencies
can be assigned. To convert the data vector into a zero-symmetric
representation, the function $y(t)$ must be modulated by $f_{\mathrm{s}}/2$
\begin{equation}
y^{*}(t)=y(t)\cdot\e^{\im\pi f_{\mathrm{s}}\cdot t}.\label{eq:2-2_Modulation}
\end{equation}
This method utilizes the shifting or damping property of the Fourier
transform as stated in Appendix~\ref{sec:Properties-of_FourierTransform}
Equation~\ref{eq:APP_Fourier_Prop_Damping}, assuming a damping factor
of $a=1$. Doing so the complete spectrum (Fig.~\ref{fig:difference-natSpec_DFT}b),
including the mean value shifts towards the half of the sampling frequency
right in the middle, as it is indicated in Figure~\ref{fig:difference-natSpec_DFT}a.

\paragraph{Application}

In an equally spaced sampling series the discrete time vector $t=n\cdot T_{\mathrm{s}}$
changes equation \eqref{eq:2-2_Modulation} to
\begin{eqnarray}
y^{*}(n\cdot T_{\mathrm{s}}) & = & \hat{y}{}_{n}\cdot\e^{\im\pi f_{\mathrm{s}}\cdot n\cdot T_{\mathrm{s}}}\nonumber \\
 & = & \hat{y}{}_{n}\cdot\underset{{\scriptstyle 1-1+1-\dots}}{\underbrace{\e^{\im\pi n}}}\nonumber \\
 & = & \hat{y}{}_{n}\cdot(-1)^{n}\label{eq:2-2-Centerd_spectrum}
\end{eqnarray}
and simplifies it. From equation \eqref{eq:2-2-Centerd_spectrum}
follows that the modulation of the data vector can be computed easily
by a multiplication of the alternating series $(-1)^{n}$. Subsequently,
all this can be extended into $n$ dimensions, so even 2D spectra
can be centered \citep[p. 154]{gonzalez2002digital}. 

The following \proglang{R}-code which is part of the \code{spec.fft()}
function explains the method. First, assume a vector \code{x} for
the spatial location where the measurements \code{y} were taken. 

\begin{CodeChunk}
\begin{Code}
nx <- length(x)
ny <- length(y)      # should equal nx
Ts <- min(diff(x))   # calculate the sampling
y <- y*(-1)^(1:ny)   # modulate the data vector
# determine the corresponding frequency vector
fx <- seq(-(nx)/(2*Ts * nx),(nx - 2)/(2*Ts * nx),length.out = nx)

# calculate and normalize the spectrum
A <- fft(y)*1/ny
\end{Code}
\end{CodeChunk}

In the calculation of the frequency vector \code{fx}, the maximum
frequency is \code{(nx - 2)/(2{*}Ts {*} nx)}, in which \code{(nx - 2)}
considers the position of $f=0$ and the intrinsic periodicity of
the signal \code{y}. Remember the examples in section~\ref{subsec:Bandlimitation},
when the last possible sampling point matches the first point of the
next signal period.

The back transform of the centered spectrum is the simple reverse
of the described procedure. The function \code{spec.fft()} supports
both methods to decompose a data vector into its centered or non-centered
spectral representation. The usage of the code is explained below. 

\begin{CodeChunk}
\begin{Code}
Ts <- 0.05 # defining the sampling
 x <- seq(0,1,by = Ts)
 x <- x[-length(x)] # removes last sample to
                    # avoid error with periodicity
 y <- function(x) sin(2*pi * 4*x)+0.5 * cos(2*pi * 2*x) + 1.5

 Y <- spec.fft(y(x), x, center = F) # the normal fft()
Yc <- spec.fft(y(x), x, center = T) # prior modulation
# spec.fft() returns a list object 
# containing the following:
# Y$fx, (Y$fy), Y$A, Y$x, Y$y, (Y$z)
\end{Code}
\end{CodeChunk}

\subsection[The analytic signal]{\label{subsec:The-analytic-signal}The analytic signal}

The concept of the analytic signal is based on the idea that a real-valued
signal, e.\,g. some measured data can only contain positive frequencies.
At least in the 1D case, the interpretation of negative frequencies
is almost impossible. Next to that, a real-world signal does only
exist for positive times, usually. This makes the signal causal, which
also must be taken into account, see Equation~\eqref{eq:ApB_Backtransform}
in Appendix~\ref{sec:Derivation_Hilbert}. The foundation of the
analytic signal is the Hilbert transform. For a short introduction
on the Hilbert transformation itself and its properties, refer to
section~\ref{par:Hilbert-transform}. Besides that, the attribute
``analytic'' originates from the Cauchy-Riemann conditions for differentiability,
which is fulfilled for analytic functions in this sense \citep{cohen1995timefrequency}.

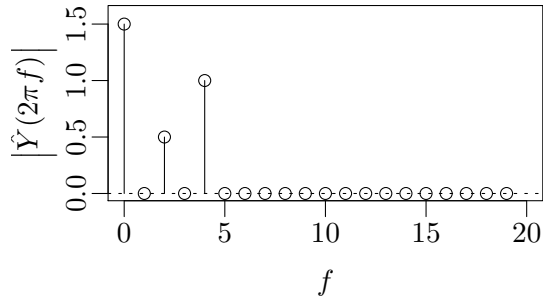
\begin{figure}
\centering
\begin{tikzpicture}[x=1pt,y=1pt]
\definecolor{fillColor}{RGB}{255,255,255}
\path[use as bounding box,fill=fillColor,fill opacity=0.00] (0,0) rectangle (198.74,115.63);
\begin{scope}
\path[clip] ( 36.00, 36.00) rectangle (198.74,109.63);
\definecolor{drawColor}{RGB}{0,0,0}

\path[draw=drawColor,line width= 0.4pt,line join=round,line cap=round] ( 42.03,102.64) circle (  2.25);

\path[draw=drawColor,line width= 0.4pt,line join=round,line cap=round] ( 49.56, 38.73) circle (  2.25);

\path[draw=drawColor,line width= 0.4pt,line join=round,line cap=round] ( 57.10, 60.03) circle (  2.25);

\path[draw=drawColor,line width= 0.4pt,line join=round,line cap=round] ( 64.63, 38.73) circle (  2.25);

\path[draw=drawColor,line width= 0.4pt,line join=round,line cap=round] ( 72.17, 81.34) circle (  2.25);

\path[draw=drawColor,line width= 0.4pt,line join=round,line cap=round] ( 79.70, 38.73) circle (  2.25);

\path[draw=drawColor,line width= 0.4pt,line join=round,line cap=round] ( 87.23, 38.73) circle (  2.25);

\path[draw=drawColor,line width= 0.4pt,line join=round,line cap=round] ( 94.77, 38.73) circle (  2.25);

\path[draw=drawColor,line width= 0.4pt,line join=round,line cap=round] (102.30, 38.73) circle (  2.25);

\path[draw=drawColor,line width= 0.4pt,line join=round,line cap=round] (109.84, 38.73) circle (  2.25);

\path[draw=drawColor,line width= 0.4pt,line join=round,line cap=round] (117.37, 38.73) circle (  2.25);

\path[draw=drawColor,line width= 0.4pt,line join=round,line cap=round] (124.91, 38.73) circle (  2.25);

\path[draw=drawColor,line width= 0.4pt,line join=round,line cap=round] (132.44, 38.73) circle (  2.25);

\path[draw=drawColor,line width= 0.4pt,line join=round,line cap=round] (139.97, 38.73) circle (  2.25);

\path[draw=drawColor,line width= 0.4pt,line join=round,line cap=round] (147.51, 38.73) circle (  2.25);

\path[draw=drawColor,line width= 0.4pt,line join=round,line cap=round] (155.04, 38.73) circle (  2.25);

\path[draw=drawColor,line width= 0.4pt,line join=round,line cap=round] (162.58, 38.73) circle (  2.25);

\path[draw=drawColor,line width= 0.4pt,line join=round,line cap=round] (170.11, 38.73) circle (  2.25);

\path[draw=drawColor,line width= 0.4pt,line join=round,line cap=round] (177.65, 38.73) circle (  2.25);

\path[draw=drawColor,line width= 0.4pt,line join=round,line cap=round] (185.18, 38.73) circle (  2.25);
\end{scope}
\begin{scope}
\path[clip] (  0.00,  0.00) rectangle (198.74,115.63);
\definecolor{drawColor}{RGB}{0,0,0}

\path[draw=drawColor,line width= 0.4pt,line join=round,line cap=round] ( 42.03, 36.00) -- (192.71, 36.00);

\path[draw=drawColor,line width= 0.4pt,line join=round,line cap=round] ( 42.03, 36.00) -- ( 42.03, 30.00);

\path[draw=drawColor,line width= 0.4pt,line join=round,line cap=round] ( 79.70, 36.00) -- ( 79.70, 30.00);

\path[draw=drawColor,line width= 0.4pt,line join=round,line cap=round] (117.37, 36.00) -- (117.37, 30.00);

\path[draw=drawColor,line width= 0.4pt,line join=round,line cap=round] (155.04, 36.00) -- (155.04, 30.00);

\path[draw=drawColor,line width= 0.4pt,line join=round,line cap=round] (192.71, 36.00) -- (192.71, 30.00);

\node[text=drawColor,anchor=base,inner sep=0pt, outer sep=0pt, scale=  1.00] at ( 42.03, 20.40) {0};

\node[text=drawColor,anchor=base,inner sep=0pt, outer sep=0pt, scale=  1.00] at ( 79.70, 20.40) {5};

\node[text=drawColor,anchor=base,inner sep=0pt, outer sep=0pt, scale=  1.00] at (117.37, 20.40) {10};

\node[text=drawColor,anchor=base,inner sep=0pt, outer sep=0pt, scale=  1.00] at (155.04, 20.40) {15};

\node[text=drawColor,anchor=base,inner sep=0pt, outer sep=0pt, scale=  1.00] at (192.71, 20.40) {20};

\path[draw=drawColor,line width= 0.4pt,line join=round,line cap=round] ( 36.00, 38.73) -- ( 36.00,102.64);

\path[draw=drawColor,line width= 0.4pt,line join=round,line cap=round] ( 36.00, 38.73) -- ( 30.00, 38.73);

\path[draw=drawColor,line width= 0.4pt,line join=round,line cap=round] ( 36.00, 60.03) -- ( 30.00, 60.03);

\path[draw=drawColor,line width= 0.4pt,line join=round,line cap=round] ( 36.00, 81.34) -- ( 30.00, 81.34);

\path[draw=drawColor,line width= 0.4pt,line join=round,line cap=round] ( 36.00,102.64) -- ( 30.00,102.64);

\node[text=drawColor,rotate= 90.00,anchor=base,inner sep=0pt, outer sep=0pt, scale=  1.00] at ( 27.60, 38.73) {0.0};

\node[text=drawColor,rotate= 90.00,anchor=base,inner sep=0pt, outer sep=0pt, scale=  1.00] at ( 27.60, 60.03) {0.5};

\node[text=drawColor,rotate= 90.00,anchor=base,inner sep=0pt, outer sep=0pt, scale=  1.00] at ( 27.60, 81.34) {1.0};

\node[text=drawColor,rotate= 90.00,anchor=base,inner sep=0pt, outer sep=0pt, scale=  1.00] at ( 27.60,102.64) {1.5};

\path[draw=drawColor,line width= 0.4pt,line join=round,line cap=round] ( 36.00, 36.00) --
	(198.74, 36.00) --
	(198.74,109.63) --
	( 36.00,109.63) --
	( 36.00, 36.00);
\end{scope}
\begin{scope}
\path[clip] (  0.00,  0.00) rectangle (198.74,115.63);
\definecolor{drawColor}{RGB}{0,0,0}

\node[text=drawColor,anchor=base,inner sep=0pt, outer sep=0pt, scale=  1.00] at (117.37,  2.40) {$f$};

\node[text=drawColor,rotate= 90.00,anchor=base,inner sep=0pt, outer sep=0pt, scale=  1.00] at (  9.60, 72.82) {$\left|\hat{Y}(2\pi f)\right|$};
\end{scope}
\begin{scope}
\path[clip] ( 36.00, 36.00) rectangle (198.74,109.63);
\definecolor{drawColor}{RGB}{0,0,0}

\path[draw=drawColor,line width= 0.4pt,line join=round,line cap=round] ( 42.03, 38.73) -- ( 42.03,102.64);

\path[draw=drawColor,line width= 0.4pt,line join=round,line cap=round] ( 57.10, 38.73) -- ( 57.10, 60.03);

\path[draw=drawColor,line width= 0.4pt,line join=round,line cap=round] ( 72.17, 38.73) -- ( 72.17, 81.34);

\path[draw=drawColor,line width= 0.4pt,dash pattern=on 1pt off 3pt ,line join=round,line cap=round] ( 36.00, 38.73) -- (198.74, 38.73);
\end{scope}
\end{tikzpicture}\caption{\label{fig:Single-sided-spectrum}Single sided spectrum of the analytic
signal of $y(t)$}
\end{figure}

Now, lets introduce the analytic signal 
\begin{equation}
a(t)=s(t)+\im\mathcal{H}\left(s(t)\right).\label{eq:analyticSignal}
\end{equation}
 It is formally defined by the sum of a signal $s(t)$ and its Hilbert
transform $\im\mathcal{H}\left(s(t)\right)$ multiplied with the imaginary
unit. So the real-valued signal $s(t)$ transforms to $a(t)$, which
now has an one-sided spectrum. Frequencies above the Nyquist-frequency
$f_{\mathrm{s}}/2$ are coerced to be zero and frequencies below that
are gained by the factor of two, so that energy is conserved. This
can be seen in Figure~\ref{fig:Single-sided-spectrum}, which shows
the result of a DFT applied on the analytic representation of the
sampled signal from Figure~\ref{fig:Over-sampled-simple-signal}.
Starting from that several applications can be derived.

First, all the calculations which utilize the signal's frequencies,
e.\,g. the estimation of the derivative introduced in section~\ref{par:derivative_of_a_function},
get simplified with respect to discrete data because the distinction
of frequencies above and below $f_{s}/2$ can be omitted. The example
of the derivative in section~\ref{par:derivative_of_a_function}
instead utilized the discrete \emph{centered} spectrum, which provides
a frequency vector with positive and negative parts, which can be
multiplied directly to the spectrum.

Second, the imaginary part of $a(t)$ equals the Hilbert transform
of its real part. The Hilbert transform's phase shifting properties
form the basis to calculate, for instance, the envelope function,
which is discussed in the next section. The analytic signal also provides
a way to estimate the instantaneous frequency of a signal. The latter
ends up in the empirical mode decomposition, which gives a time depending
spectral decomposition according to \citet{huang1998theempirical},
which is out of scope of this guide. In section~\ref{subsec:Non-stationary-processes}
it will be shown how to utilize the Fourier transform and how to overcome
its draw backs in conjunction with the analytic signal of data of
a non-stationary processes.

\paragraph{Implementation}

According to Appendix~\ref{sec:Derivation_Hilbert} the function
\code{analyticFunction(y)} provided by the \pkg{spectral} package
calculates the analytic signal. An example is given in Figure~\ref{fig:Single-sided-spectrum},
which displays the analytic representation of the sampled artificial
example function \eqref{eq:2-1_Artificial_Function} from section~\ref{subsec:Bandlimitation}.
Here the spectrum is single sided and all the amplitudes have their
correct value. Note that all components above $f_{\mathrm{s}}/2$
are zero, because the upper half of the spectrum is projected into
the lower half by this method. In consequence, the illustration in
Figure~\ref{fig:Single-sided-spectrum} is in contrast to the given
examples (e.\,g. fig~\ref{fig:difference-natSpec_DFT}) above, where
individual amplitudes only contain the half of the true value.

The program code

\begin{CodeChunk}
\begin{Code}
X <- fft(x) / length(x) # normalized DFT-spectrum

f <- 0:(length(X) - 1)  # virtual spatial vector
f <- f - mean(f)        # shifted by half the length
                        # so 0 is in the middle

   X <- X * (1 - sign( f - 0.5 )) # Hilbert transform
X[1] <- 0.5 * X[1]      # correct mean value
ht <- fft(X, inverse = T)
\end{Code}
\end{CodeChunk}
explains how the the analytic signal is calculated. Here the part
\code{(1 - sign( f - 0.5 ))} solves three things. First, this statement
shifts the phase with respect to the sign of the virtual frequency
vector \code{f}, whereby the ``negative'' frequencies are located
in the upper half of the data set. This provides a better solution
than Equation~\eqref{eq:1_Discete_Hilbert} because the evenness
of the data sets length does not matter anymore. Second, since \code{f}
is an integer vector, the subtraction \code{f - 0.5} circumvents
the problem $\mathrm{sign}(0)=0$, which avoids an error with even
length data sets at $f=0$. And third, the above code calculates the
Hilbert transform and the analytic signal in one step in the spectral
domain by combining equations \eqref{eq:1_F(H)} and \eqref{eq:2-2_Modulation}.

The code below produces the single-sided spectrum \code{Y} given
in Figure~\ref{fig:Single-sided-spectrum}. Note, that the amplitudes
correspond to the real input values of the function~\eqref{eq:2-1_Artificial_Function}
and all spectral parts above $f_{\mathrm{s}}/2$ are zero.

\begin{CodeChunk}
\begin{Code}
x <- seq(0, 1, by = 0.05)
x <- x[ -length(x) ] 

y <- function(x) sin(2*pi * 4*x) + 0.5 * cos(2*pi * 2*x) + 1.5
Y <- spec.fft( analyticFunction(y(x)), x, center = F )
\end{Code}
\end{CodeChunk}

\subsection[Calculating the envelope]{\label{subsec:Calculating-the-envelope}Calculating the envelope}

Sometimes it becomes necessary to calculate the envelope function
of data. Different approaches are possible, for instance finding all
maxima and then fiting a spline function to these points. However,
in the following the spectral way to do that is introduced.

First, remember the trigonometric identity 
\begin{equation}
\sin^{2}(x)+\cos^{2}(x)=1,
\end{equation}
which becomes the key component of the calculation later. Next to
that, assume a function
\begin{equation}
y(t)=A(t)\cdot\cos(\omega_{0}t)
\end{equation}
with $A(t)$ as the envelope function, which is modulated with the
carrier $\cos(\omega_{0}t)$. It is clear now that calculating the
envelope is equivalent to an amplitude demodulation.

Now, the amplitude function is gained by 
\begin{eqnarray*}
A(t) & = & \sqrt{y(t)^{2}+y^{*}(t)^{2}}\\
 & = & A(t)\cdot\underset{=1}{\underbrace{\sqrt{\cos^{2}(\omega_{0})+\sin^{2}(\omega_{0})}}},
\end{eqnarray*}
where $y^{*}(t)$ denotes the signal $y(t)$ phase shifted by $\pi/2$,
so that $y^{*}(t)$ equals the Hilbert transform 
\begin{eqnarray}
y^{*}(t) & = & A(t)\cdot\sin(\omega_{0}t)\\
 & = & \mathcal{H}\left(y(t)\right).\label{eq:Ideal_Phase_Shifter}
\end{eqnarray}
Under the condition of \emph{sufficient} band limitation the statement
above can be expressed as 
\begin{equation}
A(t)=\sqrt{y(t)^{2}+\mathcal{H}\left(y(t)\right)^{2}},\label{eq:Envelope}
\end{equation}
which works, because \eqref{eq:Ideal_Phase_Shifter} acts like an
ideal phase shifter as discussed in section~\ref{subsec:Integral-Transforms}
and Appendix~\ref{sec:Derivation_Hilbert}.

\paragraph{Application}

\begin{figure}
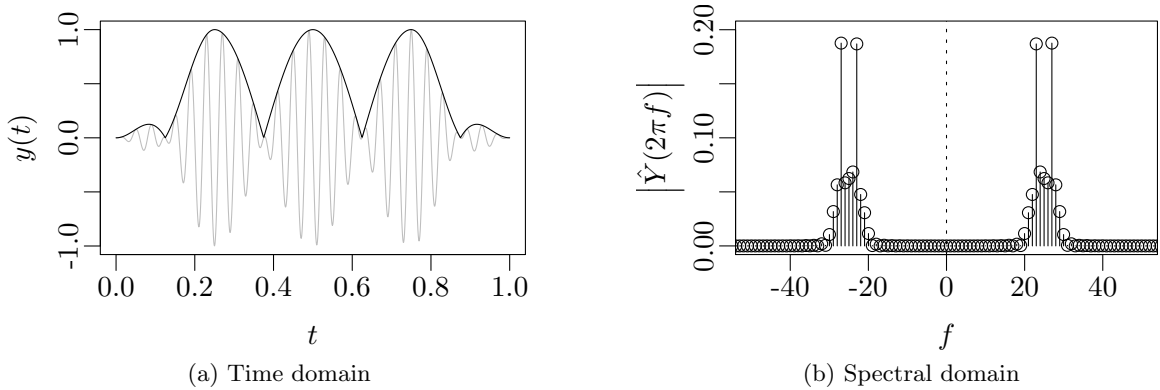

\subfloat[Time domain]{\input{Pictures/Example07_envelope.tex}

}\hfill{}\subfloat[Spectral domain]{\input{Pictures/Example07_envelope_spectrum.tex}

}

\caption{\label{fig:Envelope-function}The envelope function of a signal. The
spectrum (b) clearly indicates the band width, which is needed to
reconstruct the envelope. In case of a noisy signal the band pass
filter must have at least this width to resolve reasonable results.}
\end{figure}
The calculation is straight forward and follows Equation~\eqref{eq:Envelope}.
The \pkg{spectral} package includes the function \code{envelope()}
to perform the calculation of an 1D envelope. The only problem with
this is the band limitation, which is necessary to achieve reasonable
results. Compare Figure~\ref{fig:Envelope-function}b, here the required
bandwidth to demodulate the envelope becomes clearly visible in the
negative and positive half plane.

\subsection[Convolution]{\label{subsec:Convolution}Convolution}

The Fourier transform does not only map a spatial function into the
spectral domain, moreover it also converts several mathematical operations.
One example is the derivation of
\begin{equation}
\diff f(t)/\diff t\leftrightarrow\im\omega F(\omega),
\end{equation}
which equals a simple multiplication with the complex variable $\im\omega$
in the spectral domain. 

In the following, the convolution $\int_{-\infty}^{\infty}f(\tau)g(t-\tau)\diff\tau$
will be introduced by the identities 
\begin{eqnarray}
\mathcal{F}\left(\underset{_{=f(t)*g(t)}}{\underbrace{\int_{-\infty}^{\infty}f(\tau)g(t-\tau)\diff\tau}}\right) & = & \mathcal{F}\left(f(t)\right)\cdot\mathcal{F}\left(g(t)\right)\label{eq:2-4_folding_spatial}\\
\text{and }\mathcal{F}\left(f(t)\cdot g(t)\right) & = & \int_{-\infty}^{\infty}F(\tau)G(\omega-\tau)\diff\tau,\label{eq:2-4_folding_spectral}
\end{eqnarray}
which are valid only if the integrals $\int_{-\infty}^{\infty}\left|f(t)\right|^{2}\diff t$
and $\int_{-\infty}^{\infty}\left|g(t)\right|^{2}\diff t$ are existent.
Equation \eqref{eq:2-4_folding_spatial} describes the folding operation
(convolution) in the spacial domain, which corresponds to a simple
multiplication in spectral domain.

The working principle and the implementation of \eqref{eq:2-4_folding_spatial}
and \eqref{eq:2-4_folding_spectral} via the fast Fourier transform
(FFT) do form one of the most powerful tools in the field of numeric
computation.

\paragraph{Example \textendash{} Polynomial multiplication}

Suppose two polynomials of the degree $n=2$. Then the multiplication
of these two would be 
\begin{eqnarray}
\left(a_{0}+a_{1}x+a_{2}x^{2}\right)\cdot\left(b_{0}+b_{1}x+b_{2}x^{2}\right) & = & \hphantom{a_{0}b_{0}x^{2}}\mathllap{a_{0}b_{0}}+\hphantom{a_{0}b_{0}x^{2}}\mathllap{a_{0}b_{1}x}+\hphantom{a_{0}b_{0}x^{2}}\mathllap{a_{0}b_{2}x^{2}}+\\
 &  & \hphantom{a_{0}b_{0}x^{2}}\mathllap{a_{1}b_{0}x}+\hphantom{a_{0}b_{0}x^{2}}\mathllap{a_{1}b_{1}x^{2}}+a_{1}b_{2}x^{3}+\nonumber \\
 &  & \hphantom{a_{0}b_{0}x^{2}}\mathllap{a_{2}b_{0}x^{2}}+\hphantom{a_{0}b_{0}x^{2}}\mathllap{a_{2}b_{1}x^{3}}+\hphantom{a_{0}b_{0}x^{2}}\mathllap{a_{2}b_{2}x^{4}}\nonumber \\
 & = & \sum_{i=0}^{2n}c_{i}x^{i},
\end{eqnarray}
which finally gives an expression of degree of $2n=4$. Performing
the expansion will end in a convolution of the two coefficient vectors,
which must be $2n+1$ elements long to fit the result in the output
vector. Doing all that will spend $(n+1)^{2}=9$ floating point multiplications
on a computer. The statement
\begin{equation}
\left(\begin{array}{c}
a_{\text{0}}\\
a_{1}\\
a_{2}\\
0\\
0
\end{array}\right)*\left(\begin{array}{c}
b_{0}\\
b_{1}\\
b_{2}\\
0\\
0
\end{array}\right)=\left(\begin{array}{c}
a_{0}b_{0}\\
a_{0}b_{1}+a_{1}b_{0}\\
a_{0}b_{2}+a_{1}b_{1}+a_{2}b_{0}\\
a_{1}b_{2}+a_{2}b_{1}\\
a_{2}b_{2}
\end{array}\right)\label{eq:2-4_Poly_fold}
\end{equation}
expresses the convolution and the result, where the symbol $*$ denotes
the folding operator. Alternatively, this operation can be done with
the help of the Fourier transform. According to \eqref{eq:2-4_folding_spatial}
\[
\vec{c}=\mathcal{F}^{-1}\left(\mathcal{F}\left(\vec{a}\right)\cdot\mathcal{F}\left(\vec{b}\right)\right)
\]
requires the transform to be calculated three times.

The great advantage of using the Fourier transform, instead of the
straight forward expansion, is the scaling of $n\cdot\log_{2}n$ for
the implementation of the FFT \citep[Chap. 30]{cormen2001introduction}.
\begin{table}
\centering{}%
\begin{tabular}{ccc}
\toprule 
\emph{degree} & \emph{expansion} & \emph{FFT}\tabularnewline
\midrule
8 & 64 & 209\tabularnewline
16 & 256 & 500\tabularnewline
32 & 1024 & 1175\tabularnewline
64 & 4096 & 2714\tabularnewline
\bottomrule
\end{tabular}\caption{\label{tab:Computation-costs}Computation costs for the polynomial
multiplication. The calculation amount for the FFT is estimated by
$3\cdot\left(2n+1\right)\log_{2}\left(2n+1\right)$. }
\end{table}
As shown in table~\ref{tab:Computation-costs} the FFT accelerates
the calculation from a value of $n>32$ for this example.

The standard algorithm of the discrete FFT is limited by a vector
size of $N=2^{n}$, which can be overcome with the implementation
of the DFT method as described by \citet{frigo2005thedesign}, which
even allows prime numbered lengths without a loss of performance.

As a simple example the following code illustrates the method described
above.

\begin{CodeChunk}
\begin{CodeInput}
# defining the coefficients
a <- c(1:3, 0, 0)
b <- c(5:7, 0, 0)
# calculating the FFT
A <- fft(a)
B <- fft(b)

# convolve via multiplication of A and B
c  <- Re( fft(A * B, inverse=T) / length(a))
c2 <- convolve(a, b, conj = F) # would give the same result
print(c)
\end{CodeInput}
\begin{CodeOutput}

####### OUTPUT ########
> 5 16 34 32 21
\end{CodeOutput}
\end{CodeChunk}
The result is given in the last line of the program listing. The \emph{\proglang{R}}-command
\code{convolve(x, y, conj = F)} uses the same mechanism and produces
exactly the same results.

\subsection[Spectral filtering in conjunction with the sampling theorem]{\label{subsec:Spectral-filtering}Spectral filtering in conjunction
with the sampling theorem}

Every sampled signal has its starting point $t=0$, followed by a
point in time $T$ where it ends. The sampling procedure itself and
the nature of the signal finally define the underlying band limitation.
The intention of the first part of this section is to show the relation
between the measurement and the consequences which arises if a causal
signal is sampled. It turns out that in conjunction with the Fourier
transform several convolutions take place in the time and frequency
domain by selecting the maximum time $T$ and the maximum frequency.

\begin{figure}
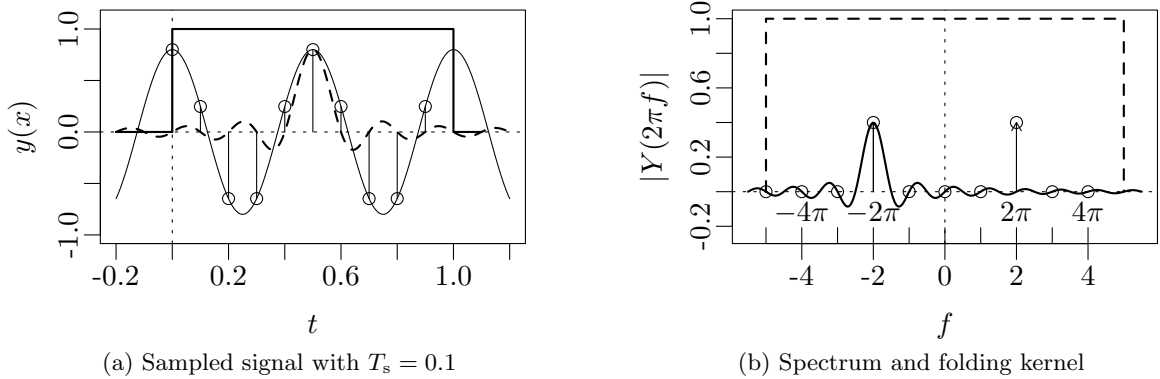

\subfloat[Sampled signal with $T_{\mathrm{s}}=0.1$]{\input{Pictures/Example06_sampled_signal.tex}

}\hfill{}\subfloat[Spectrum and folding kernel]{\input{Pictures/Example06_spectrum_and_folding_signal.tex}

}

\caption{\label{fig:2-4_First_window_Filtering}In (a) the function $y(t)$
from Equation~\eqref{eq:Spec_filt_example} is sampled in the time
range $0\leq t<1$. The finite time interval corresponds to a rectangular
weight function. The continuous spectrum (arrows in (b)) is then convoluted
with the spectral representation of that rectangular window. The black
line indicates the convolution step for one left spectral part. Note,
that only the Dirac impulse generates a value, because all other components
remain zero. The dashed line in (b) indicates the limitation to a
maximum frequency, which corresponds to a convolution in the time
domain (a). Open symbols show the sampling (a) and the result of the
DFT (b).}
\end{figure}

Regarding that, it is important to understand that the filter process
already begins with the sampling of the signal. Figure~\ref{fig:2-4_First_window_Filtering}
illustrates that for the function
\begin{equation}
y(t)=0.8\cdot\cos(2\pi\cdot2t),\label{eq:Spec_filt_example}
\end{equation}
which fits perfectly into the period of the sampling window. In principle,
this function is defined on the interval of $-\infty<t<\infty$. Then,
the data acquisition process defines a starting point at $t=0$ where
the measurement begins and an end point $t=T$ where the measurement
stops. This is illustrated by the bold rectangular window 
\begin{equation}
w(t)=\begin{cases}
1 & ,0\leq t<T=1\\
0 & \text{, elsewhere}
\end{cases}
\end{equation}
in Figure~\ref{fig:2-4_First_window_Filtering}a. This window function
picks out a range $0\leq t<T$ of $y(t)$ so that the signal 
\begin{equation}
s(t)=w(t)\cdot y(t),
\end{equation}
is now defined by $y(t)$ multiplied by $w(t)$. According to the
convolution properties of the Fourier transform (see section~\ref{subsec:Convolution}),
this operation corresponds to the convolution 
\begin{eqnarray}
S(\omega) & = & \int_{-\infty}^{\infty}W(\omega-\tau)\cdot Y(\omega)\diff\tau\nonumber \\
 & = & W(\omega)*Y(\omega)\label{eq:2-5_convolved_Spectrum}
\end{eqnarray}
of the ideal spectrum $Y(\omega)=\mathcal{F}(y(t))$ with the spectral
representation 
\begin{equation}
W(\omega)=\frac{1}{\im\omega}\left(1-\e^{-\im\omega T}\right)
\end{equation}
of the window function $w(t)$, which is in fact Shannon's sampling
function \citep{shannon1949communication}. The consequence for the
signal's spectrum $S(\omega)$ is that the ideal Dirac-spectrum $Y(\omega)$
smears out to $\sin(x)/x$ like functions, because each arrow is weighted
with $W(\omega)$. Note, the zeros of the resulting spectrum are in
an equidistant spacing of $\mathit{\delta f}=1/T$ now. At this moment
the original function $y(t)$ is just windowed and not sampled.

The next step is to limit the spectrum in Figure~\ref{fig:2-4_First_window_Filtering}b
to a maximum frequency, e.\,g. $f_{\mathrm{max}}=1/2T_{\mathrm{s}}$,
by applying an additional rectangular window in the frequency domain
\begin{equation}
V(\omega)=\begin{cases}
1 & ,\left|\omega\right|\leq\left(2T_{\mathrm{s}}\right)^{-1}\\
0 & \text{, elsewhere}
\end{cases}.
\end{equation}
The parameter $T_{\mathrm{s}}$ becomes the period of the sampling
frequency. The important link to the Nyquist frequency would be clarified
later. However, the corresponding representation of $V(\omega)$ in
the time domain
\begin{equation}
v(t)=\frac{\sin\left(\left(2T_{\mathrm{s}}\right)^{-1}\cdot t\right)}{t}
\end{equation}
is displayed by the dashed line in Figure~\ref{fig:2-4_First_window_Filtering}a.
In the context of the convolution, which now takes place in time domain,
the windowed signal 
\begin{equation}
s(t)=\left(w(t)\cdot y(t)\right)*v(t)\label{eq:2-5_pre_sampled}
\end{equation}
is finally convoluted with $v(t)$. But still, the result \eqref{eq:2-5_pre_sampled}
is not sampled yet which is done in the next step.

Sampling is a procedure where individual values of the function $s(t)$
are selected and stored. With regard to the nature of the sinc-like
functions in Figure~\ref{fig:2-4_First_window_Filtering}, the sampling
should take place at each possible zero of $v(t)$, as well as $v(t=0)$
which is the only sampling point with a non-zero value of $v(t)$.
This behavior guarantees that the convolution \eqref{eq:2-5_pre_sampled}
rejects everything, but the points of sampling. In the same moment
also the distance in between them is set to $\delta t=T_{\mathrm{s}}$,
because the function $v(t)$ provides two zeros in an interval of
$(2T_{\mathrm{s}})^{-1}\cdot t\leq2\pi$, which correspond to the
next non-zero samples of the neighboring points. In other words, by
choosing the next sample $v(t)$ must be shifted by exactly $T_{\mathrm{s}}$
to reject the former sample and to take the current value. Note, violating
this restriction, because the location of sampling points jitters,
will lead to an error in the discrete spectral representation $\hat{S}{}_{m}$. 

Finally, this conceptual description gives a second access to the
question why the sampling frequency should be at least twice the maximum
signal frequency. A detailed mathematical derivation of the discrete
Fourier transform and the properties of sampling is given in the book
of \citet[Chap 3]{debnath2015wavelet}. However, the discrete sampling
given in Figure~\ref{fig:2-4_First_window_Filtering}a and the finite
length of the sampling series causes the convolution \eqref{eq:2-5_convolved_Spectrum}
to take the only possible peaks in the spectrum at $f=\{-2,2\}$,
because all other $2\pi$-periodic values evaluate exactly to zero. 

If the signal period does not fit perfectly into the window, which
is the case in reality, the kernel function $W(\omega)$ never matches
the true Dirac, but instead of that it sums up values into the left
and right bin of the corresponding discrete Fourier vector $\hat{S}{}_{m}$.
To reduce these side effects, other window functions can be used in
advance, before the DFT takes place. The intent of the window functions
(Hamming-, Blackman- or Tukey-window) is to convolve the spectrum
with a steeper decreasing function amplitude compared to $W(\omega)$
to minimize the effect of the period's misfit. Moreover $w(t)$ has
an infinite spectrum, which is then truncated in the finite and discrete
data processing, which leads to additional side effects if the period
does not fit the sampling window. However, window functions can help
to reduce such interferences, which are generated when local non-periodic
events or infinite spectra are present.

\paragraph{Application}

\begin{figure}
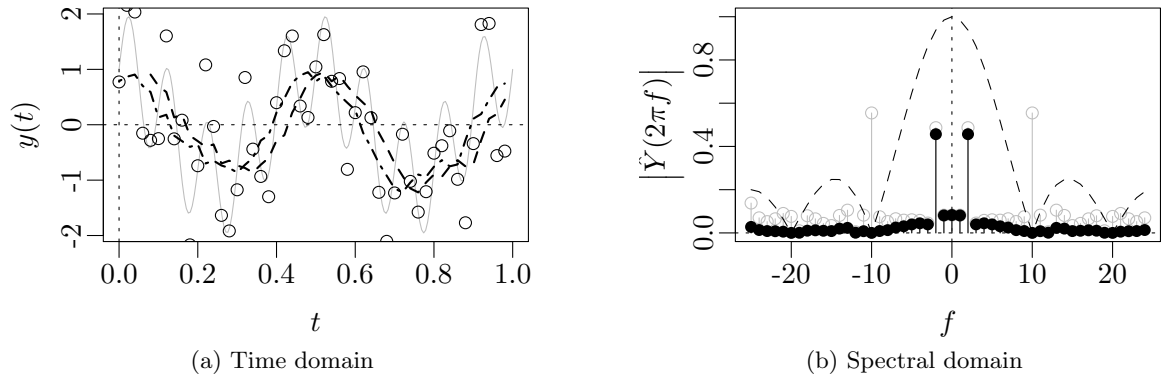

\subfloat[Time domain]{\input{Pictures/Example06_moving_average_time_domain.tex}

}\hfill{}\subfloat[Spectral domain]{\input{Pictures/Example06_moving_average_frequency_domain.tex}

}

\caption{\label{fig:moving_average}Calculating the moving average to reduce
the noise and to reject the high frequency. Panel (a) shows the original
function (gray line) and the noisy sampling points (open symbols).
The result of the moving average in the time domain is given in (a)
for the convolution (dashed) and the shifted FFT (dash-dotted). (b)
displays the spectrum of the original function (gray), the filtered
result (black) and the kernel function (dashed).}
\end{figure}
Convolution filter in the time domain, such as the moving average
filter, are easy to calculate with the help of the FFT. Given the
signal 
\[
s(t)=\left.\cos(4\pi\cdot t)+\sin(20\pi\cdot t)+\mathcal{N}(0,0.5)\right|_{0\leq t<1},
\]
which is defined in the range of $0\leq t<1$. The moving average
of the sampled signal $\hat{s}_{n}$ with the filter kernel length
of $N_{\mathrm{K}}=5$ 
\begin{equation}
\overline{\hat{s}_{n}}=\sum_{i=1}^{N_{\mathrm{K}}}\hat{k}_{i}\cdot\hat{s}_{n-i},
\end{equation}
is then calculated for the $n$\textsuperscript{th} element by taking
the weighted sum of $N_{\mathrm{K}}$ elements before that element.
The filter kernel $\hat{k}$ for the example in Figure~\ref{fig:moving_average}
is represented by the weighting coefficients 
\begin{equation}
\hat{k}=\frac{1}{5}\left(\begin{array}{c}
1\\
1\\
1\\
1\\
1
\end{array}\right).
\end{equation}

With respect to the spectrum of $\hat{k}$, it becomes clear that
the length of $\hat{k}$ is chosen in a way that the high frequency
part $\sin(20\pi\cdot t)$ is rejected completely. But it is also
evident that other frequency components still remain in the result
or were damped falsely, like the low frequency part $\cos(4\pi\cdot t)$.

Figure~\ref{fig:moving_average}a illustrates that the application
of the moving average filter in the time domain will lead to a phase-shift
in the signal. Working in the spectral domain, one will gain the possibility
to shift the signal back and reconstruct the whole definition range
as follows

\begin{eqnarray}
\mathcal{F}\left(\overline{s(t)}\right) & = & \e^{\im\omega\frac{N_{\mathrm{K}}}{2}}\cdot S(\omega)\cdot K(\omega)\\
\mathcal{F}\left(\overline{\hat{s}(n)}\right) & = & \e^{\im2\pi\frac{\left(0..(N-1)\right)\cdot N_{\mathrm{K}}}{2N}}\cdot\hat{S}_{m}\cdot\hat{K}_{m}.
\end{eqnarray}
The result of this operation illustrates the dash-dotted line in Figure~\ref{fig:moving_average}a.
The code which calculates this is presented below.

\begin{CodeChunk}
\begin{Code}
K <- fft( c(k, rep(0, length(y) - length(k))))

y_spec <- base::Re( fft( exp( 2i*pi * (1:length(y)-1) 
                             / length(y) * length(k)%/%2 
                             ) * fft( y ) / length(y) 
                         * K , inverse=T) )
\end{Code}
\end{CodeChunk}

Pay attention to the statement \code{fft( c(k, rep(0, length(y) - length(k))))}
in line 1. It first fills up $k$ with zeros until the length of the
whole data set, before the FFT is calculated. This is because the
multiplication in frequency space can only take place between vectors
of equal lengths.

With respect to the spectrum of the kernel $\hat{K}$ it becomes evident
that the moving average filter is not able to suppress higher frequencies
completely.

\paragraph{Example \textendash{} The ideal low pass filter}

Since the moving average has some drawbacks (because of its infinite
frequency response) one should think about an ideal (low pass) filter
with a finite frequency response. Such a filter has the advantage
that unwanted spectral components can be rejected completely. On the
other hand, this results in an infinite long kernel in the time domain,
which is nevertheless calculated in the spectral domain.

\begin{figure}
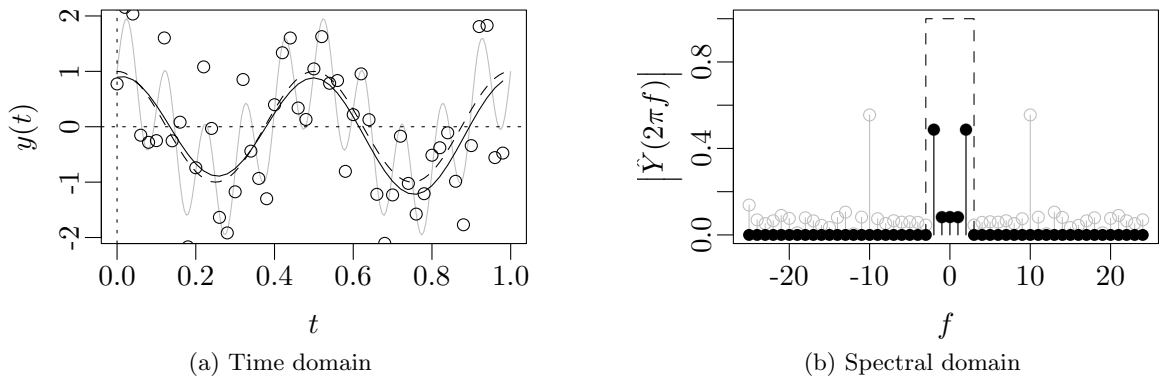

\subfloat[Time domain]{\input{Pictures/Example06_ideal_filter_time_domain.tex}

}\hfill{}\subfloat[Spectral domain]{\input{Pictures/Example06_ideal_filter_frequency_domain.tex}

}

\caption{\label{fig:Ideal-lowpass-filter}Calculating the ideal low pass filter.
Panel (a) shows the original function (gray line) and the noisy sampling
points (open symbols). The result of the ideal low pass filter in
the time domain is given in (a) by the solid line. The dashed line
in (a) illustrated the low frequency part of the signal with $f=2$.
Panel (b) displays the spectrum of the original function (gray), the
filtered result (black) and the spectral kernel function (dashed).}
\end{figure}
The basic principle of such a filter is to set each component above
a certain frequency $\left|\omega_{\mathrm{max}}\right|$ value to
zero
\begin{equation}
S(\left|\omega\right|>\omega_{\mathrm{max}})=0.
\end{equation}
Figure~\ref{fig:Ideal-lowpass-filter} illustrates the procedure.
Compared to the previous moving average example in Figure~\ref{fig:moving_average},
the output signal's shape changed to a more smooth and accurate form.
However, there remains a difference compared to the ideal low frequency
part (dashed line in Figure~\ref{fig:Ideal-lowpass-filter}a), which
is the result of the remnant spectral components around the signal
frequency $f=2$.

A code snippet to calculate the ideal low pass is presented below.

\begin{CodeChunk}
\begin{Code}
# assuming x and y to hold the time vector
# and the sampling points

Y <- spec.fft(y, x)
# doing the filtering
Y$A[abs(Y$fx) > 3] <- 0

# reconstruct the filtered signal
yf <- spec.fft(Y, inverse = T)$y
\end{Code}
\end{CodeChunk}

The \pkg{spectral} package provides the \code{filter.fft(y,x,BW,fc=0,n=3)}
function to do that in one step. It utilizes the \code{analyticFunction(y)}
to filter the one-sided spectrum and enables the user to apply a band
pass filter with the arbitrary polynomial weight function 
\begin{equation}
w\propto\omega^{n}.
\end{equation}
The parameter \code{n} sets the polynomial degree, whereas 3 is close
to the moving average. Passing higher values, say \code{n=10}, will
successively converge to the ideal bandpass solution. All this is
because some times it is desirable to have a smooth weight of frequencies
and some times not.

The used code then changes to the following example.

\begin{CodeChunk}
\begin{Code}
yf <- filter.fft(y, x, fc = 0, BP = 3, n = 10)
\end{Code}
\end{CodeChunk}

As seen in Figure~\ref{fig:Ideal-lowpass-filter} the bandwidth \code{BP}
is symmetric around the center frequency \code{fc = 0}, so the values
must be set in accordance to the desired window which should remain
in the spectrum. Utilizing the analytic signal representation avoids
mistakes when programming the weight function and keeps the code clean.

\paragraph{Example \textendash{} Noise reduction via autocorrelation}

By choosing an appropriate kernel function it becomes possible to
extract certain features of the signal. The previous example illustrated
how a low pass filter can filter out a noisy signal by zeroing all
upper frequency components in the spectrum. In the time domain this
corresponds to a folding operation of the input data with an infinite
``sinc''-like kernel function.

Now the opposite approach is to ask what the most significant periodic
components within the signal are. First of all, the continuous autocorrelation
function 
\begin{eqnarray}
acf(\tau) & = & \int_{-\infty}^{\infty}s(t)s^{*}(t-\tau)\diff\tau\label{eq:Continuous_ACF}\\
 & = & \mathcal{F}^{-1}\left(S(\omega\right)\cdot S^{*}(\omega))\label{eq:Continuous_ACF2}
\end{eqnarray}
provides a mechanism to examine a data set or function with respect
to its self-similarity. The equations above convolute the signal $s(t)$
with its complex conjugate $s^{*}(t)$. According to the convolution
properties, equation \eqref{eq:Continuous_ACF} can be expressed as
the inverse Fourier transform of the product of the signals spectra.
Note, for a real-valued signal $s(t)=s^{*}(t)$. Suppose the signal
consists of a stationary sinusoidal signal it is quite evident that
the $acf$ rejects the noise which overlays $s(t)$. The underlying
noise equals to a non-stationary process and therefore it cancels
out by utilizing the $acf$. A generalization of this statement is
the Wiener-Khinchin theorem. Its application is discussed in detail
by \citet[Chap 1.9]{cohen1995timefrequency}, \citet[Chap. 4]{marks2009handbook}
and \citet{wiener1949extrapolation}.

\begin{figure}
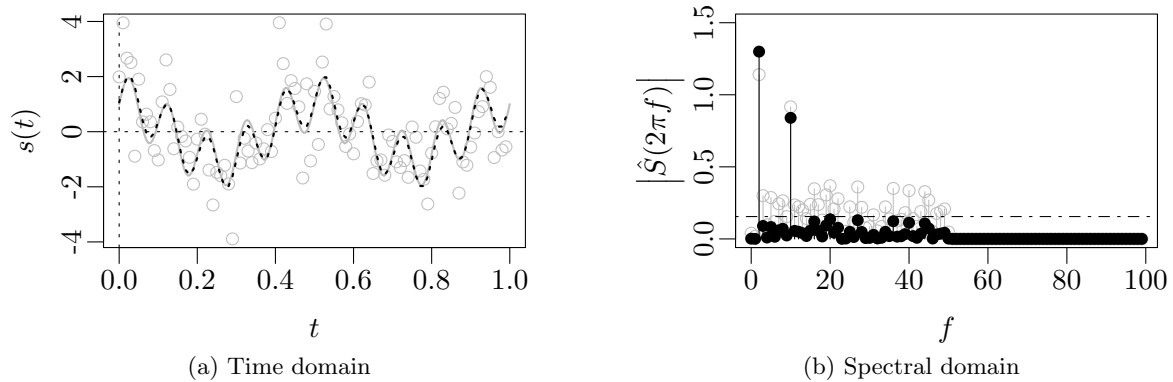

\subfloat[Time domain]{\input{Pictures/Example06_Wiener.tex}

}\hfill{}\subfloat[Spectral domain]{\input{Pictures/Example06_Wiener_spectrum.tex}

}

\caption{\label{fig:Noise-reduction-with-ACF}Noise reduction with autocorrelation.
In the left panel (a) the sampled function (gray line) is overlaid
with noise (circles). The bold line is the result of the weighted
filtering procedure. In the frequency domain (b) the single sided
spectra of the analytic functions are displayed. The input function
(gray symbols) contains much noise which is reduced due the autocorrelation
(dotted black). The horizontal line indicates the threshold below
that all spectral components are ignored. }
\end{figure}
Assuming the function
\begin{equation}
s(t)=\cos(2\pi\cdot2\cdot t)+\sin(2\pi\cdot10\cdot t)+\mathcal{N}(0,0.5)
\end{equation}
with some normal distributed noise overlaid. The signal $\hat{s}{}_{n}$
is then the sampled function $s(n\cdot T_{\mathrm{s}})$ with $T_{\mathrm{s}}=0.01$,
so it fits perfectly into one sampling period of $T=1$. This example
is very artificial, real world signals often cause aliasing effects,
so attention must be paid.

The discrete spectrum of the $acf_{n}$ can now be calculated by
\begin{equation}
ACF_{m}=\hat{S}{}_{m}\cdot\hat{S}^{*}{}_{m}.
\end{equation}
Note, the resulting spectrum is real-valued, so the phase information
will be lost. A proper noise reduction can be achieved by defining
a weighting vector
\begin{equation}
W_{m}=\begin{cases}
0 & ,\left|ACF\right|<\mathrm{sd}(\left|ACF_{m}\right|)\\
1 & ,\text{else}
\end{cases},
\end{equation}
which sets every spectral componentsmaller than the standard deviation
of the $ACF_{m}$ to zero. The resulting filtered signal 
\begin{equation}
\hat{s}_{\mathrm{f}}(n)=\mathcal{F}^{-1}\left(W_{m}\cdot\hat{S}{}_{m}\right)
\end{equation}
is given in Figure~\ref{fig:Noise-reduction-with-ACF}. For the given
example it is important to calculate everything with the analytic
function representation to preserve the energy content of the data
vectors. 

\begin{CodeChunk}
\begin{Code}
x <- seq(0, 1, by = Ts)
x <- x[ -length(x) ] # for perfect periodicity
y <- cos(2*pi * 2*x) + sin(2*pi * 10*x) + rnorm(length(x), sd = 1)
# calculating the autocorrelation
Y <- fft( analyticFunction(y) ) / length(y)
ACF <- Y * Conj(Y)

# calculating the weight vector
w <- ACF
w[abs(w) < sd( abs(w) )] <- 0 
w[w != 0] <- 1 

# backtransform of the filtered signal
yf <- Re(fft( Y * w, inverse = T) ) 
\end{Code}
\end{CodeChunk}

\subsection[Non-stationary processes -{}- spatially dependent spectral analysis]{\label{subsec:Non-stationary-processes}Non-stationary processes
\textendash{} spatially dependent spectral analysis}

Many signals in reality do not fulfill the requirement of stationarity.
It is very often the case that the measured signal is overlaid by
a trend or an ``temporal-local'' event that takes place only once.
All these in-stationary and non-periodic processes will spread out
into all frequencies in the corresponding Fourier decomposition of
the data. If a proper band limitation cannot be achieved the mapping
between physical frequencies and the spectral representation might
fail too, as stated in section~\ref{subsec:Bandlimitation}.

\begin{figure}
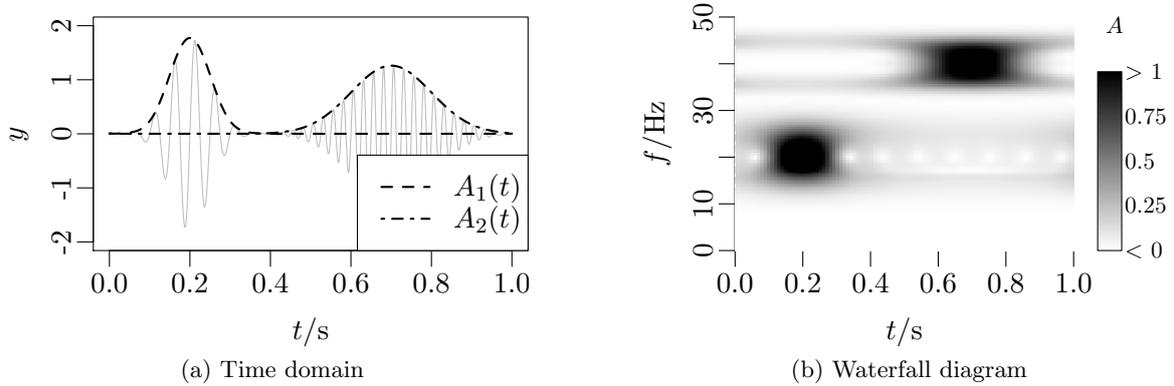

\subfloat[Time domain]{\input{Pictures/Example08_waterfall_simple.tex}

}\hfill{}\subfloat[Waterfall diagram]{\input{Pictures/Example08_waterfall_simple-spectrum.tex}}

\caption{\label{fig:Waterfall_Example1}Basic simple example of a non-stationary
signal (a) with its time depending spectral decomposition in (b).
The different line types of the envelope correspond to the first and
second event.}
\end{figure}

The equation
\begin{equation}
y(t)=\underset{A_{1}(t)}{\underbrace{\frac{1}{\sqrt{2\pi\cdot0.05}}\cdot\e^{-\frac{\left(x-0.2\right)^{2}}{2\cdot0.05^{2}}}}}\cdot\sin\left(2\pi\cdot20\cdot t\right)+\underset{A_{2}(t)}{\underbrace{\frac{1}{\sqrt{2\pi\cdot0.1}}\cdot\e^{-\frac{\left(x-0.7\right)^{2}}{2\cdot0.1^{2}}}}}\cdot\sin\left(2\pi\cdot40\cdot t\right)
\end{equation}
describes a very artificial example of a non-stationary signal, which
is illustrated in Figure~\ref{fig:Waterfall_Example1}a. 
Here two different events occur. The first one has a low frequency
at $f_{1}=\unit[20]{Hz}$, whereas the second one oscillates faster
with $f_{2}=\unit[40]{Hz}$. Both parts are modulated with an Gaussian
envelope to switch each of them on and off. Finally, the right panel
(b) shows the resulting time variant decomposition of the signal $y(t)$.
The \code{waterfall()} function from the \pkg{spectral} package
can be used to calculate such kinds of waterfall diagrams.

Note, this type of analysis produces a two-dimensional time-frequency
map like a shifting FFT, which selects only a window of the data before
doing the spectral decomposition for one time step. The amplitude
functions $A_{1}(t)$ and $A_{2}(t)$ play the role of arbitrary window
functions which mask the time series. The interpretation is that $A_{i}(t)$
contains the information in time where the finite process is located.
All this relies on the assumption that the signal of interest can
be modeled as 
\begin{equation}
y(t)=\sum_{i}A_{i}(t)\cdot\cos(\omega_{i}+\varphi_{i}).
\end{equation}

In contrast to the shifting FFT the introduced approach is to do an
amplitude demodulation by calculating the signal's envelope. The result
is the amplitude function $A(t)$ for a given center frequency $f_{\mathrm{c}}$,
which corresponds to the frequency of the physical process. Finally,
a sufficient bandwidth around $f_{\mathrm{c}}$ guarantees that $A(t)$
will be reconstructed correctly.

The first advantage of this method is that for each single frequency
of interest the whole data set is taken into account. Note, the shifting
FFT would select only a certain range of the data, so information
about low frequencies can be lost if the selected window becomes too
small. However, the overall frequency resolution decreases to the
length of the window size of the shifting FFT. In the end, two signals
that are very close in the frequency space might not be distinguishable
anymore. The second point is the necessary band pass filter at $f_{\mathrm{c}}$,
which is changed according to the actual frequency. This draws attention
to the uncertainty principle, which states that with increasing frequency
the locating of an event becomes sharper but the exact frequency gets
more incorrect.

\paragraph{Implementation}

As stated above, the key component is a band pass filter \code{BP},
which must be applied twice to calculate the two envelopes in Figure~\ref{fig:Waterfall_Example1}.
The following code shows how this can be done in \proglang{R} with
the appropriate functions provided by the \pkg{spectral} package.

\begin{CodeChunk}
\begin{Code}
A1 <- Re( envelope(Re(filter.fft(y,x,fc = 20,BW = 10,n = 10))) )
A2 <- Re( envelope(Re(filter.fft(y,x,fc = 40,BW = 10,n = 10))) )
\end{Code}
\end{CodeChunk}

The next step is to do this calculation for many frequencies. A reasonable
range is to start from $f=0$ up to $f_{\mathrm{s}}/2$. To accelerate
the code the \code{waterfall()} function uses a fast version of the
\code{envelope()} function, which combines the filtering and the
Hilbert transform as follows. 
\begin{CodeChunk}
\begin{Code}
Y.f <- seq(0,(n-1)*df,length.out=n) # Defining the frequency vector
sY.f <- (1 - sign(Y.f-mean(Y.f)))    # calculating the sign for the HT
sY.f[1] <- 1                         # correct first bin
fast_envelope <- function(y,x,fc,BW,nf)
{
   Y <- BP(Y.f,fc,BW,nf) * fft(y)/n * sY.f
  hk <- base::Mod( fft(Y + 1i*Y,inverse=T)/sqrt(2))
  return(hk)
}
\end{Code}
\end{CodeChunk}
Here the working band pass filter is implemented as a weighting vector
multiplied to the amplitudes. The second line in the \code{fast\_envelope()}
function performs the back transform. Compared to the calculation
of \code{A1} and \code{A2} this method saves two back transforms
and one real part extraction.

Finally, the bandwidth calculation is done by an empirical approach,
like
\begin{equation}
BW(f_{\mathrm{c}})=\begin{cases}
4\mathit{\delta f} & f_{\mathrm{c}}<16\mathit{\delta f}\\
f_{\mathrm{c}}/4 & 16\mathit{\delta f}\leq f_{\mathrm{c}}\leq wd\cdot\mathit{\delta f}\\
wd\cdot\mathit{\delta f} & \text{else}.
\end{cases}
\end{equation}
Here $\mathit{\delta f}=1/\Delta x$ denotes the frequency step and
$wd$ is the normalized width of the resulting band pass. The task
of $BW(f_{\mathrm{c}})$ is to widen the frequency band for higher
frequencies, whereas low frequencies take a very small band width.
This somehow pays attention to the uncertainty principle, which states
that ``a narrow waveform yields a wide spectrum and a wide waveform
yields a narrow spectrum and both the time waveform and frequency
spectrum cannot be made arbitrarily small simultaneously'' \citep{skolnik1980introduction}.

\paragraph{Application}

The two following examples illustrate how the \code{waterfall()}
function can be used for intricate signals. First of all the function
\begin{equation}
y(t)=\left|2t-1\right|\cdot\sin\left(2\pi\cdot10\cdot t\right)+\begin{cases}
0 & t\leq0.5\\
\sin\left(2\pi\cdot20\cdot t^{2}\right) & \text{else}
\end{cases}\label{eq:waterfall_example}
\end{equation}
is going to be analyzed. 
\begin{figure}
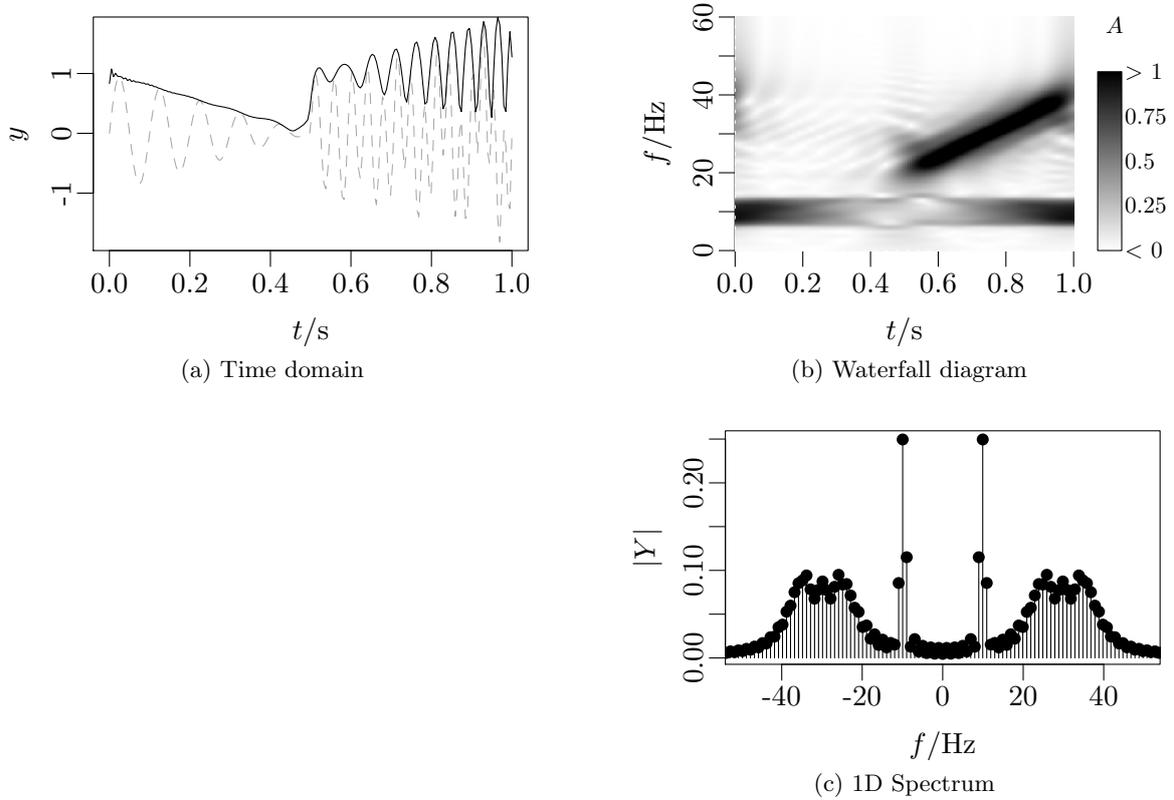

\subfloat[Time domain]{\input{Pictures/Example08_waterfall_complex.tex}

}\hfill{}\subfloat[Waterfall diagram]{\input{Pictures/Example08_waterfall_complex-spectrum.tex}}

\hfill{}\subfloat[1D Spectrum]{\input{Pictures/Example08_waterfall_complex_1D-Spectrum.tex}

}

\caption{\label{fig:Waterfall-Complex-example}Complex example for the temporal-dependent
spectral decomposition. (a) shows the time series and (b) its decomposition.
Panel (c) demonstrates that the simple 1D spectrum would produce misleading
results. Here one can only identify the $\unit[10]{Hz}$ carrier,
but the frequency drift is hidden behind the wide band spectrum at
around $\unit[30]{Hz}$.}
\end{figure}
Figure~\ref{fig:Waterfall-Complex-example} illustrates the results.
The waterfall diagram in panel (b) displays all the features of Equation~\eqref{eq:waterfall_example}.
Note that even the $\left|x\right|$ function and the $\sin\left(x^{2}\right)$
term can be distinguished, whereas the normal time-invariant Fourier
spectrum in Figure~\ref{fig:Waterfall-Complex-example}c hides this
property completely. Here the $\unit[10]{Hz}$ carrier is the only
``correctly'' visible signal component.

\begin{figure}
\subfloat[Outflow temperature.]{\input{Pictures/Example09_Output.tex}

}\hfill{}\subfloat[Ambient temperature.]{\input{Pictures/Example09_environment.tex}

}

\subfloat[Spectrum of the outflow temperature]{
\begin{tikzpicture}[x=1pt,y=1pt]
\definecolor{fillColor}{RGB}{255,255,255}
\path[use as bounding box,fill=fillColor,fill opacity=0.00] (0,0) rectangle (397.48,130.09);
\begin{scope}
\path[clip] (  0.00,  0.00) rectangle (397.48,130.09);
\definecolor{drawColor}{RGB}{0,0,0}

\path[draw=drawColor,line width= 0.4pt,line join=round,line cap=round] ( 36.00, 36.00) -- (355.48, 36.00);

\path[draw=drawColor,line width= 0.4pt,line join=round,line cap=round] ( 36.00, 36.00) -- ( 36.00, 30.00);

\path[draw=drawColor,line width= 0.4pt,line join=round,line cap=round] ( 89.25, 36.00) -- ( 89.25, 30.00);

\path[draw=drawColor,line width= 0.4pt,line join=round,line cap=round] (142.49, 36.00) -- (142.49, 30.00);

\path[draw=drawColor,line width= 0.4pt,line join=round,line cap=round] (195.74, 36.00) -- (195.74, 30.00);

\path[draw=drawColor,line width= 0.4pt,line join=round,line cap=round] (248.99, 36.00) -- (248.99, 30.00);

\path[draw=drawColor,line width= 0.4pt,line join=round,line cap=round] (302.24, 36.00) -- (302.24, 30.00);

\path[draw=drawColor,line width= 0.4pt,line join=round,line cap=round] (355.48, 36.00) -- (355.48, 30.00);

\node[text=drawColor,anchor=base,inner sep=0pt, outer sep=0pt, scale=  1.00] at ( 36.00, 20.40) {11};

\node[text=drawColor,anchor=base,inner sep=0pt, outer sep=0pt, scale=  1.00] at ( 89.25, 20.40) {12};

\node[text=drawColor,anchor=base,inner sep=0pt, outer sep=0pt, scale=  1.00] at (142.49, 20.40) {13};

\node[text=drawColor,anchor=base,inner sep=0pt, outer sep=0pt, scale=  1.00] at (195.74, 20.40) {14};

\node[text=drawColor,anchor=base,inner sep=0pt, outer sep=0pt, scale=  1.00] at (248.99, 20.40) {15};

\node[text=drawColor,anchor=base,inner sep=0pt, outer sep=0pt, scale=  1.00] at (302.24, 20.40) {16};

\node[text=drawColor,anchor=base,inner sep=0pt, outer sep=0pt, scale=  1.00] at (355.48, 20.40) {17};

\path[draw=drawColor,line width= 0.4pt,line join=round,line cap=round] ( 36.00, 44.55) -- ( 36.00,130.09);

\path[draw=drawColor,line width= 0.4pt,line join=round,line cap=round] ( 36.00, 44.55) -- ( 30.00, 44.55);

\path[draw=drawColor,line width= 0.4pt,line join=round,line cap=round] ( 36.00, 56.77) -- ( 30.00, 56.77);

\path[draw=drawColor,line width= 0.4pt,line join=round,line cap=round] ( 36.00, 68.99) -- ( 30.00, 68.99);

\path[draw=drawColor,line width= 0.4pt,line join=round,line cap=round] ( 36.00, 81.21) -- ( 30.00, 81.21);

\path[draw=drawColor,line width= 0.4pt,line join=round,line cap=round] ( 36.00, 93.43) -- ( 30.00, 93.43);

\path[draw=drawColor,line width= 0.4pt,line join=round,line cap=round] ( 36.00,105.65) -- ( 30.00,105.65);

\path[draw=drawColor,line width= 0.4pt,line join=round,line cap=round] ( 36.00,117.87) -- ( 30.00,117.87);

\path[draw=drawColor,line width= 0.4pt,line join=round,line cap=round] ( 36.00,130.09) -- ( 30.00,130.09);

\node[text=drawColor,rotate= 90.00,anchor=base,inner sep=0pt, outer sep=0pt, scale=  1.00] at ( 27.60, 44.55) {10};

\node[text=drawColor,rotate= 90.00,anchor=base,inner sep=0pt, outer sep=0pt, scale=  1.00] at ( 27.60, 68.99) {30};

\node[text=drawColor,rotate= 90.00,anchor=base,inner sep=0pt, outer sep=0pt, scale=  1.00] at ( 27.60, 93.43) {50};

\node[text=drawColor,rotate= 90.00,anchor=base,inner sep=0pt, outer sep=0pt, scale=  1.00] at ( 27.60,117.87) {70};

\path[draw=drawColor,line width= 0.4pt,line join=round,line cap=round] ( 36.00, 36.00) --
	(355.48, 36.00) --
	(355.48,130.09) --
	( 36.00,130.09) --
	( 36.00, 36.00);
\end{scope}
\begin{scope}
\path[clip] (  0.00,  0.00) rectangle (397.48,130.09);
\definecolor{drawColor}{RGB}{0,0,0}

\node[text=drawColor,anchor=base,inner sep=0pt, outer sep=0pt, scale=  1.00] at (195.74,  2.40) {$t$/hrs};

\node[text=drawColor,rotate= 90.00,anchor=base,inner sep=0pt, outer sep=0pt, scale=  1.00] at (  9.60, 83.04) {$T$/min};
\end{scope}
\begin{scope}
\path[clip] ( 36.00, 36.00) rectangle (355.48,130.09);
\node[inner sep=0pt,outer sep=0pt,anchor=south west,rotate=  0.00] at ( 35.73,  35.92) {
	\pgfimage[width=320.02pt,height= 94.24pt,interpolate=false]{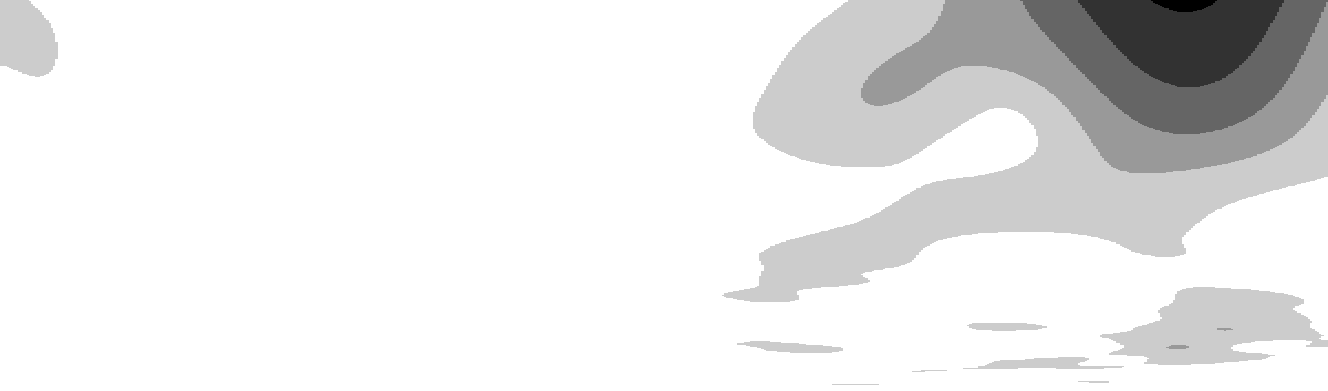}};
\definecolor{drawColor}{RGB}{255,255,255}

\path[draw=drawColor,line width= 0.4pt,dash pattern=on 1pt off 3pt ,line join=round,line cap=round] ( 36.00, 32.33) -- (355.48, 32.33);
\end{scope}
\begin{scope}
\path[clip] (  0.00,  0.00) rectangle (397.48,130.09);
\definecolor{fillColor}{RGB}{255,255,255}

\path[fill=fillColor] (364.48, 36.00) rectangle (373.48, 47.81);
\definecolor{fillColor}{gray}{0.80}

\path[fill=fillColor] (364.48, 47.81) rectangle (373.48, 59.63);
\definecolor{fillColor}{gray}{0.60}

\path[fill=fillColor] (364.48, 59.63) rectangle (373.48, 71.44);
\definecolor{fillColor}{RGB}{101,101,101}

\path[fill=fillColor] (364.48, 71.44) rectangle (373.48, 83.26);
\definecolor{fillColor}{RGB}{50,50,50}

\path[fill=fillColor] (364.48, 83.26) rectangle (373.48, 95.07);
\definecolor{fillColor}{RGB}{0,0,0}

\path[fill=fillColor] (364.48, 95.07) rectangle (373.48,106.89);
\definecolor{drawColor}{RGB}{0,0,0}

\path[draw=drawColor,line width= 0.4pt,line join=round,line cap=round] (364.48, 36.00) rectangle (373.48,106.89);

\node[text=drawColor,anchor=base west,inner sep=0pt, outer sep=0pt, scale=  0.90] at (374.88, 38.81) {<  0};

\node[text=drawColor,anchor=base west,inner sep=0pt, outer sep=0pt, scale=  0.90] at (374.88, 50.62) {0.1};

\node[text=drawColor,anchor=base west,inner sep=0pt, outer sep=0pt, scale=  0.90] at (374.88, 62.44) {0.2};

\node[text=drawColor,anchor=base west,inner sep=0pt, outer sep=0pt, scale=  0.90] at (374.88, 74.25) {0.3};

\node[text=drawColor,anchor=base west,inner sep=0pt, outer sep=0pt, scale=  0.90] at (374.88, 86.07) {0.4};

\node[text=drawColor,anchor=base west,inner sep=0pt, outer sep=0pt, scale=  0.90] at (374.88, 97.88) {>0.5};
\end{scope}
\begin{scope}
\path[clip] (  0.00,  0.00) rectangle (397.48,130.09);
\definecolor{drawColor}{RGB}{0,0,0}

\node[text=drawColor,anchor=base west,inner sep=0pt, outer sep=0pt, scale=  0.90] at (364.48,121.29) {$\frac{A}{^\circ\mathrm{C}}$};
\end{scope}
\end{tikzpicture}

}

\caption{\label{fig:Outflow-temperature}Outflow temperature of a $\unit[300]{kW}$
cooling system. Panel (a) shows the performance of the outflow temperature.
At about 4~p.m. the system reaches its control limits, because the
ambient temperature (b) exceeds significantly the set-point ($20^{\circ}\mathrm{C}$)
of the outflow temperature. The high eigenfrequencies of the hydraulic
system become visible from about 2:30~p.m. on.}
\end{figure}
A second example is given in Figure~\ref{fig:Outflow-temperature}a.
Here the given data represents the outflow temperature of a cooling
system with about $\unit[300]{kW}$ power. It consists of a heat exchanger
outside the building and a secondary water loop inside from which
to draw the heat. A sprinkler system improves the performance of the
outside heat exchanger, so that outflow temperatures below the ambient
temperature become possible. Panel (b) illustrates the ambient temperature
near the external air heat exchanger. The impact of the suns radiation
becomes visible in a signal amplitude with a very long period of about
$T_{\mathrm{sun}}\approx\unit[80]{min}$. In terms of period lengths
the high eigenfrequencies of the hydraulic system become visible at
the bottom, approximately within a range of $\unit[10]{min}$ to $\unit[20]{min}$,
when the control loop reaches its limit.

Finally, the following code shows how to invert the frequencies to
periods. Note, that a re-sampling of the matrix must be performed,
because the unequally spaced vector $1/f$ now maps to the corresponding
spectrum. This task is done by the \code{plot.fft()} function, which
also plots objects with the attribute \code{mode = \textquotedbl{}waterfall\textquotedbl{}}. 

\begin{CodeChunk}
\begin{Code}
# calculating the waterfall diagram
wf <- waterfall(temp,time,nf = 3)

# change to periods = 1/f
wf$fx <- 1/wf$fx
# avoid "infinite" values
wf$fx[1] <- 2*max(wf$fx[!is.infinite(wf$fx)],na.rm = T)

# plot graph with hour-scale on the x-axis
wf$x <- wf$x/60

plot(wf,xlim=c(11,17),ylim=c(3,80),zlim=c(0,0.5))
\end{Code}

\end{CodeChunk}

\subsection[Fragmented and irregularly sampled data]{\label{subsec:Gapped-and-irregularly-sampled-data}Fragmented and
irregularly sampled data}

Sometimes the nature of measurement prevents an equally spaced sampling.
Observations of astrophysical bodies like the sun or the moon are
examples of objects only visible at a certain time of day. Concerning
a definite property of these bodies, the corresponding time series
would be fragmented, what makes a spectral analysis with a standard
Fourier transform almost impossible. A nice application example can
be found in ``The Lick Planet Search: Detectability and Mass Thresholds''
by \citet{cumming1999thelick}. Another issue would be the measurement
with randomly sampled data, for instance the occurrence of an event
like the appearance of a malfunctioning product in a production lane,
or even a measurement with a high jitter.

The time instances $t_{n}$ when a sample is taken now depend on the
sample number $n$. This means, the time interval
\[
\Delta t=t_{n-1}-t{}_{n}\neq\mathrm{const.}
\]
is not constant anymore, so the sampling time becomes an additional
data vector. This enables one advantage in case of randomly sampled
data. Given a stationary process, it becomes possible to estimate
signal amplitudes for frequencies which are in the order of $f_{\max}\approx\mathcal{O}\left(\min(\Delta t)^{-1}\right)$,
with the minimal distance between two points. In fact, this is related
to an average Nyquist frequency $\overline{f}_{\mathrm{c}}$, where
\begin{equation}
f_{\max}>\frac{1}{2}\frac{N}{\max(t)-\min(t)}=\overline{f}_{\mathrm{c}},
\end{equation}
which can be defined by the time range of the time series and the
total number of samples.

Lets introduce the Lomb-Scargle periodogram as a consistent estimator
for spectral properties of discrete data. The estimation in that sense
is not a mathematical transform like the Fourier transform. The frequencies
of interest can be selected freely, so the whole method analyses a
data set against amplitude and phase. The associated mathematical
details of this statistical approach can be found in the papers by
\citet{mathias2004algorithms}, \citet{hocke2009gapfilling}, \citet{scargle1982studies}
and \citet{lomb1976leastsquares}, who originally developed the method.

In principle, the concept of the procedure is about a least square
optimization of the parameters $A_{i}$ and $B_{i}$ for the model
function
\begin{equation}
y(t)=\sum_{i}A_{i}\cdot\cos\left(\omega_{i}\cdot(t-\tau_{i})\right)+B_{i}\cdot\sin\left(\omega_{i}\cdot(t-\tau_{i})\right),\label{eq:LS-Modelfunction}
\end{equation}
which is fitted to the data. Remember, the introductory chapter~\ref{subsec:Frequency-analysis}
shows a slightly different approach in conjunction with the trigonometric
identities \eqref{eq:trig_ident_1} and \eqref{eq:trig_ident_2},
which also become the key component here. While the traditional calculation
of a Newton algorithm takes several iteration steps until the result
converges, the optimal parameters can alternatively be estimated by
one single matrix inversion, see \citet{mathias2004algorithms}. This
can be done only if the orthogonal property of the trigonometric model
function is taken into account. The mathematical proof can be found
in ``A generalized inverse for matrices'' by \citet{penrose1955ageneralized}.
As a necessary condition to obtain an optimal result for $A_{i}$
and $B_{i}$, the relation
\begin{equation}
\sum_{n=1}^{N}\cos\left(\omega_{i}\cdot(t_{n}-\tau_{i})\right)\cdot\sin\left(\omega_{i}\cdot(t_{n}-\tau_{i})\right)=0\label{eq:tau_1}
\end{equation}
must be fulfilled. Given that, it can be shown that the error of the
resulting fit against the normal quadrature approach from chapter~\ref{subsec:Frequency-analysis}
is minimized. From \eqref{eq:tau_1} the calculation of $\tau_{i}$
can be derived
\begin{equation}
\tan\left(2\omega_{i}\cdot\tau_{i}\right)=\frac{\sum_{n}\sin(2\omega_{i}\cdot t_{n})}{\sum_{n}\cos(2\omega_{i}\cdot t_{n})}.\label{eq:tau-2}
\end{equation}

Next to that, the parameters $R$, $I$, $C$ and $S$ will be defined
as follows:
\begin{align}
R\left(\omega_{i}\right) & \equiv\sum_{n}y_{n}\cos\left(\omega_{i}\cdot\left(t_{n}-\tau_{i}\right)\right),\\
I\left(\omega_{i}\right) & \equiv\sum_{n}y_{n}\sin\left(\omega_{i}\cdot\left(t_{n}-\tau_{i}\right)\right),\\
C\left(\omega_{i}\right) & \equiv\sum_{n}\cos^{2}\left(\omega_{i}\cdot\left(t_{n}-\tau_{i}\right)\right)\\
\text{and }S(\omega_{i}) & \equiv\sum_{n}\sin^{2}\left(\omega_{i}\cdot\left(t_{n}-\tau_{i}\right)\right).
\end{align}

Finally, the power spectral density $P$, the absolute amplitude $A$
and the phase $\varphi$ 
\begin{align}
P(\omega_{i}) & =\frac{1}{2\sigma^{2}}\left(\frac{R(\omega_{i})^{2}}{C(\omega_{i})}+\frac{I(\omega_{i})^{2}}{S(\omega_{i})}\right),\\
A(\omega_{i}) & =\sqrt{\frac{2}{N}\left(\frac{R(\omega_{i})^{2}}{C(\omega_{i})}+\frac{I(\omega_{i})^{2}}{S(\omega_{i})}\right),}\\
\varphi(\omega_{i}) & =-\left(\tan^{-1}\left(\frac{I}{R}\right)+\omega_{i}\tau_{i}\right)
\end{align}
can be calculated from these coefficients. Hereby $\sigma$ describes
the standard deviation of the discrete data vector $y_{n}$. In comparison
to equation \eqref{eq:1_QDT_error2} from chapter~\ref{subsec:Frequency-analysis},
the equations above work quite similar. But now the modified function
argument \textendash{} extended by $\tau_{i}$, the correction terms
$C(\omega_{i})$ and $S(\omega_{i})$ \textendash{} leads to the optimal
least square fit solution. 

In the limit of an infinite number of equally spaced samples, the
Lomb-Scargle estimator converges to the Fourier transform, so it becomes
a consistent estimator \citep{mathias2004algorithms}. The last statement
is very important, because with an increasing number of samples the
error reduces until the result converges to the true amplitude and
phase.

To value the significance of the estimated amplitude the ``false
alarm probability'' (FAP)
\begin{equation}
p\left(P<P_{0}\right)=1-\left(1-\e^{-P}\right)^{M}\label{eq:FAP}
\end{equation}
is defined. Here the probability $p$, that there is no other larger
amplitude $P$ than $P_{0}$, is expressed in terms of the exponential
function above. For small values of $p$ the approximation 
\begin{equation}
p\left(P<P_{0}\right)\approx\left.M\cdot\e^{-P}\right|_{p\ll1}
\end{equation}
can be used. The free parameter $M$ counts the \emph{independent}
frequencies in the data. These are difficult to measure \emph{a priori},
but it turns out that with $M=N/2$ sufficient results can be achieved.
A brief discussion on this issue can be read in the work of \citet[Cap. 6.2]{townsend2010fastcalculation},
\citet{zechmeister2009thegeneralised}and \citet{cumming1999thelick}. 

\paragraph{Implementation}

The calculation of the phase is tricky. Invoking the \code{arctan2()}
function is mandatory. Nevertheless, to prevent errors the phase $\varphi$
is calculated by
\begin{equation}
\varphi=-\left(\text{arctan2}\left(\frac{I}{N},\frac{R}{N}\right)+\omega_{i}\tau_{i}\right),
\end{equation}
which take the normalized components $I$ and $R$.

Next to that, if the suggestions made by \citet{townsend2010fastcalculation}
are taken into account then the above algorithm can be shortened.
The problem is that the straight forward implementation runs \emph{two}
times over the whole data set, while calculating $\tau$ prior the
rest of the parameters. A keen refactoring of the equations above
will minimize the computational cost. For a certain frequency $\omega_{i}$
the amplitudes 
\begin{equation}
A(\omega_{i})=\sqrt{\frac{2}{N}\left(\frac{\left(c_{\tau}\cdot XC+s_{\tau}\cdot XS\right)^{2}}{c_{\tau}^{2}\cdot CC+2c_{\tau}\cdot s_{\tau}\cdot CS+s_{\tau}^{2}\cdot SS}+\frac{\left(c_{\tau}\cdot XS-s_{\tau}\cdot XC\right)^{2}}{c_{\tau}^{2}\cdot SS+2c_{\tau}\cdot s_{\tau}\cdot CS+s_{\tau}^{2}\cdot CC}\right)}
\end{equation}
can be calculated out of the parameters
\begin{alignat}{2}
XC & =\sum_{j}y_{j}\cdot\cos\left(\omega_{i}\cdot t_{j}\right), & XS & =\sum_{j}y_{j}\cdot\sin\left(\omega_{i}\cdot t_{j}\right),\\
CC & =\sum_{j}\cos^{2}\left(\omega_{i}\cdot t_{j}\right), & SS & =\sum_{j}\sin^{2}\left(\omega_{i}\cdot t_{j}\right),\\
CS & =\sum_{j}\cos\left(\omega_{i}\cdot t_{j}\right)\cdot\sin\left(\omega_{i}\cdot t_{j}\right) & \tau_{\mathrm{L}} & =\tan^{-1}\left(\frac{2\cdot CS}{2\cdot CC-1}\right)\\
c_{\tau} & =\cos\left(\omega\cdot\tau_{\mathrm{L}}\right) & s_{\tau} & =\sin\left(\omega_{i}\cdot\tau_{\mathrm{L}}\right)
\end{alignat}
in one single loop. A vectorized \proglang{R}-code example is given
below. Here the data and the corresponding frequencies form the matrix
\code{omega.x}, which is processed successively. 

\begin{CodeChunk}
\begin{Code}
# put everything in a matrix. One frequency per column
# x correspondes to the time
# y_ corresponds to the mean-free data vector
omega.x <- x %*% t(omega)
co <- cos(omega.x); si <- sin(omega.x)
co2 <- co^2; si2 <- 1 - co2 # use trigonometric identities
si <- sqrt(1-co2)

  CC <- colSums(co2); SS <- colSums(si2)
 YCS <- colSums(y_ * co); YSS <- colSums(y_ * si)
  CS <- colSums(si*co)
tauL <- atan(2 * CS / (2*CC - 1) / (2*omega))
 ct  <- cos(omega * tauL); st   <- sin(omega * tauL)
 ct2 <- ct^2; st2 <- 1-ct2

ctstCS <- 2*ct*st*CS
R <- (ct*YCS + st*YSS); I<-(ct*YSS - st*YCS)

l <- sqrt(R^2+I^2) # the trick to reduce numeric error in atan2()
A <- sqrt( 2/nt* ( R^2/(ct2 * CC + ctstCS + st2 * SS) 
         + I^2/(ct2 * SS - ctstCS + st2 * CC) ) )
phi <- - omega*tauL - atan2((I/l),(R/l))
\end{Code}
\end{CodeChunk}

The \proglang{R}-code illustrates how to use the specialties of the
language. Instead of programming a \code{for} statement the faster
vector and matrix operations of \proglang{R} can be invoked.

\paragraph{Application}

\begin{figure}
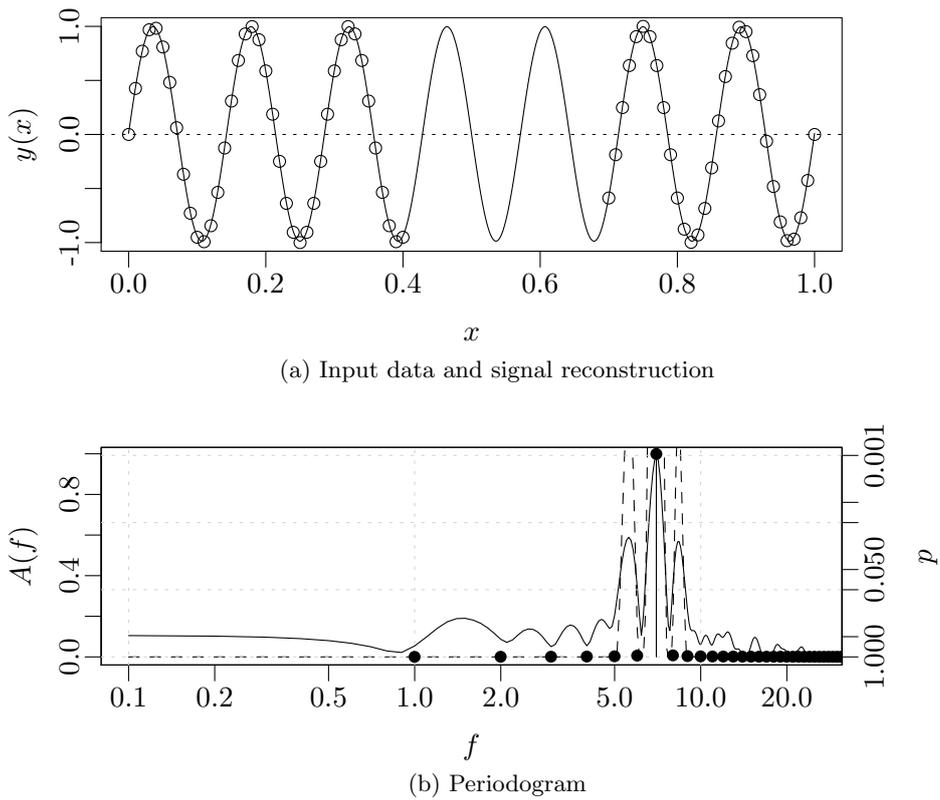

\begin{centering}
\subfloat[Input data and signal reconstruction]{\begin{centering}
\input{Pictures/Example10_lomb_simple.tex}
\par\end{centering}
}
\par\end{centering}
\begin{centering}
\subfloat[Periodogram]{\begin{centering}
\input{Pictures/Example10_lomb_simple_spec.tex}
\par\end{centering}
}
\par\end{centering}
\caption{\label{fig:Simple-Lomb-Scargle-periodogram}Simple Lomb-Scargle periodogram.
The open symbols in the top panel (a) display the input signal. The
solid line shows the reconstruction out of the spectrum. Panel (b)
focuses on the periodogram. Here, the solid line corresponds to the
spectral amplitude and the black points indicate the results of a
DFT, as if there where no gap in the data. The dashed line represents
the false alarm probability $p$. Note, the value of 1 is located
at the bottom, whereas smaller values are displayed above.}
\end{figure}
To show the power of the method let us first assume a simple example.
The function
\begin{equation}
y(x)=\sin\left(2\pi\cdot7\cdot x\right)
\end{equation}
is going to be sampled with $N=101$ equally spaced samples. Remember
the last point of the data vector equals the first point of the period
of the signal. As discussed in the application section of chapter~\ref{subsec:Bandlimitation},
the violation of the periodicity would lead to an error in the result
of the Fourier transform. In addition to that, approximately 30\%
of the data where deleted.

Subsequently, Figure~\ref{fig:Simple-Lomb-Scargle-periodogram} shows
the fragmented signal and the corresponding periodogram, which is
calculated with the \code{spec.lomb(x, y, f)} function. 

\begin{CodeChunk}
\begin{Code}
x <- seq(0,1,by=0.01)
x.new <- seq(0,1,by=1e-3) 
 yorg <- function(x) return(sin(2*pi*7*x))

cond <- !(x > 0.4 & x < 0.7) # create the gap
   x <- x[cond]; y <- yorg(x)

 l <- spec.lomb(x=x, y=y, f=seq(0,25,by=0.1))
lf <- filter.lomb(l,newx=x.new,phase="lin",threshold=3)
\end{Code}
\end{CodeChunk}

The code above shows how Figure~\ref{fig:Simple-Lomb-Scargle-periodogram}
can be created. Note that the frequency vector \code{f} contains
250 different frequencies to analyze the data. Compared to the single
sided spectrum of the \emph{unfragmented} data's analytic signal representation
(black dots in fig.~\ref{fig:Simple-Lomb-Scargle-periodogram}b)
the Lomb-Scargle periodogram produces large side band amplitudes to
the left and right of the main amplitude. This is quite typical if
the data is non-uniformly sampled or even patchy. The dashed line
represents the false alarm probability, which tends to zero if the
corresponding amplitude is significant.

The spectral package also provides a \code{filter.lomb()} function,
with which the most significant amplitudes can be extracted for reconstruction.
Provided the continuous sampling vector \code{x.new}, the result
is a new data set in which the remaining gaps are filled.

\begin{figure}
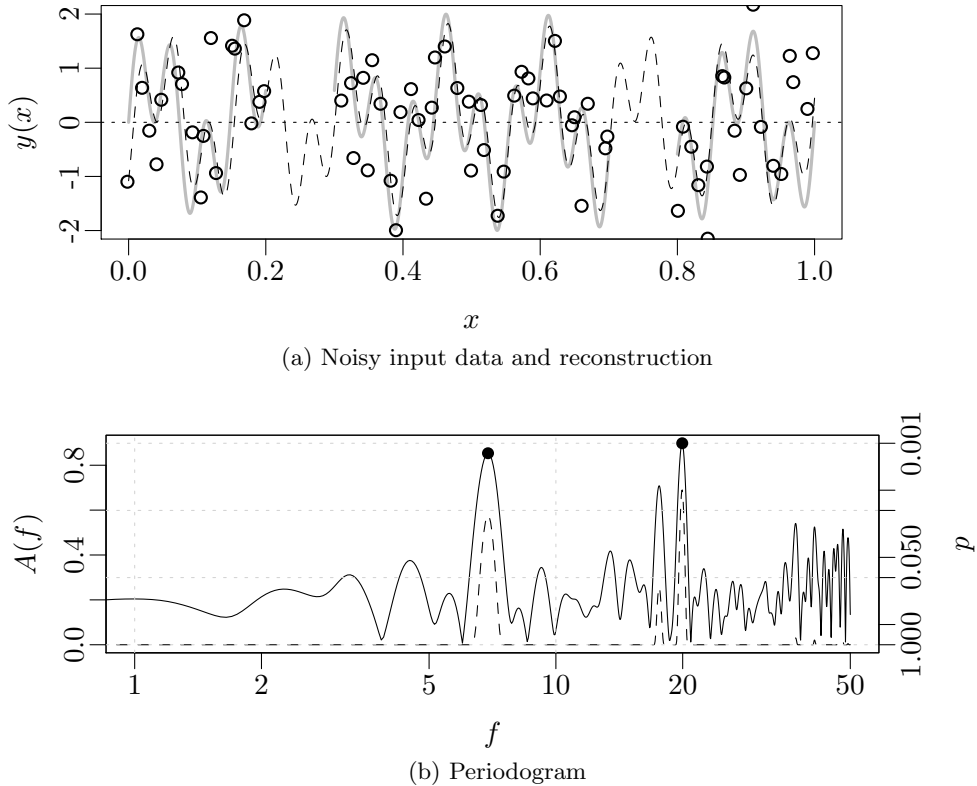

\begin{centering}
\subfloat[Noisy input data and reconstruction]{\begin{centering}
\input{Pictures/Example10_lomb_signal.tex}
\par\end{centering}
}
\par\end{centering}
\centering{}\subfloat[Periodogram]{\centering{}\input{Pictures/Example10_lomb_spec.tex}}\caption{\label{fig:Randomly-sampled-data}Randomly sampled data with noise
and gaps. The bold gray line in (a) is the discontinuous signal. The
circles represent the sampling points, which are overlaid with some
strong noise and jitter. The dashed line represents the reconstruction.
In (b) the corresponding Lomb-Scargle periodogram is given. The black
line is the amplitude spectrum and the dashed line indicates $p$.
The black points show the amplitude maxima, which the \protect\code{filter.lomb()}
function chose for reconstruction.}
\end{figure}
A more complex example is given in Figure~\ref{fig:Randomly-sampled-data}.
Here the function
\begin{equation}
y\left(x\right)=\sin\left(2\pi\cdot(x+\mathcal{N}(0,\Delta x/4))\right)+\sin\left(2\pi\cdot20\cdot(x+\mathcal{N}(0,\Delta x/4))\right)+\mathcal{N}(0,1)
\end{equation}
consists of two $\sin$-terms, the location of sampling jitters as
well as the amplitude itself. In the simple test case here, normal
distributed noise is assumed and the resulting data vectors $x_{n}$
and $y_{n}$ are $N=101$ elements in length. However, for reconstruction
the most significant amplitude values are selected again, so that
the dashed line in panel (a) will fill the gaps correspondingly.

For further reading about non-uniformly sampled data the adaptive
approach introduced by \citet{stoica2009spectral} is recommended.
This method has a better signal to noise ratio, but will have much
more computational cost.

\section{Conclusion}

Today, spectral analysis is one of the key methods in data processing.
It helps to filter signals, to accelerate calculations or to identify
processes. While the complex mathematics is already documented in
the given references and the common text books, the present work focuses
on the \emph{application} of the basic spectral methods in one dimension.
Hereby working principles of the introduced methods and their possible
misunderstandings are of interest. Small examples support the explanations
about the work of operation and show the ``do's and don'ts''. The
idea of the present work is to show how the complex underlying theory
of spectral methods can be applied to solve real world tasks. A dedicated
tool supporting this intent is the \pkg{spectral} package in the
statistic environment \proglang{R}, which was developed contemporaneous
to this paper. Further work will compose part II of this paper, which
will cover the application of spectral methods in two and more dimensions.

\newpage{}

\bibliography{guide_to_spectral_methods}

\newpage{}

\appendix

\section*{Appendix}

\section[The origin of fs/2]{\label{sec:The-origin}The origin of $\mathbf{\mathit{\boldsymbol{f}}_{s}/2}$}

\setcounter{equation}{0}To derive the band limitation we have to
assume a function $y(t)$ for which the Fourier transform $Y(\omega)$
exist. There might be a natural maximum frequency $\omega_{\mathrm{max}}\leq\omega_{\mathrm{B}}$
in the signal, so the integration boundaries of the back transform
\begin{equation}
y(t)=\frac{1}{2\pi}\int_{-\omega_{\mathrm{B}}}^{\omega_{\mathrm{B}}}Y(\omega)\cdot\e^{\im\omega t}\diff\omega
\end{equation}
can be limited to $\left|\omega_{\mathrm{B}}\right|$. In the next
step this function is sampled at discrete time instances $t=n\cdot T_{\mathrm{s}}$,
with a sampling frequency $f_{\mathrm{s}}=1/T_{\mathrm{s}}$. It follows
\begin{align}
y(n\cdot T_{\mathrm{s}}) & =\frac{1}{2\pi}\int_{-\omega_{\mathrm{B}}}^{\omega_{\mathrm{B}}}Y(\omega)\cdot\e^{\im\omega\frac{n}{f_{\mathrm{s}}}}\diff\omega,\nonumber \\
 & =\frac{1}{2\pi}\int_{-2\pi f_{\mathrm{B}}}^{2\pi f_{\mathrm{B}}}Y(\omega)\cdot\e^{\im2\pi\frac{f}{f_{\mathrm{s}}}n}\diff\omega.\label{eq:APP_fs-2}
\end{align}
By substituting $\diff\omega\leftrightarrow2\pi\diff f$ the modified
equation \eqref{eq:APP_fs-2} can be written as
\begin{equation}
y(n\cdot T_{\mathrm{s}})=\int_{-f_{\mathrm{B}}}^{f_{\mathrm{B}}}Y(2\pi f)\cdot\e^{\im2\pi\frac{f}{f_{\mathrm{s}}}n}\diff f.
\end{equation}
The pointer $\e^{\im\varphi}$ is periodic with respect to $2\pi$,
so the span of $\Delta f/f_{\mathrm{s}}$ must not exceed unity to
reject duplication. It follows
\[
\frac{\Delta f}{f_{\mathrm{s}}}\leq1
\]
so that the maximum integration boundaries
\begin{equation}
f_{\mathrm{B}}\leq0.5\cdot f_{\mathrm{s}}
\end{equation}
can be concluded in the first instance.

Due to the requirement that the \emph{frequency span} must be less
than the sampling frequency, the under-sampling of functions becomes
possible. This changes the meaning of a typical Nyquist limit to a
limitation of bandwidth which is necessary to obtain uniqueness. \newpage{}

\section[Properties of the Fourier operator]{\label{sec:Properties-of_FourierTransform}Properties of the Fourier
operator}

\setcounter{equation}{0}In the following the properties of the Fourier
operator are listed. 
\begin{description}
\item [{Linearity}] 
\begin{equation}
\mathcal{F}\left(a\cdot f(t)+b\cdot g(t)\right)=a\cdot F(\omega)+b\cdot G(\omega)
\end{equation}
\item [{Scaling}] 
\begin{equation}
\mathcal{F}\left(f(t/a)\right)=|a|\cdot F(a\cdot\omega)\qquad a\in\mathbb{R}\text{ and }a\neq0
\end{equation}
\item [{Translation}] 
\begin{equation}
\mathcal{F}\left(f(a\cdot t+b)\right)=\frac{1}{a}\e^{\im\omega\cdot b/a}F(\omega/a)\qquad a,b\in\mathbb{R}\text{ and }a\neq0
\end{equation}
\item [{Damping}] 
\begin{align}
\mathcal{F}\left(\e^{\im b\cdot t}f(a\cdot t)\right) & =\frac{1}{a}F\left((\omega-b)/a\right)\qquad a\in\mathbb{R}^{+}\label{eq:APP_Fourier_Prop_Damping}
\end{align}
\item [{n\textsuperscript{th}~Derivation}] the function $f(t)$ must
be absolutely integrable and $\lim_{t\rightarrow\pm\infty}f(t)=0$
\begin{equation}
\mathcal{F}\left(f^{(n)}(t)\right)=(\im\omega)^{n}\mathcal{F}\left(f(t)\right)
\end{equation}
\item [{Integration}] Only if $\int_{-\infty}^{\infty}f(t)\diff t=0$ then
\begin{equation}
\mathcal{F}\left(\int_{-\infty}^{t}f(t)\right)=\frac{1}{\im\omega}F(\omega).
\end{equation}
\item [{Folding}] Only if the integrals $\int_{-\infty}^{\infty}\left|f(t)\right|^{2}\diff t$
and $\int_{-\infty}^{\infty}\left|g(t)\right|^{2}\diff t$ are existent,
the two sided folding of $f(t)$ and $g(t)$
\begin{equation}
f(t)*g(t)=\int_{-\infty}^{\infty}f(\tau)g(t-\tau)\diff\tau
\end{equation}
can be calculated with the help of the Fourier transform. In the frequency
domain the folding operation reduces to a multiplication. This is
expressed as follows: 
\begin{equation}
\mathcal{F}\left(f(t)*g(t)\right)=\mathcal{F}\left(f(t)\right)\cdot\mathcal{F}\left(g(t)\right).
\end{equation}
\newpage{}
\end{description}

\section[Derivation of the discrete Fourier-transform ]{\label{sec:Derivation_Fourier}Derivation of the discrete Fourier-transform}

\setcounter{equation}{0}The derivation of the discrete Fourier transform
(DFT) is described below in detail. The already introduced and sketched
out properties of the DFT will become clear with the discrete formulation
of the Fourier transform. We will see that the spectral information
becomes periodic and in the 1D case redundant.

First, let $y$ be a real-valued function of time, with
\begin{equation}
y=x\left(t\right):\left\{ t\in\mathbb{R}\to x\left(t\right)\in\mathbb{R}\right.,
\end{equation}
which is going to be sampled later on. For that the Dirac-function
must be defined as a distribution as follows: 
\begin{align}
\delta\left(t\right) & =\left\{ \begin{array}{c}
\infty,\ t=0\\
0,\ t\ne0
\end{array}\right.,\\
\text{with }1 & =\int_{-\infty}^{\infty}\delta\left(t\right)\diff{t}.\label{eq:FT_Distribution}
\end{align}
The Dirac pulse is defined as $\delta(t\neq0)=0$ for all values of
$t$ except $t=0$. At this point the Dirac function returns an infinite
value, $\delta(t=0)\rightarrow\infty$. However, the behavior \eqref{eq:FT_Distribution}
turns the Dirac function into a distribution. The finite value of
this integral enables subsequent calculations to sample a point from
the continuous function $y=x(t)$. This is called ``sifting'' property
and can be written as 
\begin{equation}
\int_{-\infty}^{\infty}\delta\left(t\right)\phi\left(t\right)\diff{t}=\phi\left(0\right).\label{eq:A4_Ausblend}
\end{equation}
In conjunction with the Fourier transform \eqref{eq:A4_Ausblend}
leads to a necessary simplification for the derivation of the DFT.

Next to that the continuous function $x(t)$ is going to be sampled
with the help of the sampling function $\sum_{n=-\infty}^{\infty}\delta(t-nT_{\mathrm{s}})$.
The sampling series
\begin{align}
A\left(t\right) & =x\left(t\right)\sum_{-\infty}^{\infty}\delta\left(t-nT_{\mathrm{s}}\right)\nonumber \\
 & =\sum_{-\infty}^{\infty}x\left(nT_{\mathrm{s}}\right)\delta\left(t-nT_{\mathrm{s}}\right)\label{eq:App_sampling_series}
\end{align}

evaluates $x(t)$ at $n$ reference points in time by multiplying
the sampling function to $x(t)$. Because of the properties of the
Dirac pulse the sum \eqref{eq:App_sampling_series} converges to the
value $A(t)\rightarrow\infty$.

\begin{figure}
\includegraphics[width=6cm]{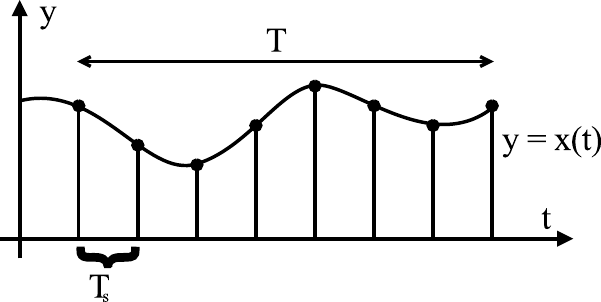}

\caption[Sampled signal]{Common band limited and real-valued function. The reference points
are equally spaced with the distance $T_{s}$.}
\end{figure}

In the next step, the sampling series $A(t)$ is going to be inserted
in the Fourier integral
\begin{equation}
X\left(\omega\right)=\int_{-\infty}^{\infty}x\left(t\right)\e^{-\im\omega\cdot t}\diff{t}.\label{eq:App_fourier_of_x}
\end{equation}
This transforms the function $x(t)$ from time or spatial domain into
frequency domain, expressed by the complex valued spectrum $X(\omega)$.
The capital letter $X(\omega)$ supports the result, which is a bijective
mapping between $x(t)\leftrightarrow X(\omega)$. At this point both
functions $x(t)$ and its spectrum $X(\omega)$ are still continuous.
For real signals $x(t)$ is causal, real-valued and finite in time,
with
\[
0<t\leq T.
\]
This means the signal is defined in a certain range of positive $t$
and stops existing after a maximum time $T$. With this behavior the
boundaries of integration in \eqref{eq:App_fourier_of_x} change.
The result is the discrete spectrum of the Fourier series\foreignlanguage{german}{
}
\begin{equation}
X\left(m\right)=\frac{1}{T}\int_{0}^{T}x\left(t\right)\e^{-\im m\cdot\omega_{0}\cdot t}\diff{t}\label{eq:App_discont_spec}
\end{equation}
with $m\in\mathbb{N}$. Inserting the sampling series $A(t)$ in \eqref{eq:App_discont_spec},
now discrete points enter the discrete spectrum: 
\begin{equation}
X\left(m\right)=\frac{1}{T}\int_{0}^{T}{\underbrace{\sum_{n=-\infty}^{\infty}x\left(nT_{\mathrm{s}}\right)\delta\left(t-nT_{\mathrm{s}}\right)}_{A\left(t\right)}\e^{-\im m\cdot\omega_{0}\cdot t}}\diff{t}.
\end{equation}

The distribution character \eqref{eq:FT_Distribution} and the sifting
property of the Dirac function help to rearrange the equation into
the DFT form \citet{bronvstejn2001taschenbuch,hoffmann1998signalanalyse},
\begin{equation}
X\left(m\right)=\frac{1}{T}\sum_{n=-\infty}^{\infty}x\left(nT_{\mathrm{s}}\right)\e^{-\im m\cdot\omega_{0}\cdot nT_{\mathrm{s}}}.
\end{equation}
For a finite data set of $N$ equidistant data points, $\omega_{0}=2\pi/T$
and a total length of $T=N\cdot T_{s}$, this can be written to: 
\begin{eqnarray}
X\left(m\right) & = & \frac{1}{NT_{\mathrm{s}}}\sum_{n=0}^{N-1}x\left(nT_{\mathrm{s}}\right)\e^{-\im m\cdot2\pi\frac{nT_{\mathrm{s}}}{N\cdot T_{s}}}\nonumber \\
 & = & \frac{1}{N}\sum_{n=0}^{N-1}x\left(n\right)\e^{-\im m\cdot2\pi\cdot\frac{n}{N}}.\label{eq:App_DFT}
\end{eqnarray}
Here $T_{s}$ denotes the inverse sampling frequency $T_{s}=f_{s}^{-1}$
and $m\in\mathbb{N}$ measures the normalized discrete frequencies
on which the spectrum is being evaluated.

\begin{figure}
\includegraphics[width=6cm]{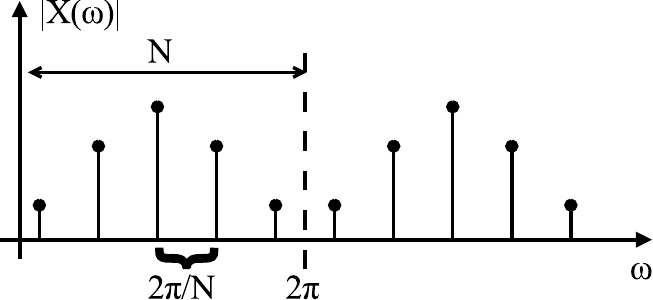}

\caption{DFT with periodic spectrum}
\end{figure}

This discrete form of the Fourier transform provides several properties
which define the spectrum $X(m)$.
\begin{enumerate}
\item The spectrum is periodic with respect to $\frac{2\pi}{N}$. This means
at least $m=N$ frequencies are necessary to describe a unique mapping
of $x(n)\leftrightarrow X(m)$
\item $X(\omega)$ can be interpreted as a line spectrum because it is only
evaluated at $\omega=\frac{2\pi}{N}m$ discrete frequencies.
\end{enumerate}
The main feature, which arises from \eqref{eq:App_DFT}, is the periodic
spectrum. In consequence of that, frequencies which are larger than
half of the sampling frequency are projected into the lower half of
the spectrum. In case of a signal which is not band limited, an explicit
mapping between a certain frequency and the underlying physical process
is not possible. Although, an additional information is provided which
enables a fold back. An example is given in chapter \ref{subsec:Bandlimitation}.

\selectlanguage{german}%
\newpage{}
\selectlanguage{english}%

\section[Derivation of the analytic signal and the Hilbert transform]{\label{sec:Derivation_Hilbert}Derivation of the analytic signal
and the Hilbert transform}

\selectlanguage{german}%
\setcounter{equation}{0}\foreignlanguage{english}{Given a real-valued
time-dependent signal $s(t)$, the Fourier transform can be calculated
by
\begin{equation}
S\left(\omega\right)=\frac{1}{\sqrt{2\pi}}\int_{-\infty}^{\infty}s(t)\cdot\e^{-\im\omega\cdot t}\diff t.
\end{equation}
The question is: What would happen if a function $a(t)$ maps to a
one-sided spectrum of $s(t)$ consisting only of the positive frequencies
in $S(\omega)$? In terms of a back transform this can be written
in the form: 
\begin{equation}
a(t)=\frac{2}{\sqrt{2\pi}}\int_{0}^{\infty}A\left(\omega\right)\e^{\im\omega\cdot t}\diff\omega.\label{eq:ApB_Backtransform}
\end{equation}
Supplying the previous spectrum $S\left(\omega\right)=A\left(\omega\right)$
to \eqref{eq:ApB_Backtransform} will give
\begin{align}
a(t) & =2\cdot\left(\frac{1}{\sqrt{2\pi}}\right)^{2}\int_{0}^{\infty}\left(\int_{-\infty}^{\infty}s(t')\cdot\e^{-\im\omega\cdot t'}\diff t'\right)\e^{\im\omega\cdot t}\diff\omega\\
 & =\frac{1}{\pi}\int_{-\infty}^{\infty}\int_{0}^{\infty}s(t')\e^{\im\omega\cdot(t-t')}\diff t'\diff\omega.
\end{align}
With the identity $\int_{0}^{\infty}\e^{\im\omega\cdot t}\diff\omega=\pi\cdot\delta(t)+\frac{\im}{t}$
the analytic signal $a(t)$ can be determined as follows:
\begin{align}
a(t) & =\frac{1}{\pi}\int_{-\infty}^{\infty}s(t')\left(\pi\cdot\delta(t-t')+\frac{\im}{t-t'}\right)\diff t'\nonumber \\
 & =s(t)+\underset{\mathcal{H}\left(s(t')\right)}{\underbrace{\frac{\im}{\pi}\int_{-\infty}^{\infty}\frac{s(t')}{t-t'}\diff t'}}.\label{eq:App_analy_Sig}
\end{align}
}

\selectlanguage{english}%
Equation \eqref{eq:App_analy_Sig} shows the analytic signal definition
for $a(t)$. It consists of the real-valued signal $s(t)$ and its
complex extension in form of the Hilbert transform $\im\mathcal{H}\left(s(t)\right)$. 

The convolution
\begin{equation}
\mathcal{H}\left(s(t)\right)=\frac{1}{\pi}\int_{-\infty}^{\infty}\frac{s(t')}{t-t'}\diff t'
\end{equation}
can be calculated in the frequency domain by applying the Fourier
transform to the Hilbert transform \citet[Chap. 3]{duoandikoetxeazuazo2001fourier}
\begin{equation}
\mathcal{F}\left(\mathcal{H}\left(s(t)\right)\right)=-\im\cdot\mathrm{sign}(\omega)\cdot\mathcal{F}\left(s(t)\right).
\end{equation}
Now, the reader sees that $\mathcal{H}\left(f(t)\right)$ shifts the
phase by 90 degrees. In that sense the Hilbert transform represents
an ideal phase shifter. 

The analytic signal and the Hilbert transform have certain properties,
which are discussed in detail in Chapter \ref{subsec:The-analytic-signal}
and in \citet{cohen1995timefrequency,duoandikoetxeazuazo2001fourier}.

\paragraph*{Note:}

The introduced identity
\[
\int_{0}^{\infty}\e^{\im\omega\cdot t}\diff\omega=\pi\cdot\delta(t)+\frac{\im}{t}
\]
is not so obvious to resolve. The following derivation will explain
this in detail. First, the exponent of the left hand side is extended
with $-\varepsilon\cdot\omega$, which is then taken in the limit
to zero.
\[
\int_{0}^{\infty}\e^{\im\omega\cdot t}\diff z=\lim_{\varepsilon\rightarrow0}\int_{0}^{\infty}\e^{\im\omega\cdot t-\varepsilon\cdot\omega}\diff z
\]
This is now being successively reduced:
\begin{eqnarray*}
\lim_{\varepsilon\rightarrow0}\int_{0}^{\infty}\e^{\im\omega\cdot t-\varepsilon\cdot\omega}\diff\omega & = & \lim_{\varepsilon\rightarrow0}\int_{0}^{\infty}\e^{-\omega\cdot(-\im t+\varepsilon)}\diff\omega,\\
 & = & \lim_{\varepsilon\rightarrow0}\,\frac{1}{-\im t+\varepsilon}\cdot\left.-\e^{-\omega\cdot(-\im t+\varepsilon)}\right|_{0}^{\infty},\\
 & = & \lim_{\varepsilon\rightarrow0}\,\frac{1}{-\im t+\varepsilon}\cdot(0+1),\\
 & = & \lim_{\varepsilon\rightarrow0}\,\frac{1}{-\im t+\varepsilon}=\lim_{\varepsilon\rightarrow0}\,\frac{\im}{t+\im\varepsilon}.
\end{eqnarray*}

The last term is now expanded by the conjugate denominator $t-\im\varepsilon$.
It follows:
\begin{eqnarray*}
\lim_{\varepsilon\rightarrow0}\,\frac{\im}{t+\im\varepsilon} & = & \lim_{\varepsilon\rightarrow0}\,\underset{\rightarrow\pi\cdot\delta(t)}{\underbrace{\frac{\varepsilon}{t^{2}+\varepsilon^{2}}}}+\frac{\im t}{t^{2}+\varepsilon^{2}},\\
 & = & \pi\cdot\delta(t)+\frac{\im}{t}.
\end{eqnarray*}
The last term expresses the Poisson kernel, which is the solution
of the Laplace equation in the upper half plane \citep{stein1975introduction}.
The expression 
\[
\lim_{\varepsilon\rightarrow0}\,\frac{1}{\pi}\frac{\varepsilon}{t^{2}+\varepsilon^{2}}=\delta(t)
\]
also equals to the Lorentz distribution and converges to the Dirac
function in the limit of $\varepsilon\rightarrow0$.
\end{document}